\documentclass[iop,apj,emulateapj-rtx4]{emulateapj}

\usepackage{graphicx,latexsym,multirow}
\usepackage{color}
\usepackage{natbib}
\usepackage{rotating,ams}
\usepackage{appendix}
\usepackage[ps2pdf,linktocpage]{hyperref}
\hypersetup{%
  colorlinks = false,
  bookmarksopen = true,
  bookmarksnumbered = true,
  pdfstartpage = {1},  
  pdftitle = {Stochastic IMFs: Models and impact on cluster parameter determination},
  pdfauthor = {P. Anders, R. de Grijs, R. Kotulla, J. Wicker} ,
  pdfkeywords = {star clusters, stochastic IMF, parameter determination},
  pdfproducer = {LaTeX with hyperref}
}
\newcommand{\msun}{\mbox{$M_{\odot}$}}

\newcommand{\msunmsun}{\mbox{$M_{\odot} \,$}}

\newcommand{\Mcl}{\mbox{$M_{\rm cl} \,$}}

\newcommand{\feh}{$[$Fe/H$] \,$}

\def\spose#1{\hbox to 0pt{#1\hss}}
\def\lta{\mathrel{\spose{\lower 3pt\hbox{$\mathchar"218$}}
     \raise 2.0pt\hbox{$\mathchar"13C$}}}
\def\gta{\mathrel{\spose{\lower 3pt\hbox{$\mathchar"218$}}
     \raise 2.0pt\hbox{$\mathchar"13E$}}}

\newcommand{\be}{\begin{enumerate}} 
\newcommand{\ee}{\end{enumerate}} 
\newcommand{\bi}{\begin{itemize}} 
\newcommand{\ei}{\end{itemize}} 

\def\spose#1{\hbox to 0pt{#1\hss}}
\def\lta{\mathrel{\spose{\lower 3pt\hbox{$\mathchar"218$}}
     \raise 2.0pt\hbox{$\mathchar"13C$}}}
\def\gta{\mathrel{\spose{\lower 3pt\hbox{$\mathchar"218$}}
     \raise 2.0pt\hbox{$\mathchar"13E$}}}

\shorttitle{Stochastic clusters: models and analysis}
\shortauthors{Anders et al.}

\begin{document}

\title{Stochastic stellar cluster IMFs: Models and impact on integrated cluster parameter determination}

\author{P. Anders}
\affil{Kavli Institute for Astronomy and Astrophysics, Peking University, Yi He Yuan Lu 5, Hai Dian District, Beijing 100871, China \& National Astronomical Observatories, Chinese Academy of Sciences, 20A Datun Road, Chaoyang District, Beijing 100012, China}
\email{anders@bao.ac.cn}

\author{R. Kotulla}
\affil{Center for Gravitation and Cosmology, Department of Physics, University of Wisconsin - Milwaukee, 1900 E Kenwood Blvd., Milwaukee WI, 53211, USA}

\author{R. de Grijs}
\affil{Kavli Institute for Astronomy and Astrophysics \& Department of Astronomy, Peking University, Yi He Yuan Lu 5, Hai Dian District, Beijing 100871, China}

\and

\author{J. Wicker}
\affil{National Astronomical Observatories, Chinese Academy of Sciences, 20A Datun Road, Chaoyang District, Beijing 100012, China}

\begin{abstract}
Stellar clusters are regularly used to study the evolution of their host galaxy. Except for a few nearby galaxies, these studies rely on the interpretation of integrated cluster properties, especially integrated photometry observed using multiple filters (i.e. the Spectral Energy Distribution SED). To allow interpretation of such observations, we present a large set of GALEV cluster models using the realistic approach of adopting stochastically-sampled stellar IMFs. We provide models for a wide range of cluster masses ($10^3$ $-$ $2 \times 10^5$ \msun), metallicities ($-$2.3 $\le$ \feh $\le$ $+$0.18 dex), foreground extinction, and 184 regularly used filters. We analyze various sets of stochastic cluster SEDs by fitting them with non-stochastic models, which is the procedure commonly used in this field. We identify caveats and quantify the fitting uncertainties associated with this standard procedure. We show that this can yield highly unreliable fitting results, especially for low-mass clusters.
\end{abstract}

\keywords{astronomical databases: miscellaneous --- galaxies: star clusters: general --- methods: data analysis --- stars: mass function --- stars: statistics --- techniques: photometric}

\section{Introduction}

Stellar clusters represent an essential tool to study the stellar populations of galaxies and their evolution. Only stellar clusters can be studied individually out to distances of 100 Mpc or slightly beyond (see e.g. \citealt{2003NewA....8..155D}), far beyond distances which allow studies of resolved stellar populations.

Multi-band integrated photometry of stellar clusters is regularly obtained to study large cluster samples. The clusters' Spectral Energy Distributions (SEDs) are compared with model predictions (\citealt{1999ApJS..123....3L, 2003A&A...401.1063A, 2003MNRAS.344.1000B, 2009MNRAS.396..462K, 2009AJ....138.1724P}) to provide estimates of the clusters' physical parameters (see e.g. \citealt{2004MNRAS.347..196A,2005MNRAS.359..874D}). The derived parameters are cluster age, mass, metallicity, and foreground extinction. While observed clusters represent populations with finite numbers of stars, the model predictions commonly used are based on fully-sampled stellar initial mass functions (IMFs). Stochastic fluctuations around this smooth IMF are consistently taken into account only in the MASSCLEAN models (\citealt{2009AJ....138.1724P}). Other studies have been performed (\citealt{2010A&A...521A..22F, 2011A&A...525A.122P,2011A&A...529A..25S}), although the stochastic cluster models they obtained are not publicly available.

It is generally believed that a galaxy's field star population originates from stellar clusters, which have either dissolved quickly after their formation or slowly dissolved due to long-term evolution. By combining the stochastic cluster models with a galaxy's SFH, stochastic effects in field star formation can be taken into account. Non-universal IMFs are found from the analysis of SDSS galaxies, predominantely for faint galaxies (\citealt{2008ApJ...675..163H}). This effect is related to the IGIMF, the Integrated Galactic IMF, which proposes the IGIMF originates from dissolved stellar clusters (see e.g. \citealt{2004MNRAS.348..187W, 2005ApJ...625..754W}).

In this paper we provide a new publicly available model suite of stellar cluster models including the impact of stochastic IMF sampling. This model set is an extension of the GALEV models (\citealt{2009MNRAS.396..462K}). The cluster photometry obtained is used to derive physical cluster parameters, i.e. cluster age, mass, metallicity, and foreground extinction. The stochastic models are described in Sect. \ref{sec:models}, where the resulting magnitudes and colors are presented and explained. Sect. \ref{sec:paradet} describes the principles of cluster parameter determination based on $\chi^2$ minimization and Sect. \ref{sec:results} shows the results of stochastically-sampled clusters analyzed with fully-sampled (i.e. non-stochastic) cluster models. The paper wraps up with the conclusions in Sect. \ref{sec:conclusions}.

\section{Models}
\label{sec:models}

The integrated stellar cluster photometry is derived using the GALEV population synthesis models (\citealt{2009MNRAS.396..462K}). Population synthesis models combine stellar evolution data (usually stellar isochrone data), spectral libraries, and an IMF (to weigh the contributions of stars of different masses) to provide an integrated spectrum of the stellar cluster being studied. Convolving the integrated spectrum with filter response curves provides model predictions of integrated magnitudes. Such population synthesis models form the input for analysis and interpretation of observed integrated cluster photometry. 

The standard GALEV models use an analytic description of the IMF, similar to most other population synthesis models (i.e. ``fully-sampled models''). The cluster mass \Mcl acts solely as a scaling factor of the integrated magnitude predictions (see \citealt{2009MNRAS.396..462K}). The derived cluster colors are independent of cluster mass. The fully-sampled approach uses fractional stars to account for the total cluster mass. While this is applicable for the description of idealized clusters, it results in deviations for realistic clusters.

In real stellar clusters, a stellar cluster's IMF is derived by stochastically sampling stars from the analytic IMF description up to the total stellar mass \Mcl (i.e. ``stochastically-sampled models''). In this paper we use stochastically sampled \citet{2001MNRAS.322..231K} IMFs in the mass range 0.1 $-$ 100 \msunmsun (with d$N/$d$m \propto m^{-1.3}$ for stellar masses 0.1 \msunmsun $\le$ m $\le$ 0.5 \msun, d$N/$d$m \propto m^{-2.3}$ for stellar masses 0.5 \msunmsun $<$ m $\le$ 100.0 \msun)  for a range of total cluster masses ($10^3$ $-$ $2 \times 10^5$ \msun). Observations have indicated the existence of an upper stellar mass limit (USML, as derived by \citealt{2007ApJ...671.1550P}) as a function of cluster mass. For clusters with total masses of $10^3$ \msunmsun we also present models with the appropriate USML of 42.7 \msun. However, other observations (see e.g. \citealt{2010ApJ...719L.158C,2013ApJ...767...51A}) found no strong evidence for an USML as a function of cluster mass exceeding the expectations from stochastic IMF sampling. \citet{2010ApJ...719L.158C} claim that their observations are only 1-2$\sigma$ away from the USML models, with the observations in between the USML models and models with a cluster-mass-independent stellar mass limit, hence they cannot unambiguously evaluate the existence of an USML. Therefore, models with and without USML are provided.

\begin{figure}
\begin{center}
   \includegraphics[angle = 270,width = 0.7\linewidth]{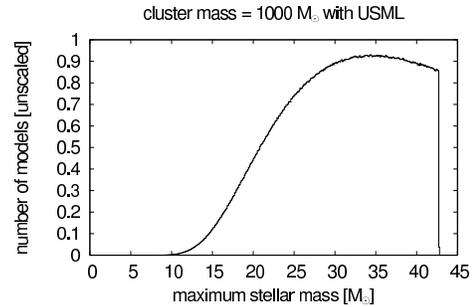}
\end{center}
\caption{Binned distribution of maximum stellar masses within a cluster. Clusters have masses of $10^3$ \msunmsun and the USML criterion is applied.}
\label{fig:maxmass}
\end{figure}

\citet{2007ApJ...671.1550P} found a maximum stellar mass $m_{max}$ as a function of stellar cluster mass. Instead we use their $m_{max}$ as an upper stellar mass limit, which can be reached, but not exceeded. Therefore a stochastic distribution of the maximum stellar mass present in a stellar cluster is established, in contrast to \citet{2007ApJ...671.1550P}. The distribution of drawn $m_{max}$, based on $2 \times 10^7$ test IMFs with 6.96 \msunmsun $\le m_{max} \le$ 42.7 \msun, is shown in Fig. \ref{fig:maxmass}. In this respect the maximum stellar mass becomes a stochastic quantity, which also applies in general to the model clusters' IMFs.

The IMFs of model clusters are populated with individual stars drawn stochastically from the \citet{2001MNRAS.322..231K} IMF. This process is stopped when the total cluster mass just exceeds the target mass. Due to the piece-wise power-law description of the \citet{2001MNRAS.322..231K} IMF and to simplify the implementation of the IMF sampling for the entire mass range, the stochastic sampling process consists of three steps:
\be
\item Determination of the mass fraction ($MF$) below the \citet{2001MNRAS.322..231K} IMF break-point at 0.5 \msun;
\item Performance of stochastic sampling of stars with $M \le 0.5$ \msun, until the total cluster mass reaches $MF \times$ \Mcl. The last sampled star causes a slight excess over the mass $MF \times$ \Mcl.
\item Stochastic sampling of stars with $M > 0.5$ \msun, until the total cluster mass reaches \Mcl. The last sampled star causes a slight excess over the total cluster mass \Mcl.
\item Stars drawn with masses outside the sampled mass range are not considered.
\ee
This procedure results in the excess of the total cluster mass by a fraction of a single star (see below), though it resembles the observed IMF accurately. The independent sampling of stars with masses above and below 0.5 \msunmsun prevents cluster models with a bias towards an excess of low-mass or high-mass stars. Although the stochastic sampling is performed for the whole stellar mass range, stochastic fluctuations for the most massive stars at a given age dominate the fluctuations in the integrated cluster light. Stochastic fluctuations for low-mass stars have a negligible impact due to the large total number of low-mass stars.

To test the IMF sampling, we designed an alternative sampling method: first a random number\footnote{All random numbers used are based on the RAND random number generator in PERL.} determines whether the stellar mass should be above or below the \citet{2001MNRAS.322..231K} IMF break-point at 0.5 \msun. A second random number determines the stellar mass using the appropriate relations for the mass range requested by the first random number. Using this sampling technique, clusters with a strong bias towards high-mass or low-mass stars could appear. We created $10^4$ test models with both IMF sampling methods. The resulting magnitudes and colors (i.e. GALEX satellite $FUV$ magnitude to $K$-band magnitude, using solar metallicity and foreground extinction [$E(B-V)=$ 0.0mag, 1.0mag and 2.0mag] of clusters with $10^3$ \msunmsun and $10^4$ \msunmsun without USML) of both IMF sampling methods were evaluated using a KS test. The KS test results proved the consistency of the results derived from both IMF sampling methods (with p-values $>$ 0.1), except for the H-band of $10^3$ \msunmsun clusters. The H-band is experiencing the impact of stochastic sampling of both red supergiants and AGB stars, which are bright but rare stars in such a low-mass cluster. $10^4$ test models might have been a too small sample to evaluate these rare cases in its full extend.

We analyzed the resulting cluster masses for anticipated masses of $10^3$ \msunmsun (without USML), $10^4$ \msun, and $10^5$ \msun. The stochastic sampling process leads to a slight excess of the cluster mass, both for low-mass stars ($M \le 0.5$ \msun) and for high-mass stars ($M > 0.5$ \msun) due to the last sampled star. The size of this excess in cluster mass is dominated by the adopted IMF and is largely independent of the anticipated cluster mass. For high-mass stars a slight tendency for increasing excess with decreasing cluster mass is found: low-mass clusters have a greater chance to be affected by stochastic draws with exceptionally large stellar masses, which cannot be balanced by a large number of further draws. This effect dominates the excess of the total cluster mass. The fraction of low-mass stars is exceeded by at least 0.12 \msunmsun / 0.30 \msunmsun / 0.43 \msunmsun for 50\% / 10\% / 1\% of all clusters. The fraction of high-mass stars is exceeded by at least 1.00 \msunmsun / 14.92 \msunmsun / 58.24 \msunmsun for 50\% / 10\% / 1\% of all clusters. Applying these fractions to the lowest modelled cluster mass of $10^3$ \msunmsun shows: The majority of model clusters experience a mass excess $\le$ 0.15\%, while for 1\% of model clusters the mass excess reaches $\ge$ 6\% of the intended cluster mass. This fractional mass excess decreases with increasing intended cluster mass. This mass excess is caused by a single star for both mass ranges due to the stochastic sampling process. As the excess is caused by a single star whose mass is determined by the stellar IMF adopted we do not expect a systematic impact on the modelling results.

The stochastic IMF's shape follows the underlying analytic \citet{2001MNRAS.322..231K} IMF. The stochastic realizations of the IMF deviate from the smooth fully-sampled IMF, with mass ranges containing more stars and mass ranges lacking stars in comparison with the fully-sampled IMF prediction. The deviations are strongest for high-mass stars and increase with decreasing \Mcl, both caused by the smaller number of high-mass stars compared to their lower-mass counterparts. The present-day mass function evolves from the IMF due to stellar evolution effects, i.e. the death of stars, and dynamical effects, such as the loss of low-mass stars due to tidal interactions. The impact of dynamical cluster evolution on integrated cluster magnitudes has been studied by \citet{2009A&A...502..817A}. Since effects due to dynamical interactions and stochastic IMF sampling are largely independent of each other we do not include the former and focus on the latter effect in this paper.

The individual stellar masses from the stochastically-sampled IMF are used as input data for the GALEV models. The stellar evolution of these stars is taken from isochrones of \citet{2008A&A...482..883M} (these isochrones do not include pre-main-sequence models), based solely on the effective temperature, surface gravity, metallicity, and scaled by the bolometric luminosity. Therefore, AGB star magnitude corrections and circumstellar dust are not required. A spectrum from the catalogs of \citet{1997A&AS..125..229L,1998A&AS..130...65L} is assigned to each star. The individual contributions of all stars are summed up to provide the integrated spectrum. Convolving the integrated spectra with a variety of filter functions provides the integrated photometry, for both the fully-sampled models and the stochastically-sampled models. Models are created for all ages for which isochrones are provided: log(age [yr]) range = 6.6 $-$ 10.1 (i.e. age range = 4 Myr $-$ 12.6 Gyr) with steps of $\Delta$log(age [yr]) = 0.05. In total, predictions for 71 age steps have been calculated. The isochrones for log(age) = 9.20 and 9.25 are strongly dominated by AGB stars (see \citealt{2008A&A...482..883M}) for metallicities $-$0.125 $\le$ \feh $\le$ $+$0.1 dex, causing a significant distortion in the color evolution. These distortions are intrinsic to the isochrones and not related to the chosen time steps (i.e. not related to binning effects of the stellar tracks). Models for the remaining ages are not affected. These ages have been removed to allow a better analysis and presentation of the results.

\begin{figure*}
\begin{center}
  \begin{tabular}{ccc}
   \includegraphics[angle = 270,width = 0.3\linewidth]{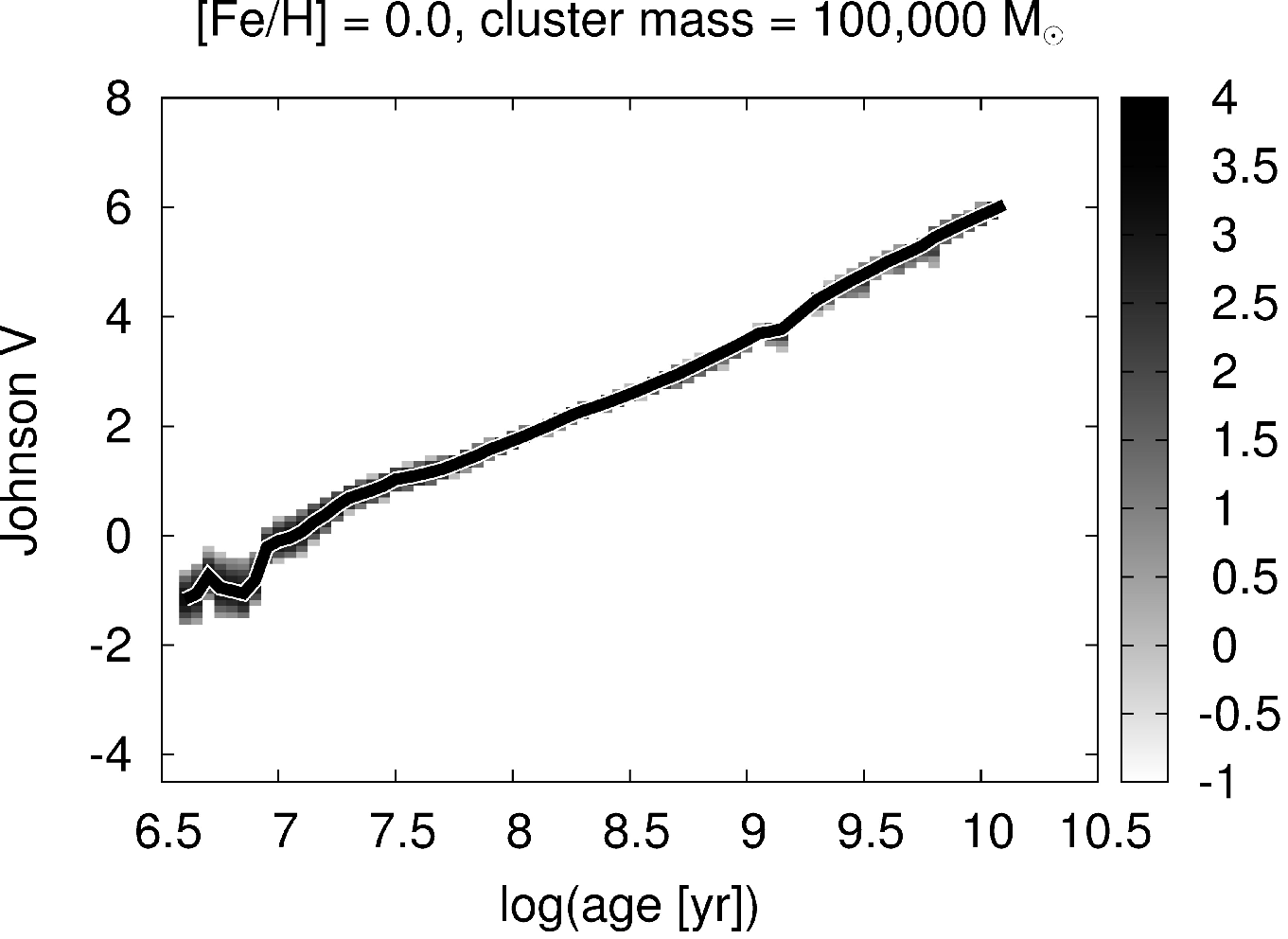} & 
   \includegraphics[angle = 270,width = 0.3\linewidth]{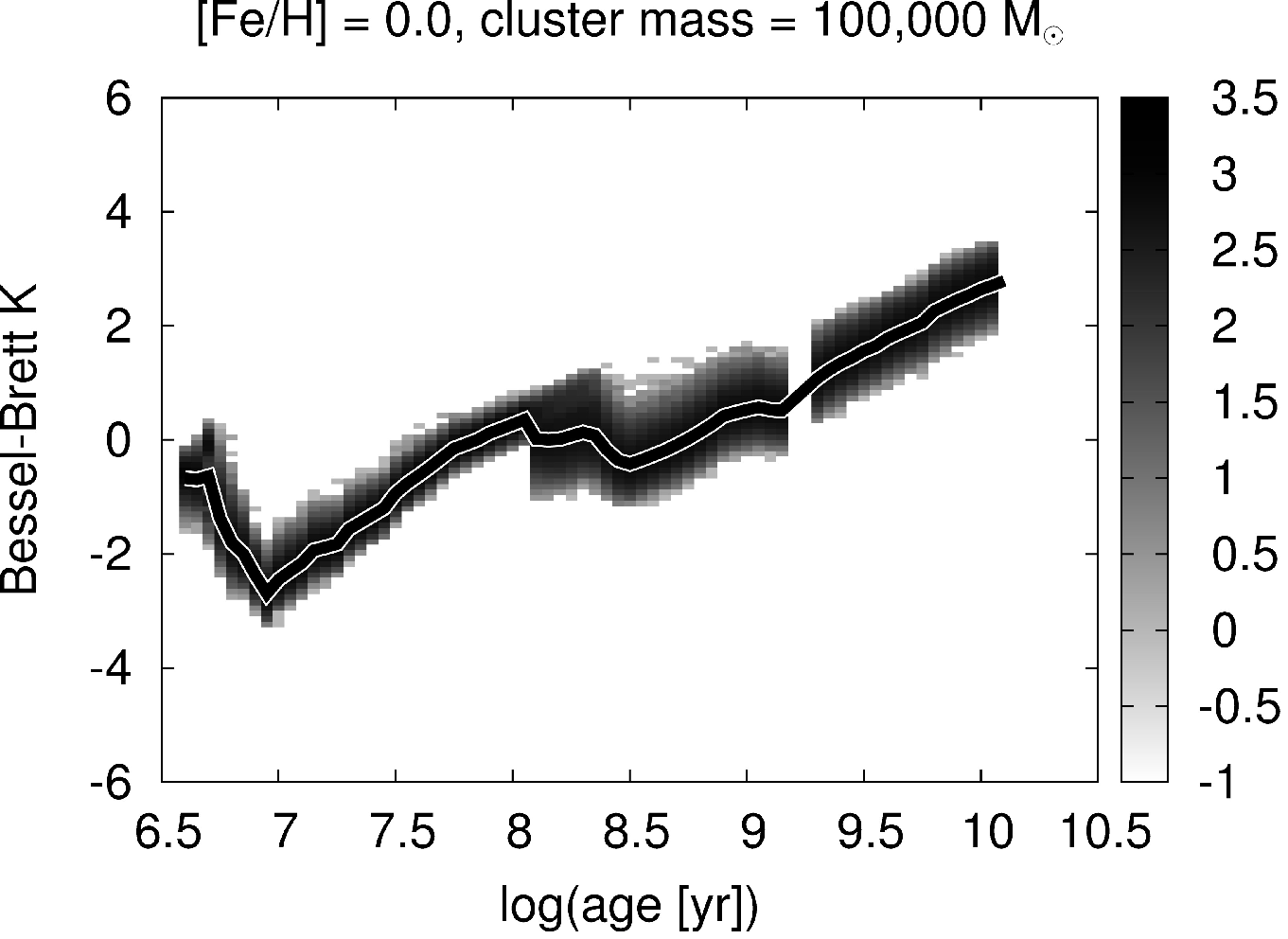} & 
   \includegraphics[angle = 270,width = 0.3\linewidth]{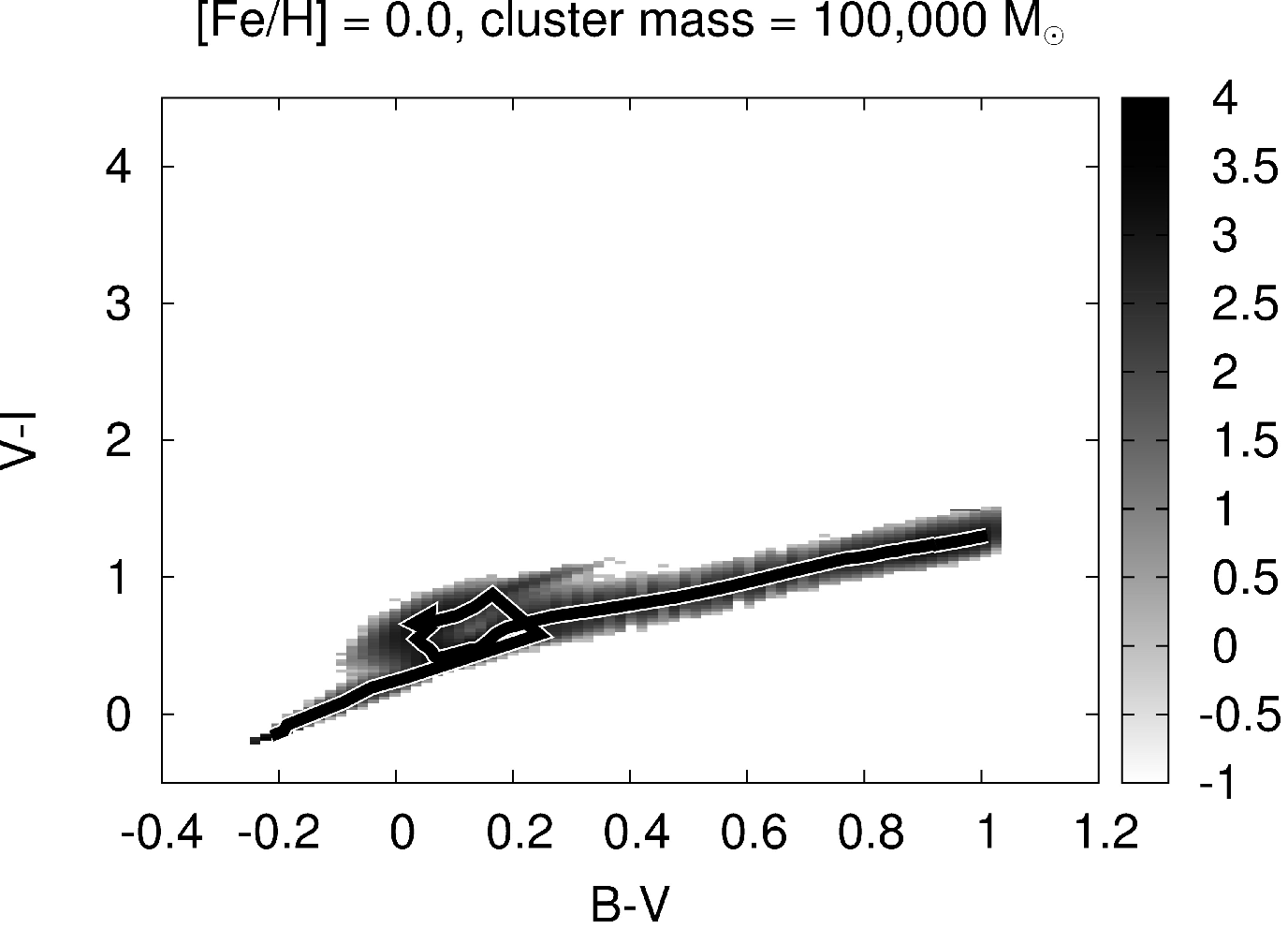} \\
   \includegraphics[angle = 270,width = 0.3\linewidth]{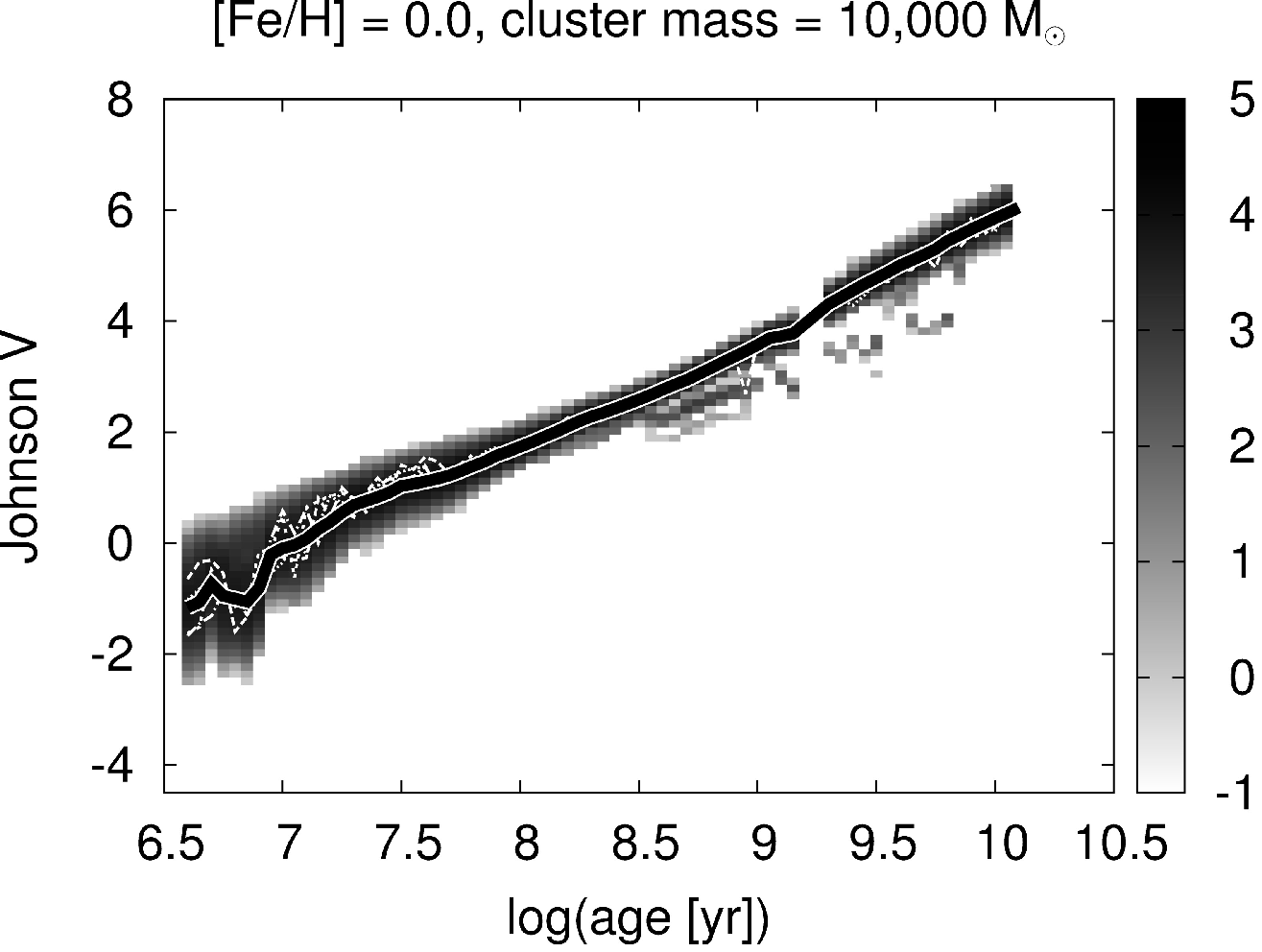} & 
   \includegraphics[angle = 270,width = 0.3\linewidth]{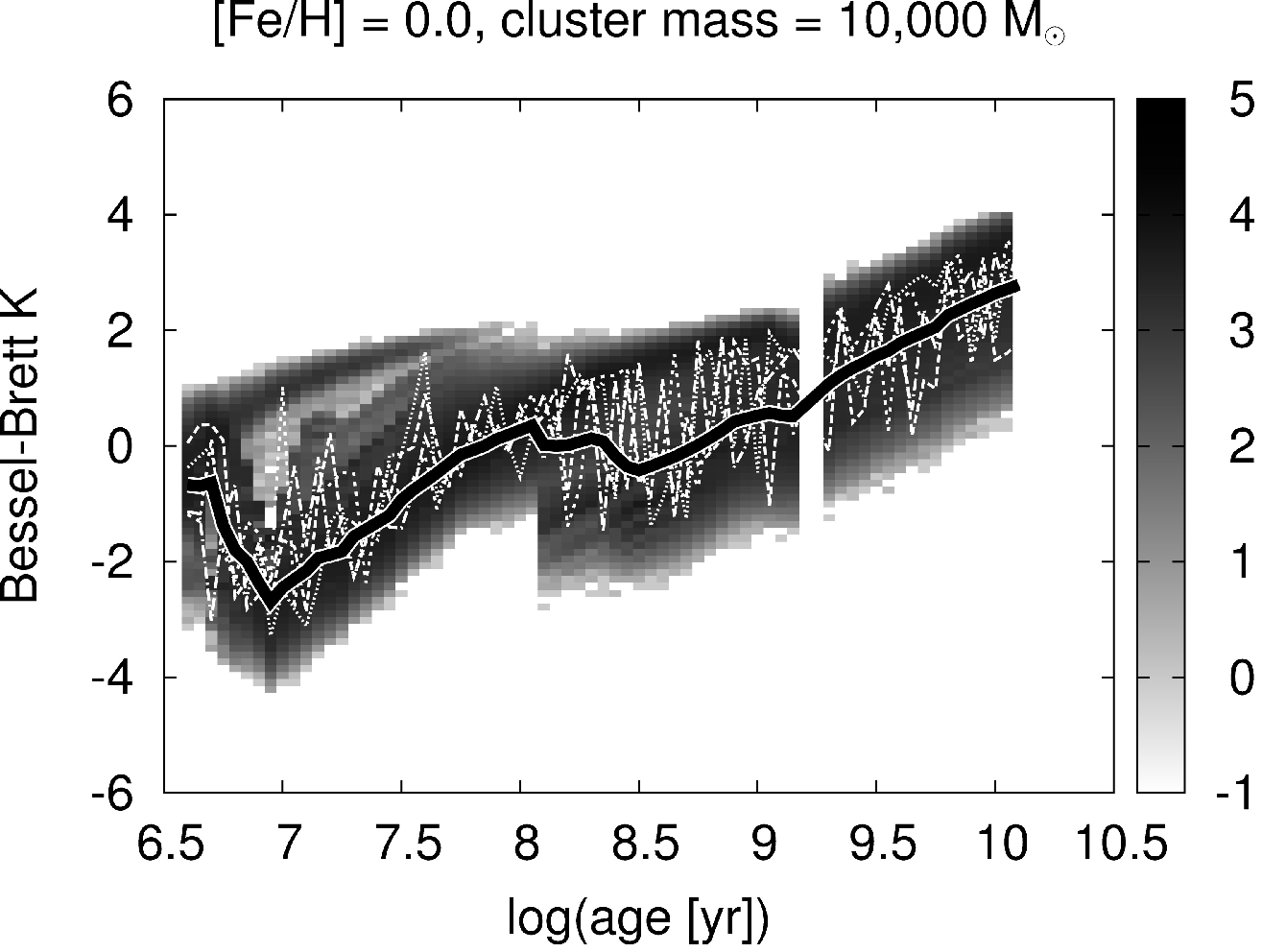} & 
   \includegraphics[angle = 270,width = 0.3\linewidth]{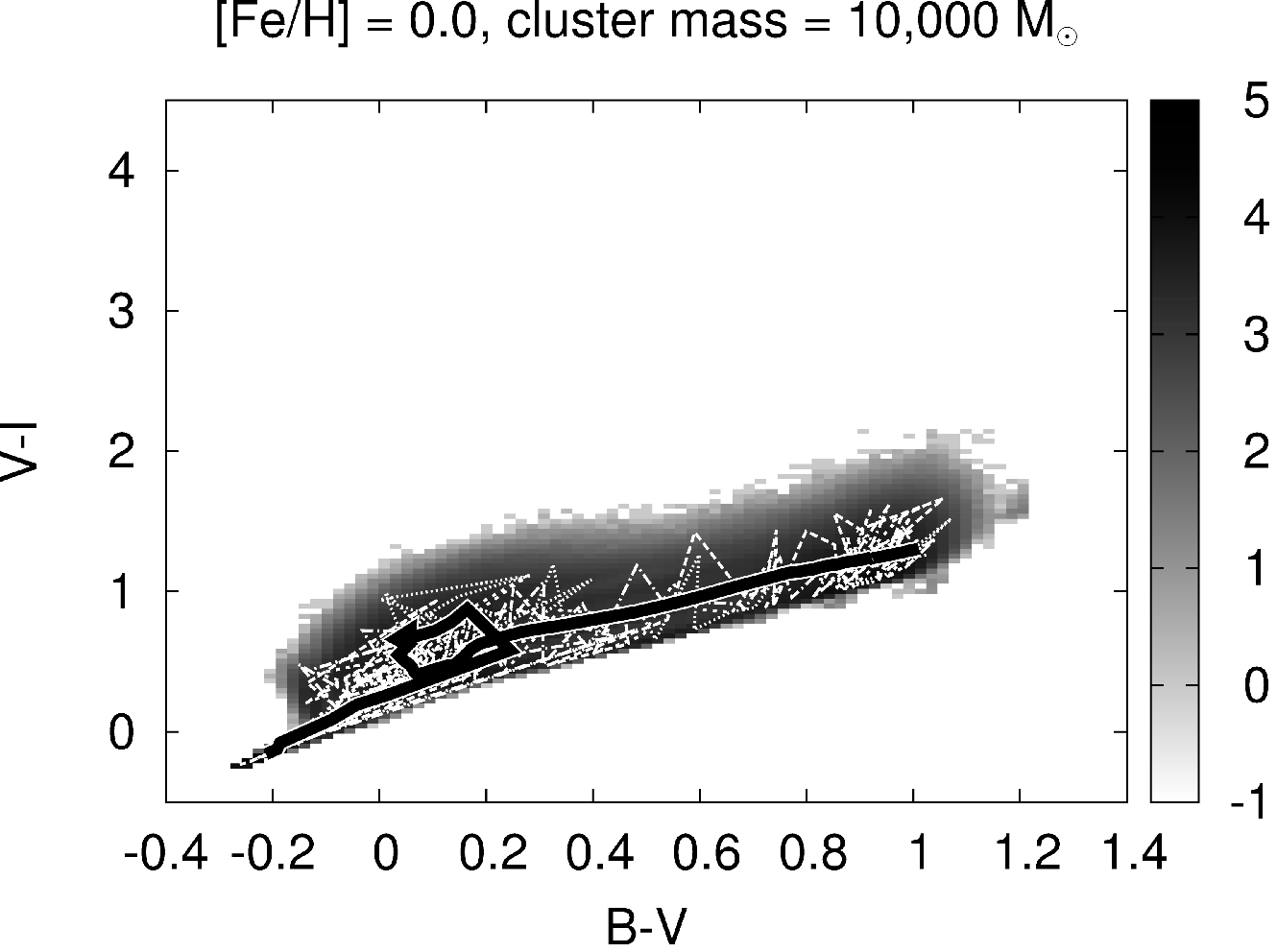} \\
   \includegraphics[angle = 270,width = 0.3\linewidth]{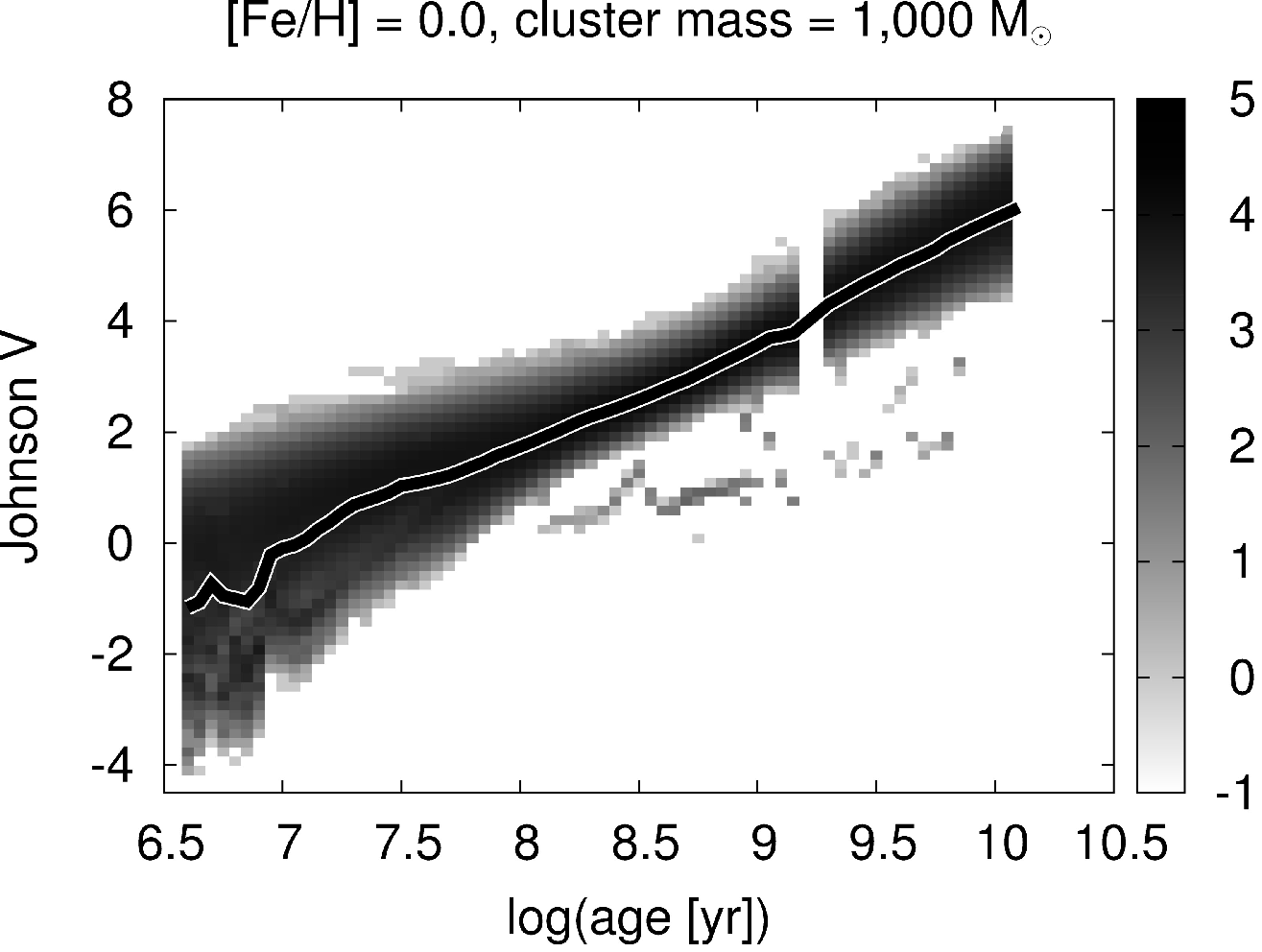} & 
   \includegraphics[angle = 270,width = 0.3\linewidth]{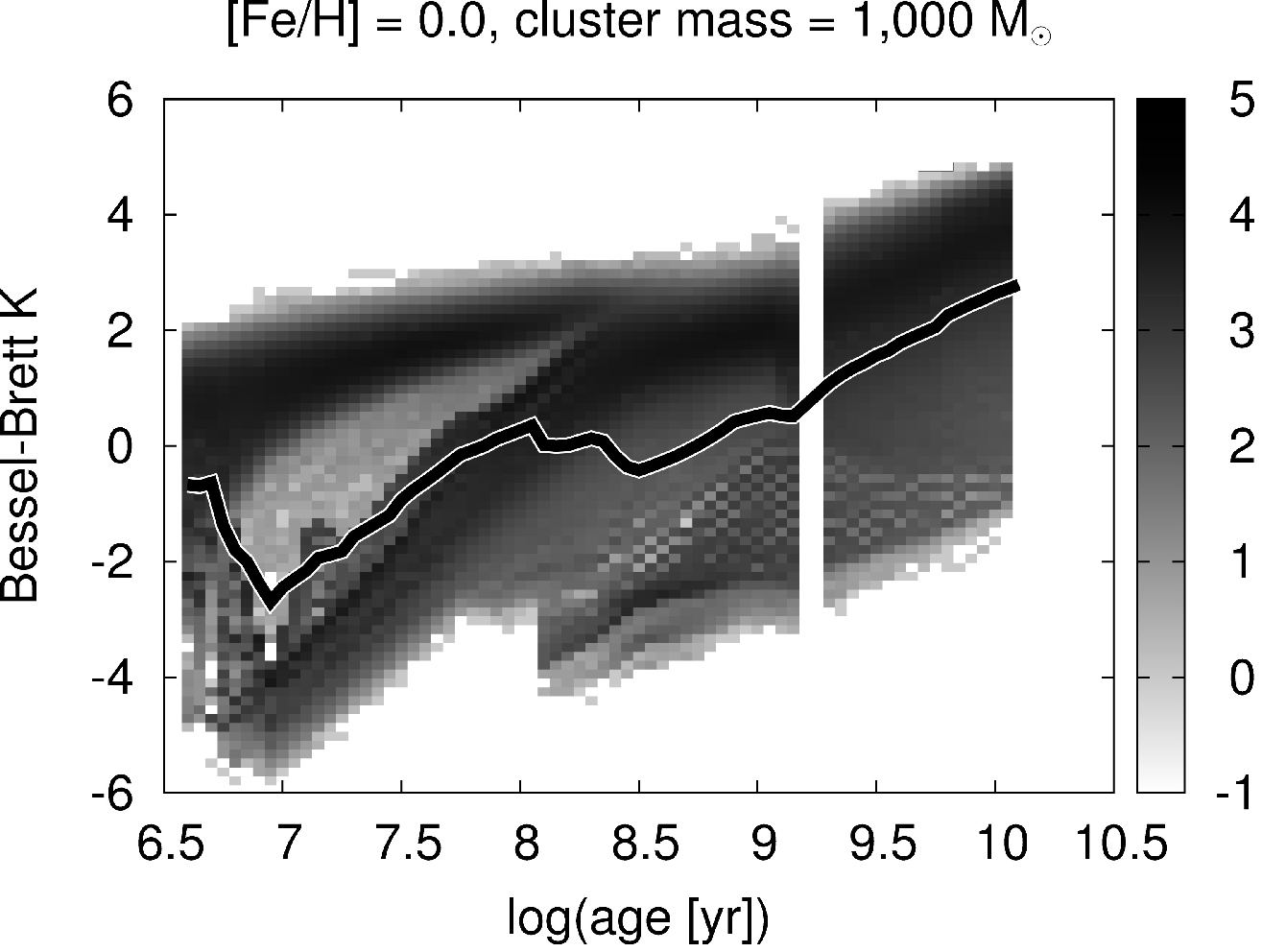} & 
   \includegraphics[angle = 270,width = 0.3\linewidth]{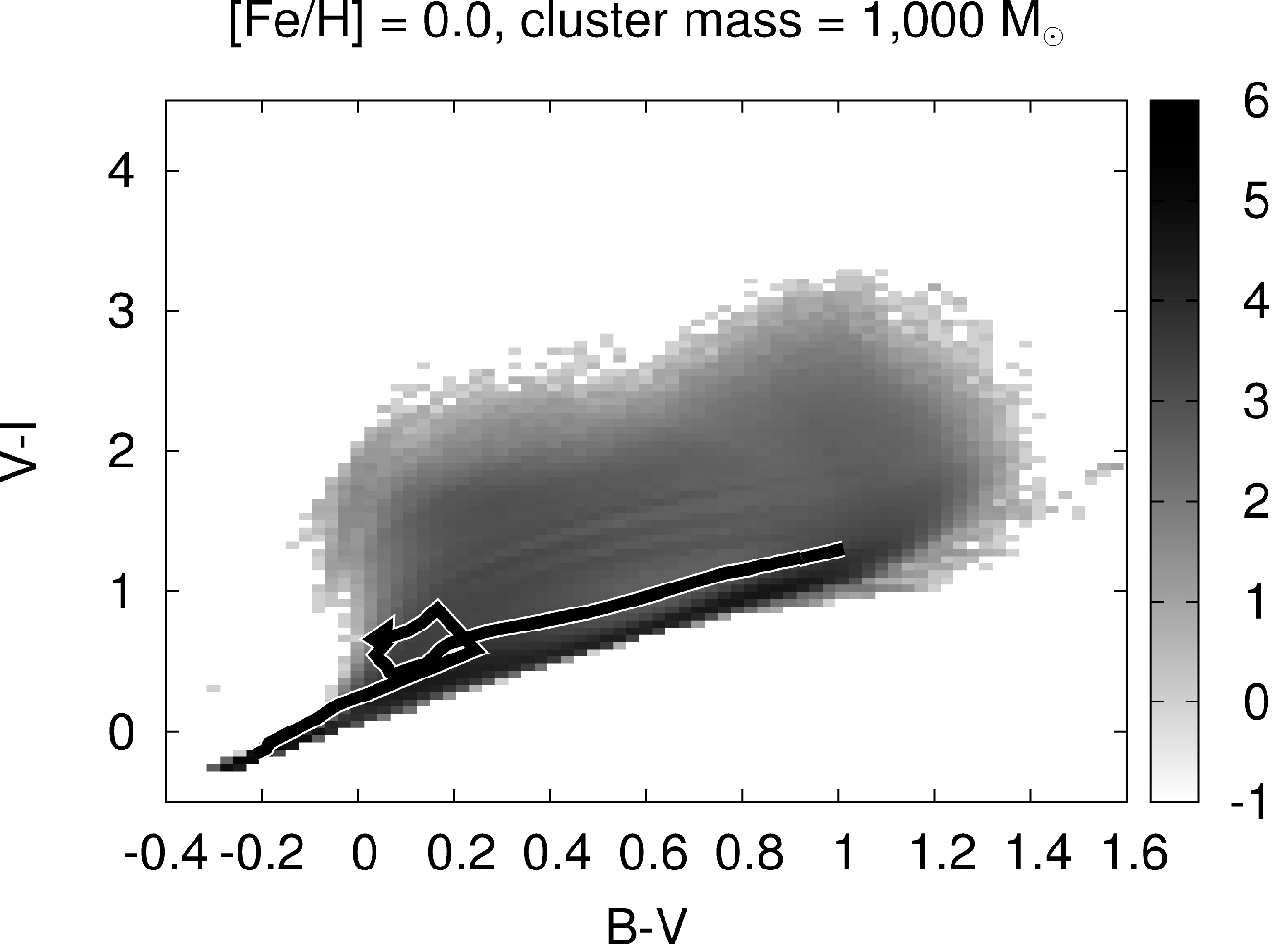} \\
 \end{tabular}
\end{center}
\caption{Examples of time evolution of various magnitudes and color-color plots, smoothed to enhance presentational clarity. Shown are cluster samples with cluster masses of $10^5$ \msunmsun (top row), $10^4$ \msunmsun (middle), and $10^3$ \msunmsun (bottom). Properties shown are the magnitudes in the Johnson $V$-band (left column) and Bessel-Brett $K$-band (middle). In addition, color-color plots are presented for $B-V$ vs. $V-I$ (right). Greyscale represents the log(number of models per bin). The models have solar metallicity and no foreground extinction. The solid lines (with white shadow) represent the corresponding fully-sampled models. For cluster masses of $10^4$ \msun, 4 models of single clusters have been added to give an impression of their evolution.}
\label{fig:example_mags}
\end{figure*}

\begin{figure*}
\begin{center}
  \begin{tabular}{cc}
   \includegraphics[angle = 270,width = 0.4\linewidth]{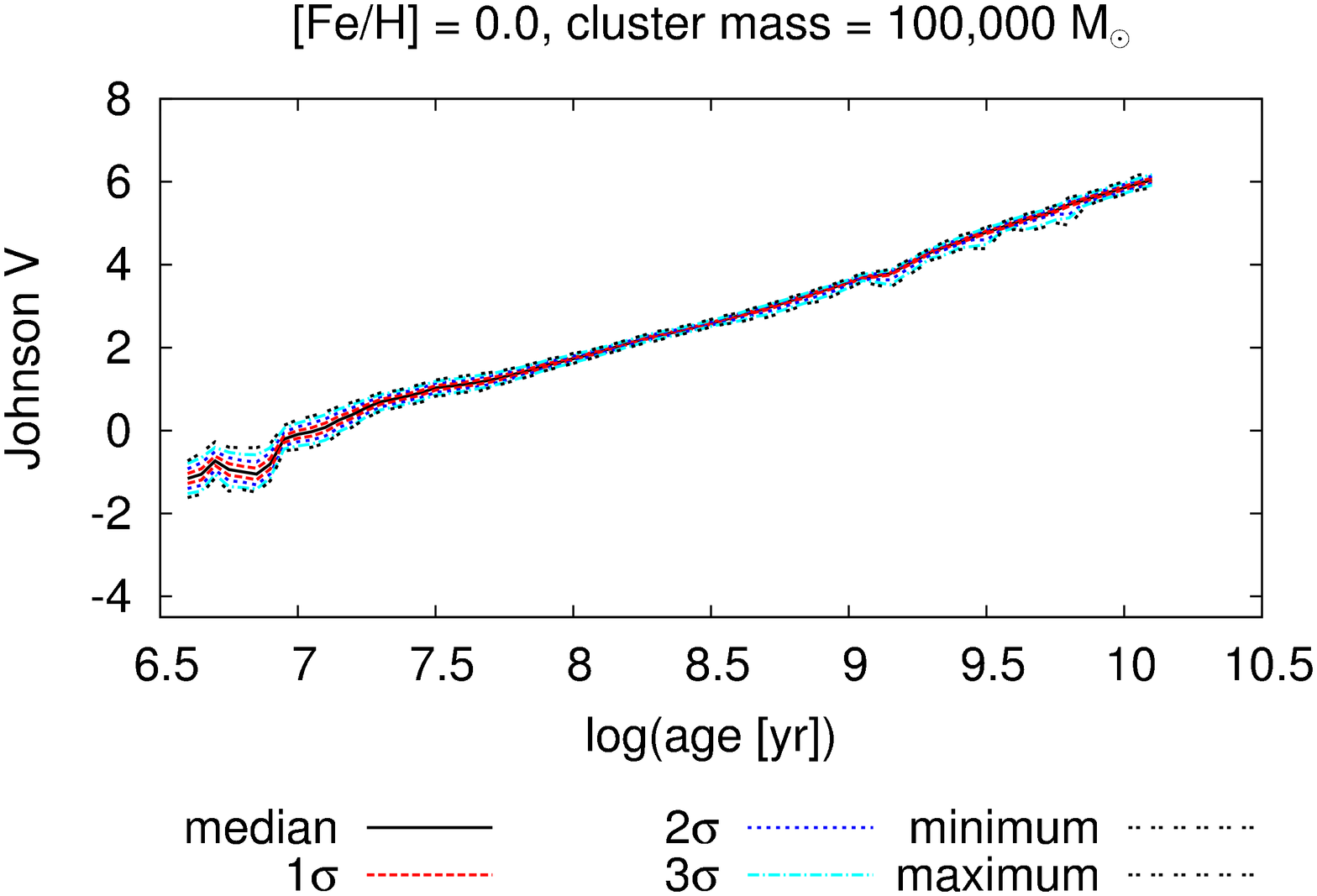} & 
   \includegraphics[angle = 270,width = 0.4\linewidth]{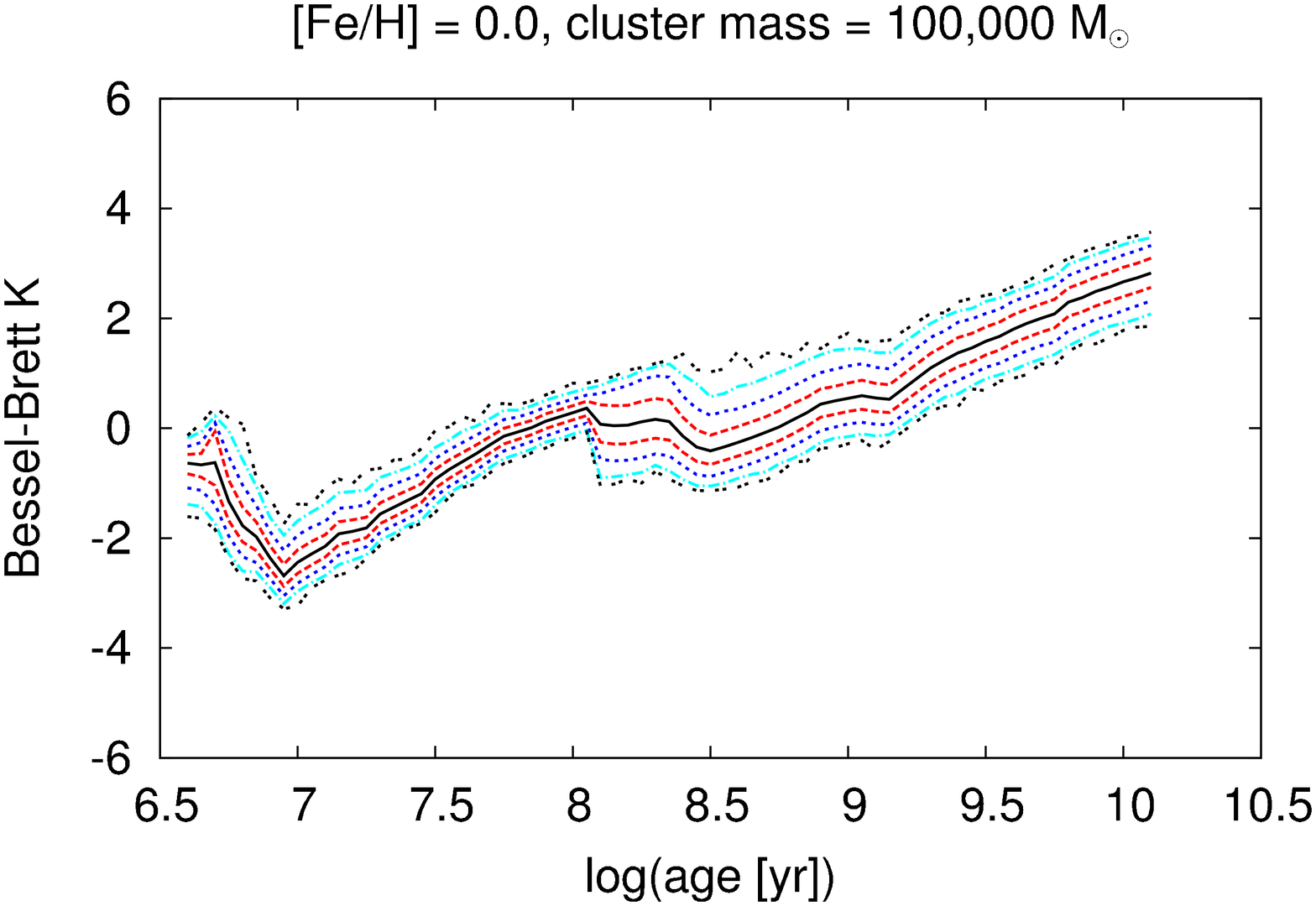} \\ 
   \includegraphics[angle = 270,width = 0.4\linewidth]{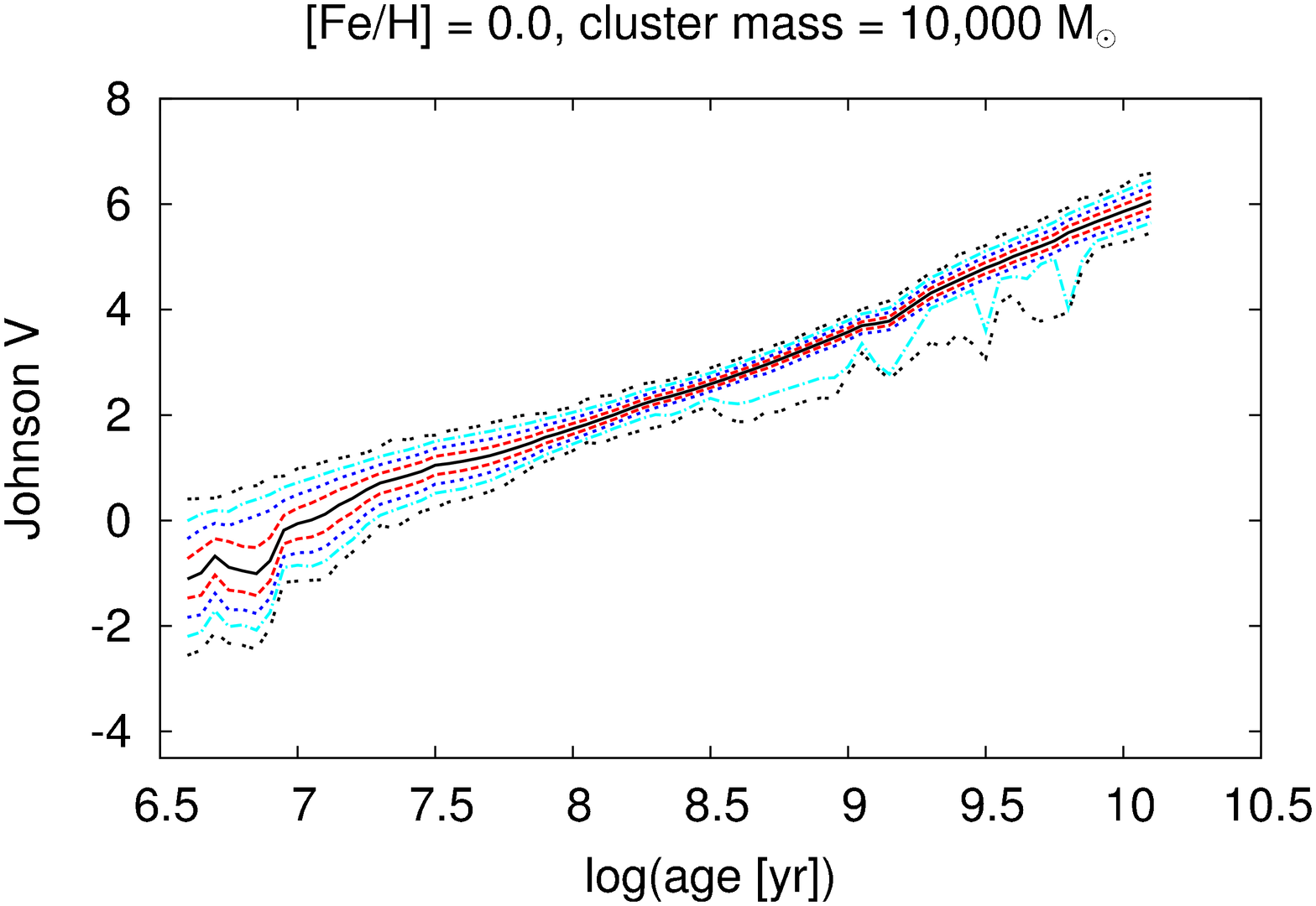} & 
   \includegraphics[angle = 270,width = 0.4\linewidth]{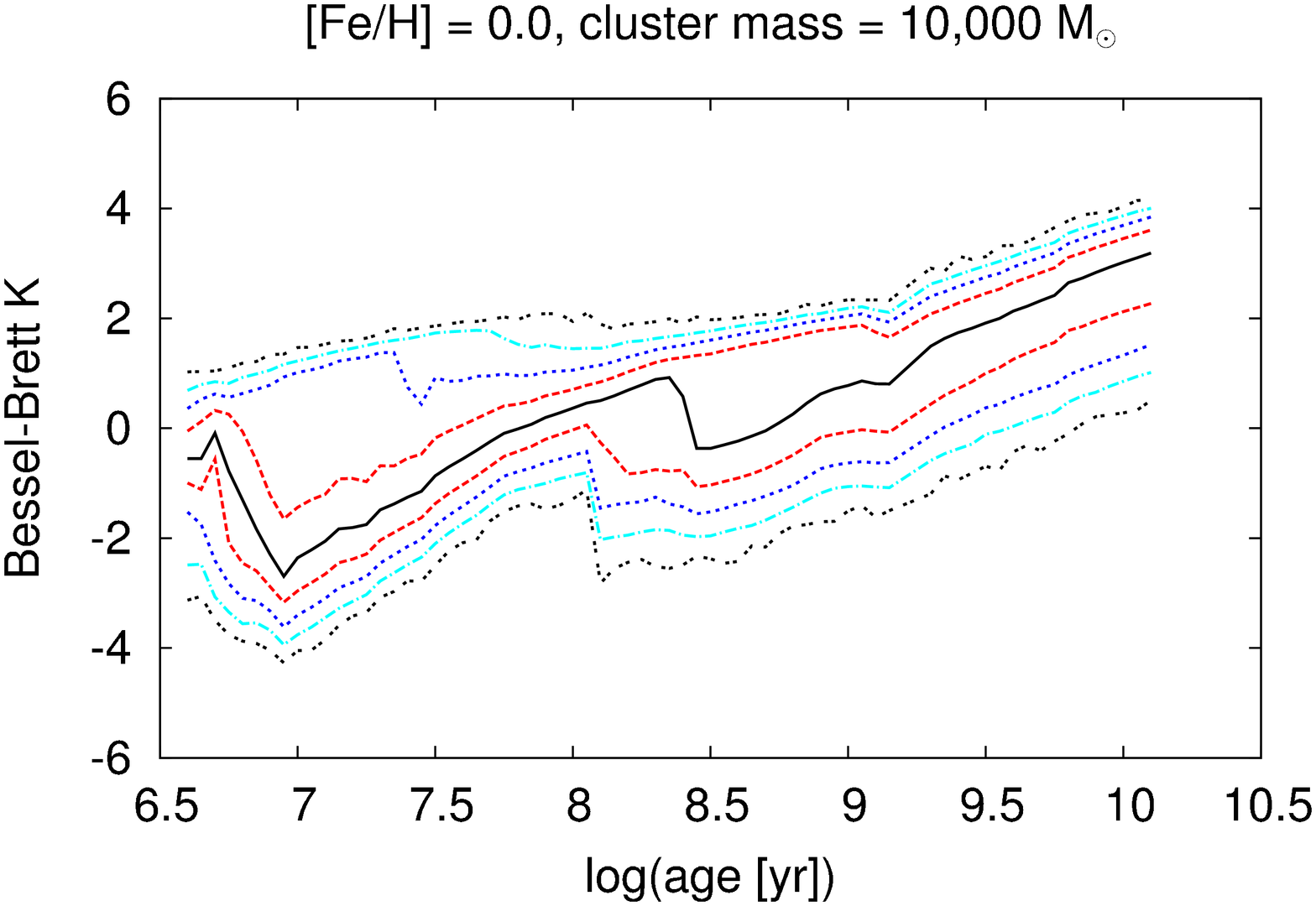} \\ 
   \includegraphics[angle = 270,width = 0.4\linewidth]{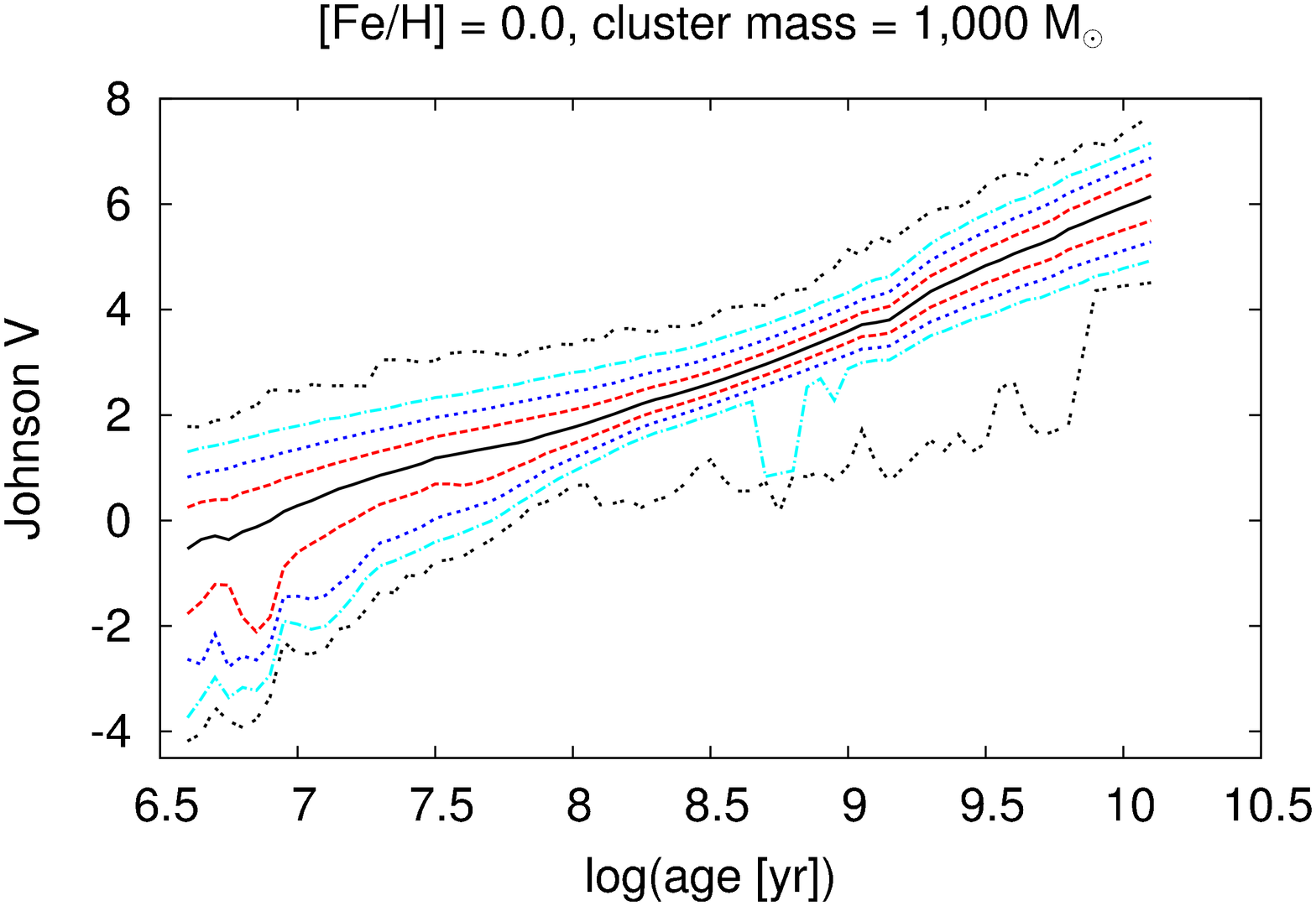} & 
   \includegraphics[angle = 270,width = 0.4\linewidth]{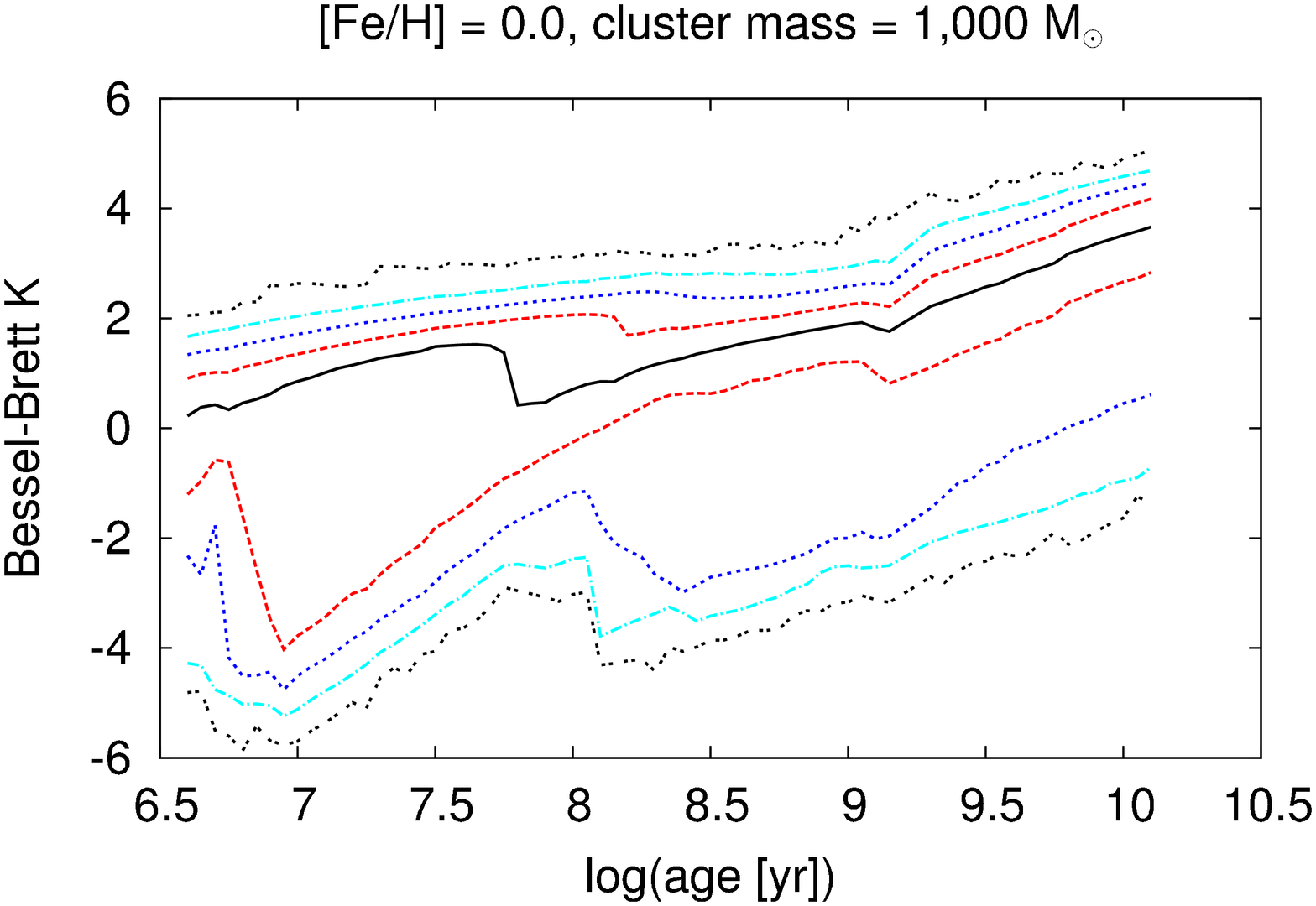} \\ 
 \end{tabular}
\end{center}
\caption{Examples of time evolution of various magnitudes. Shown are cluster samples with cluster masses of $10^5$ \msunmsun (top row), $10^4$ \msunmsun (middle), and $10^3$ \msunmsun (bottom). Properties shown are the magnitudes in the Johnson $V$-band (left column) and Bessel-Brett $K$-band (right). 
Shown are the median, 1$\sigma$, 2$\sigma$, and 3$\sigma$ equivalent quantiles as well as the minimum and maximum values for every input age. The models have solar metallicity and no foreground extinction.}
\label{fig:example_mags2}
\end{figure*}

\begin{figure*}
\begin{center}
  \begin{tabular}{ccc}
   \includegraphics[angle = 270,width = 0.3\linewidth]{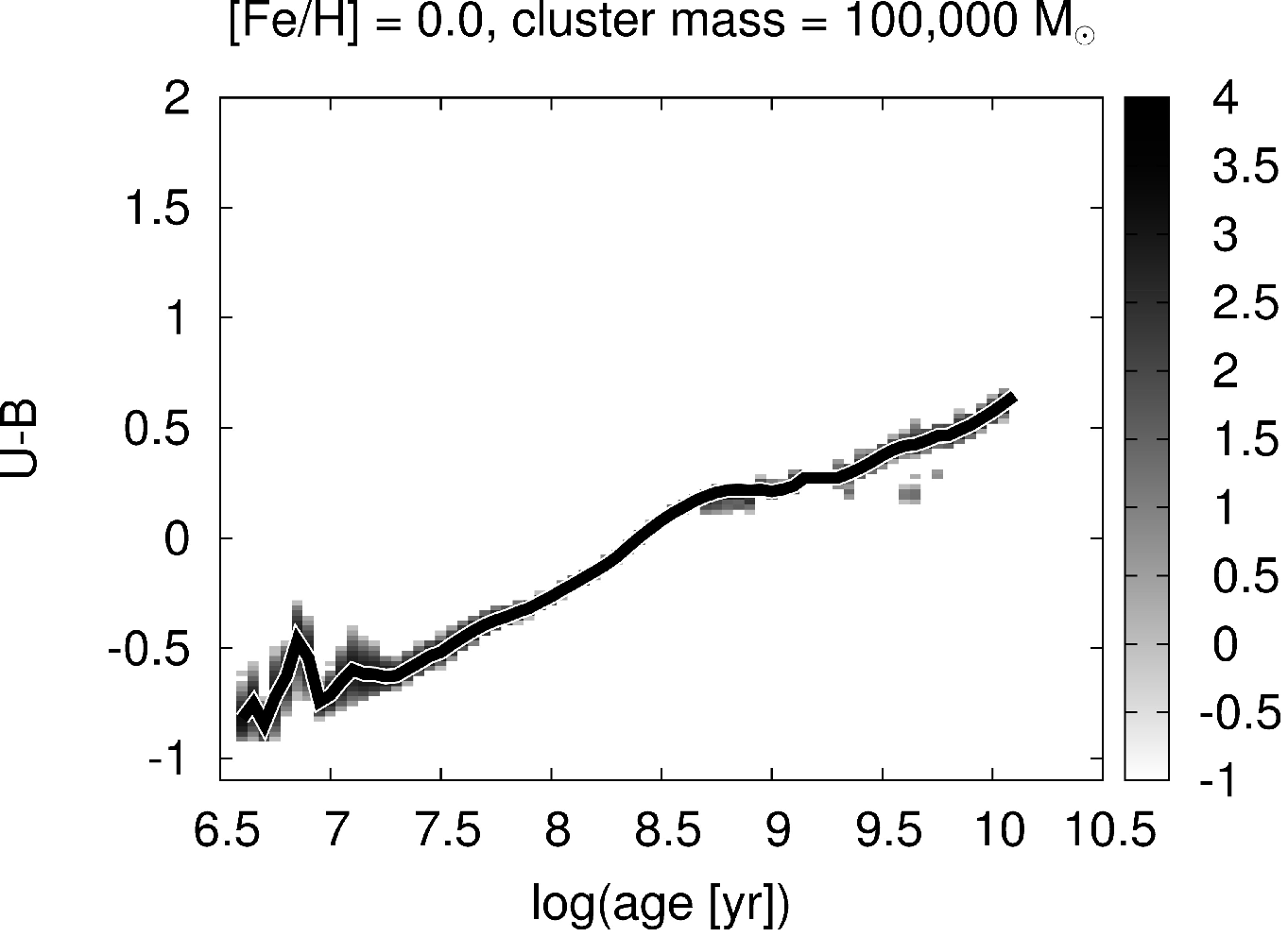} & 
   \includegraphics[angle = 270,width = 0.3\linewidth]{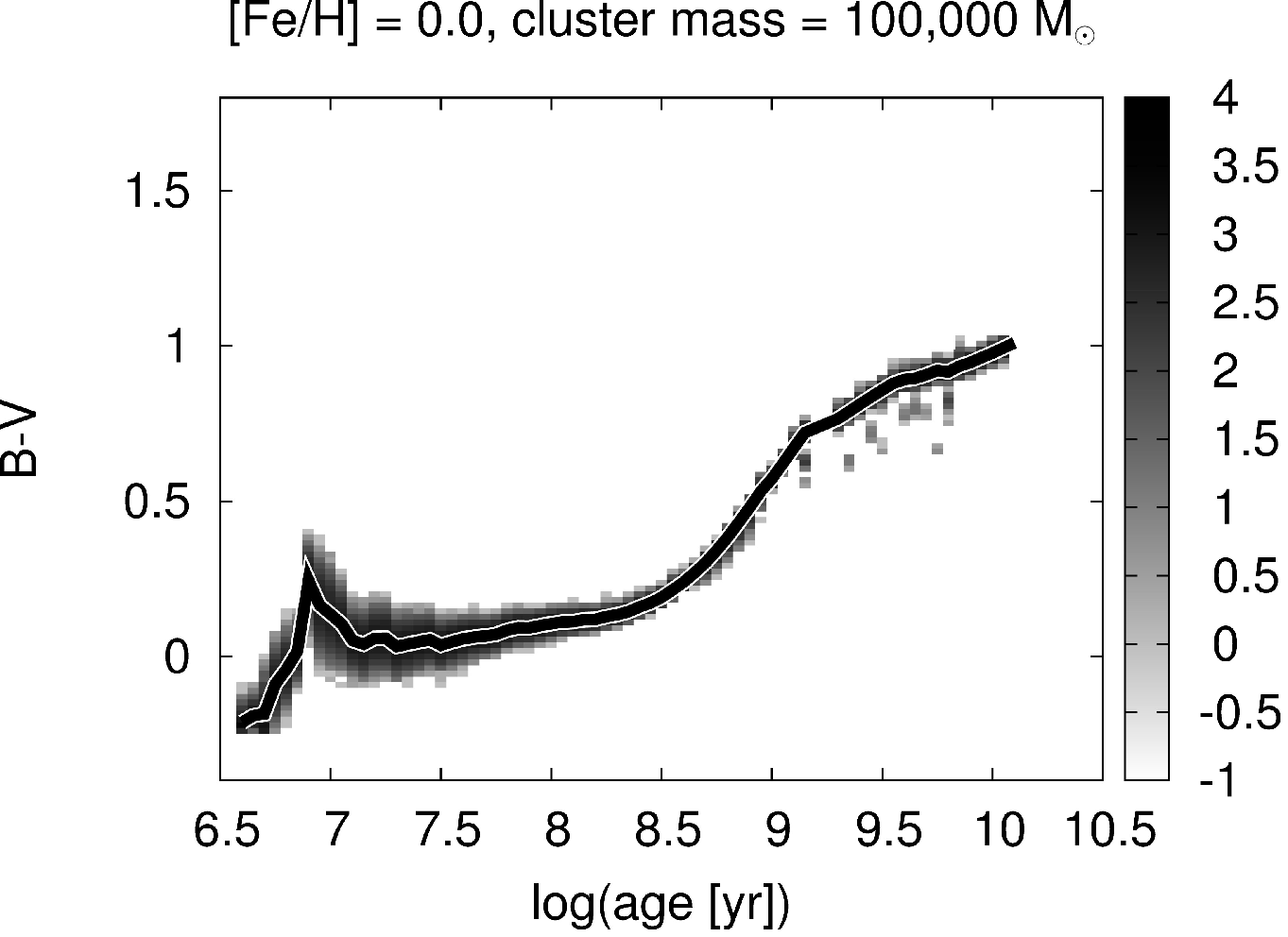} & 
   \includegraphics[angle = 270,width = 0.3\linewidth]{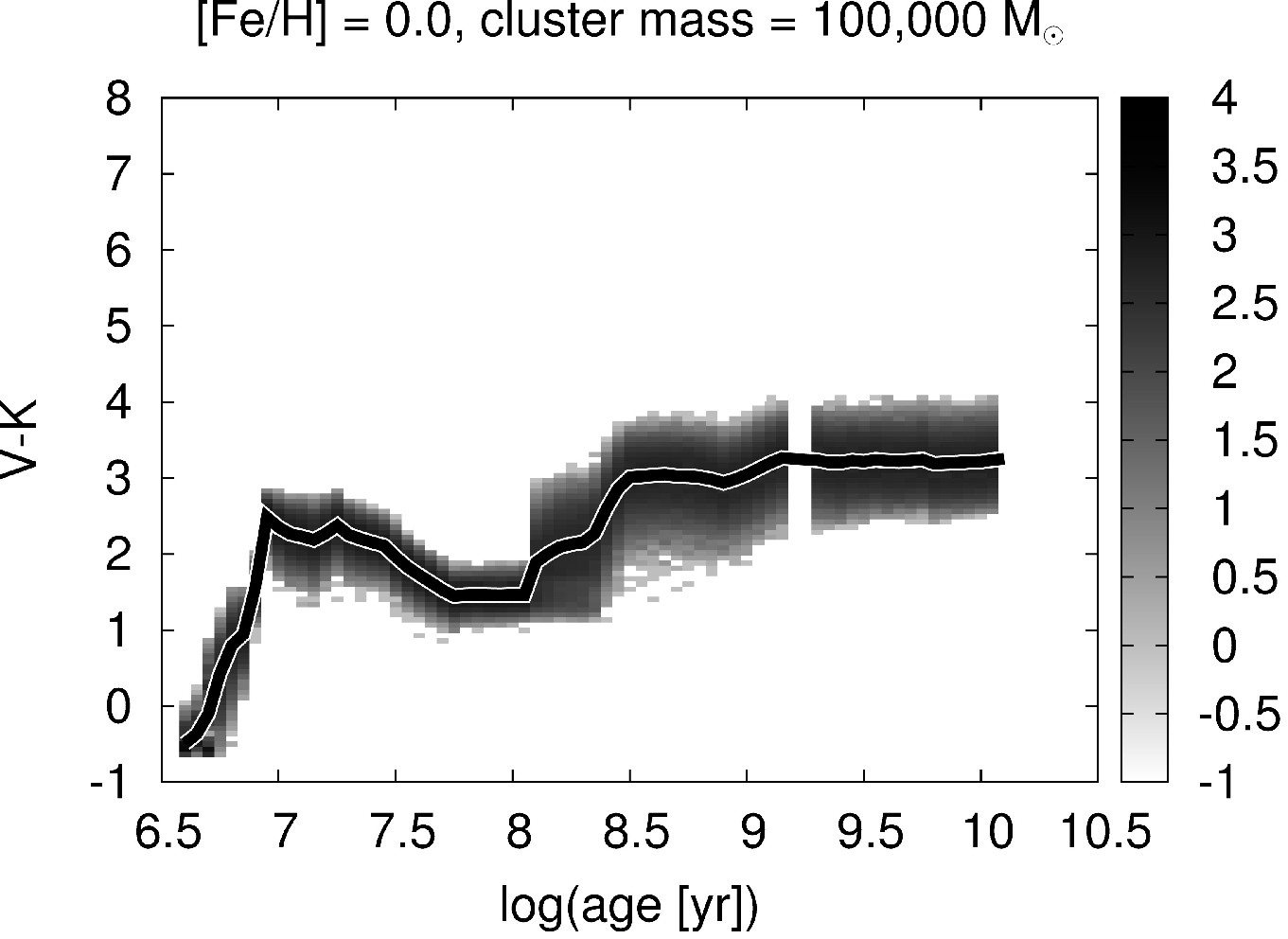} \\
   \includegraphics[angle = 270,width = 0.3\linewidth]{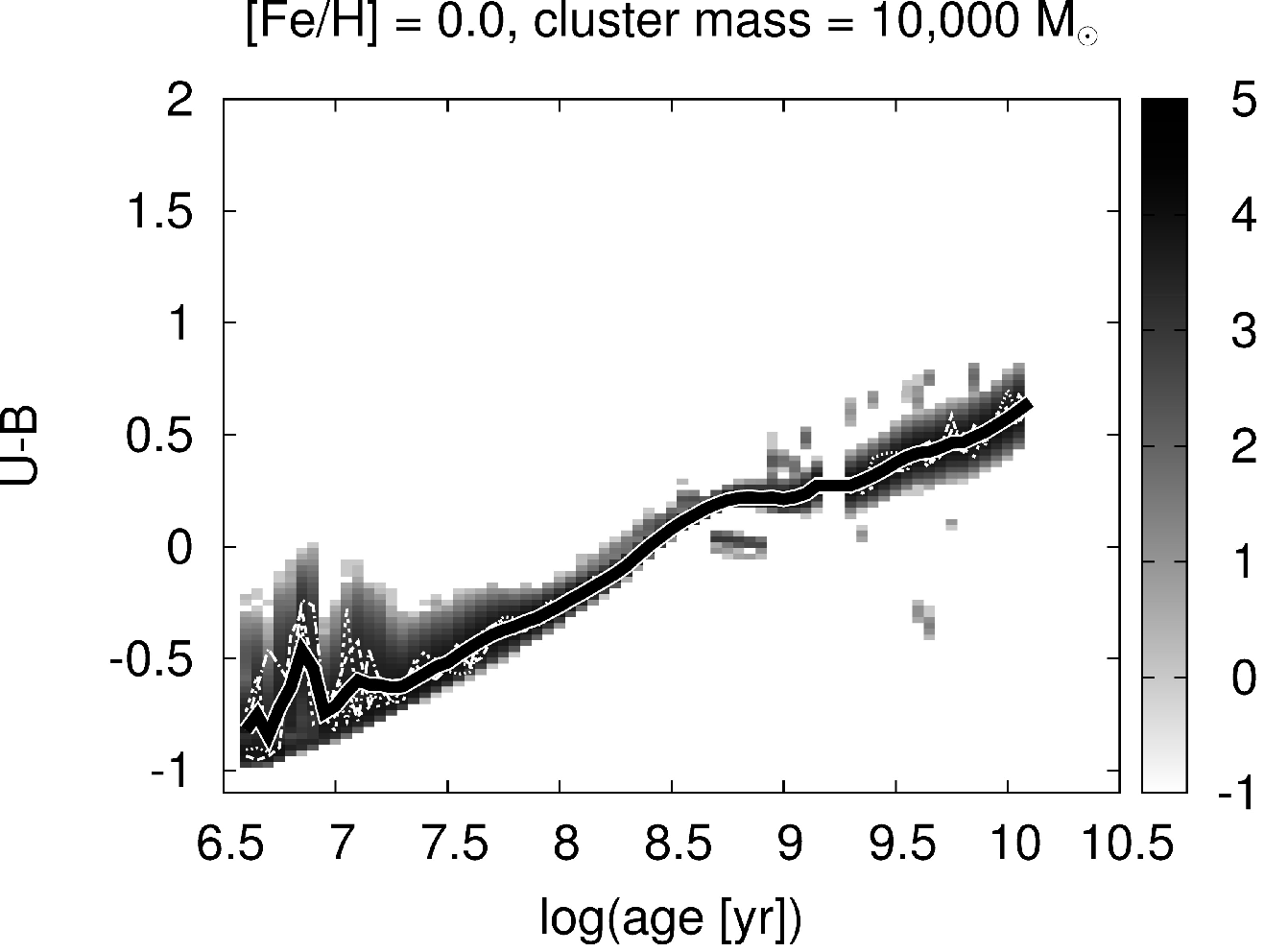} & 
   \includegraphics[angle = 270,width = 0.3\linewidth]{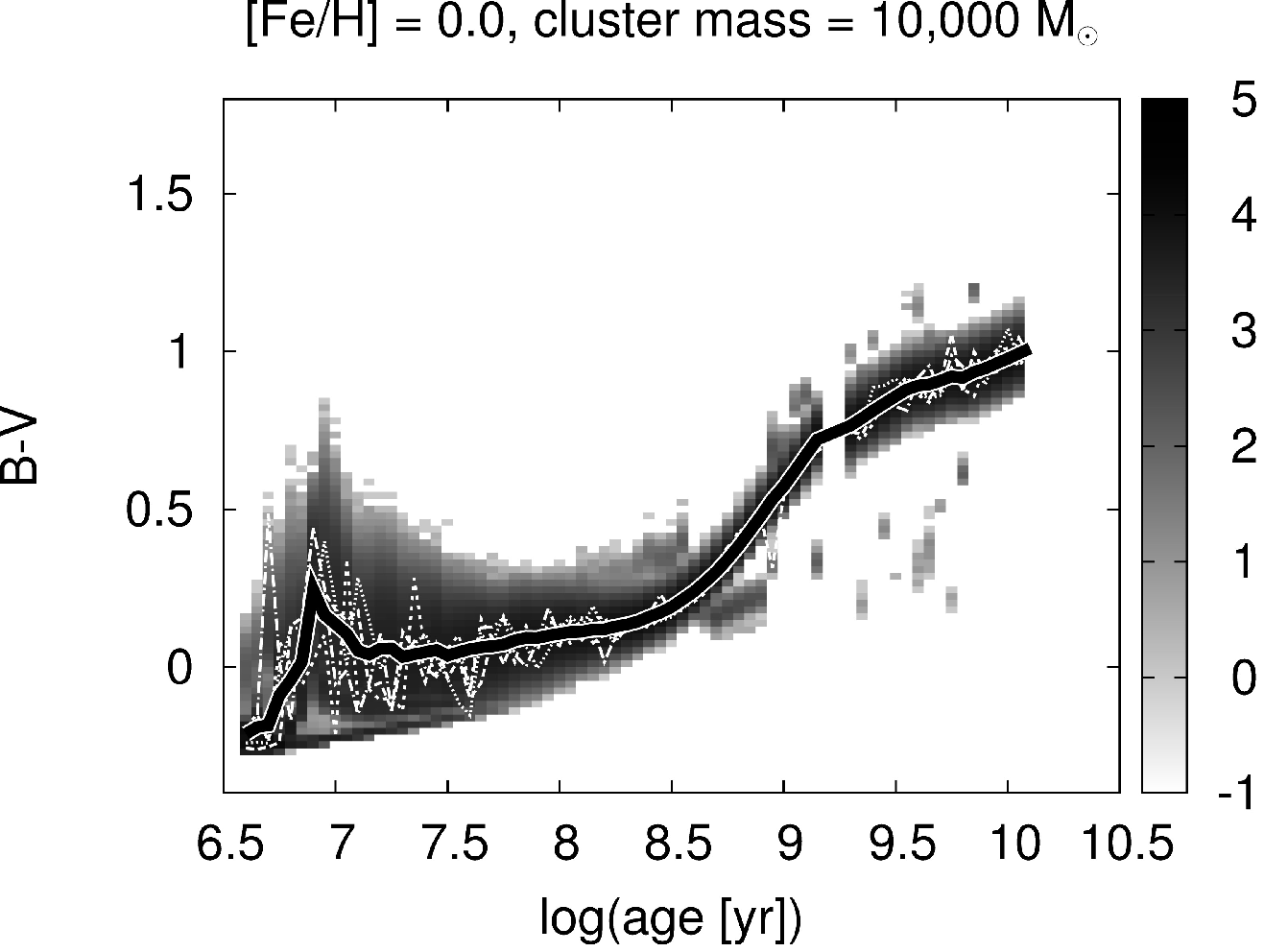} & 
   \includegraphics[angle = 270,width = 0.3\linewidth]{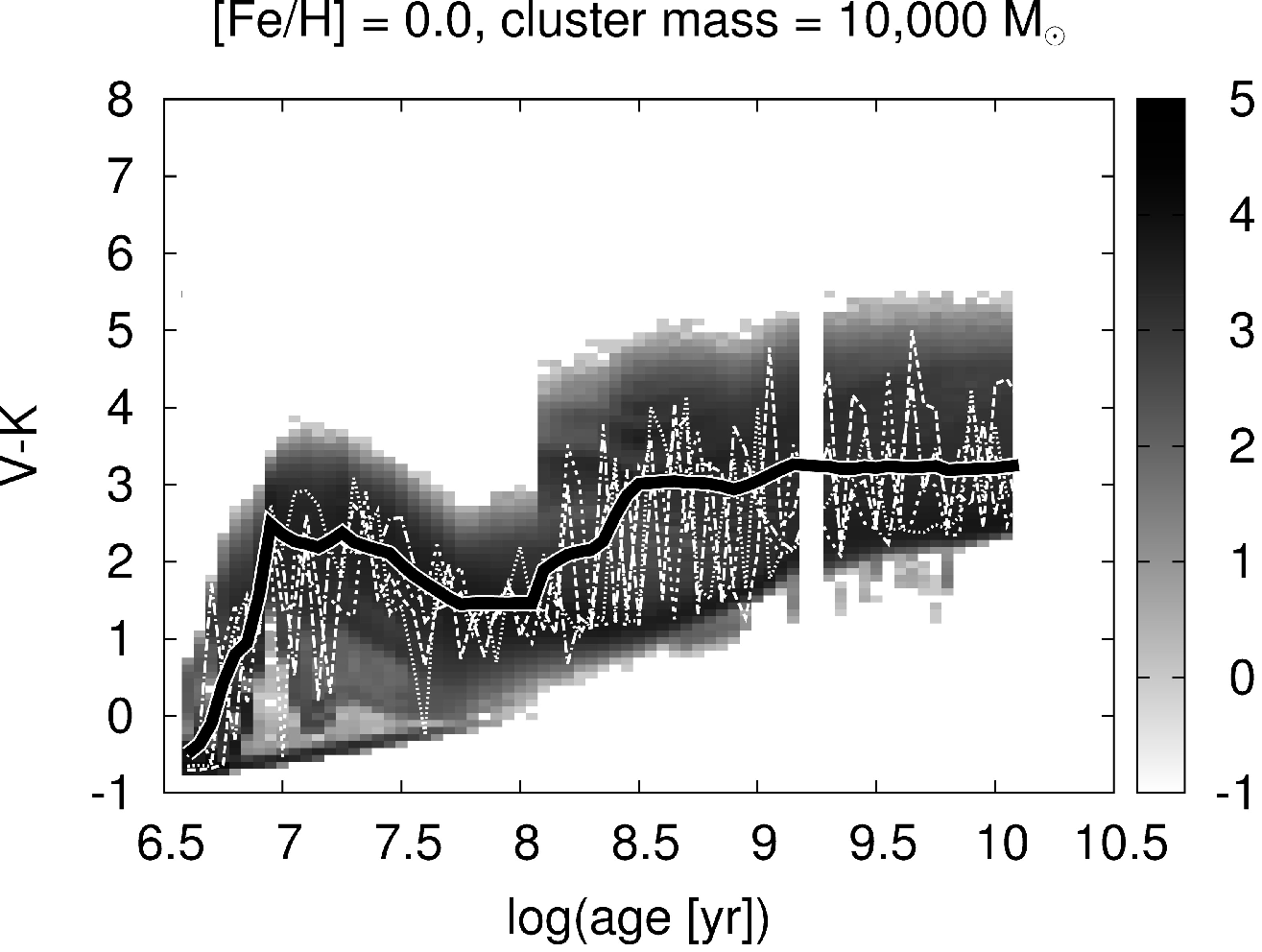} \\
   \includegraphics[angle = 270,width = 0.3\linewidth]{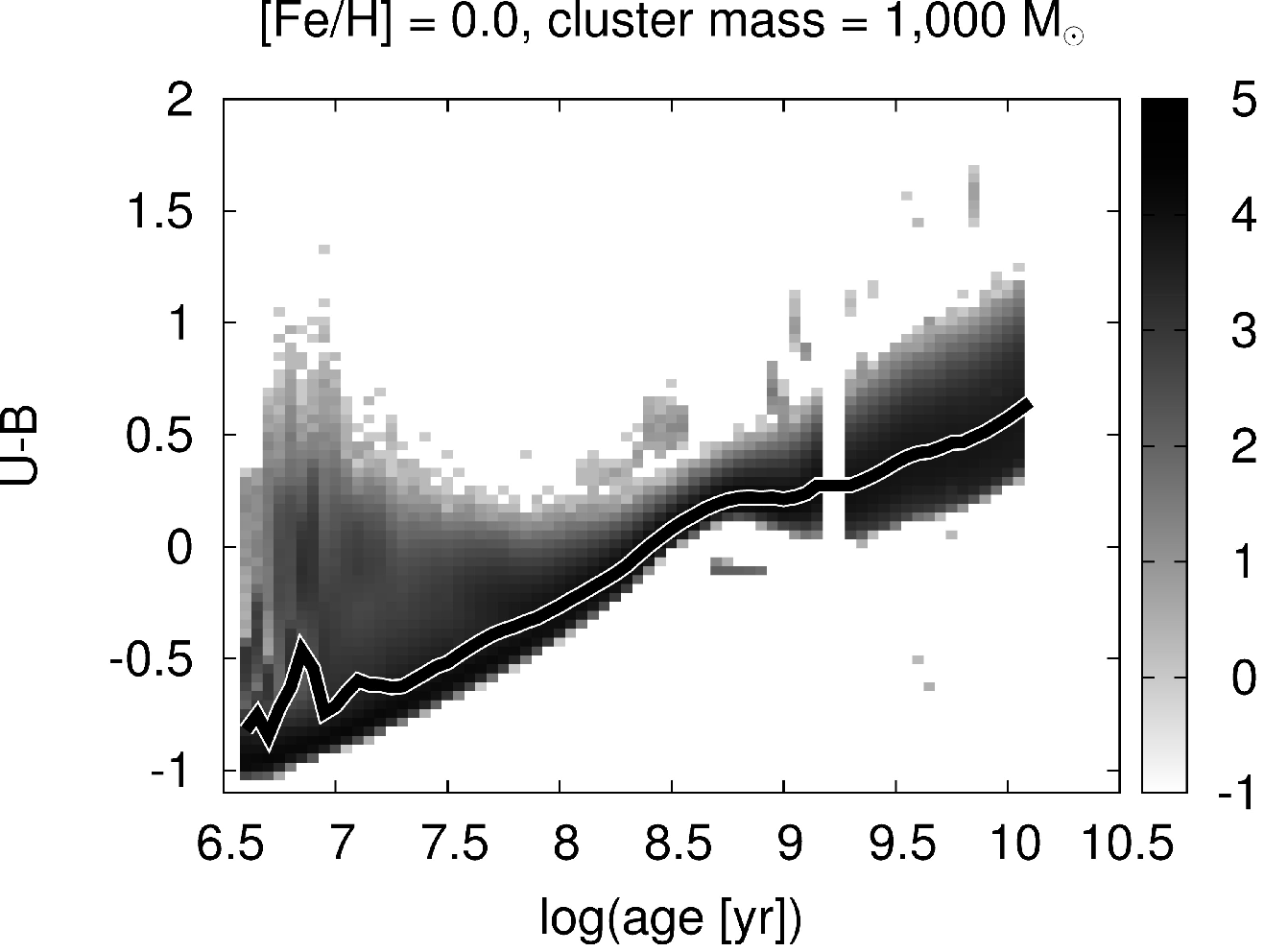} & 
   \includegraphics[angle = 270,width = 0.3\linewidth]{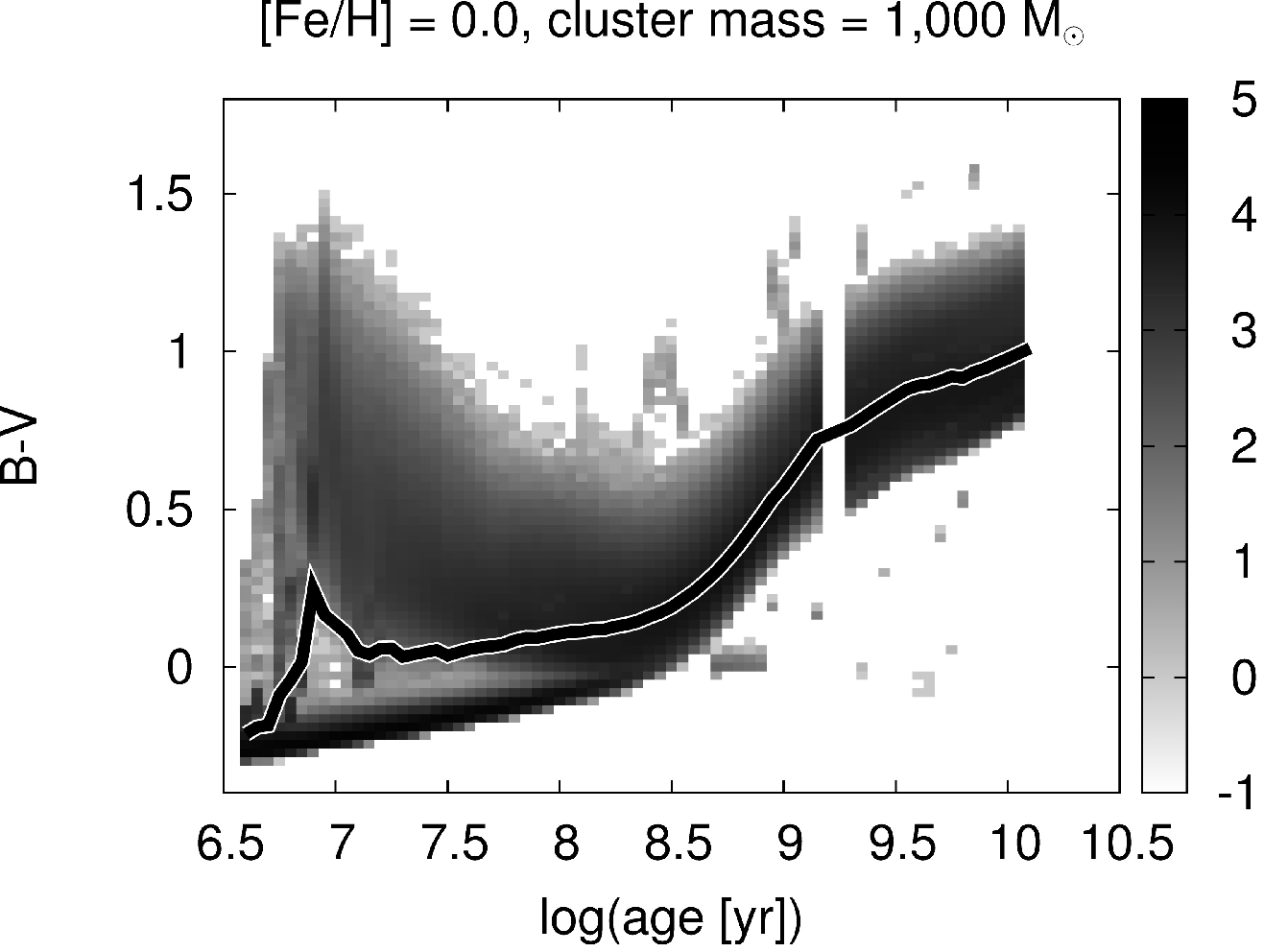} & 
   \includegraphics[angle = 270,width = 0.3\linewidth]{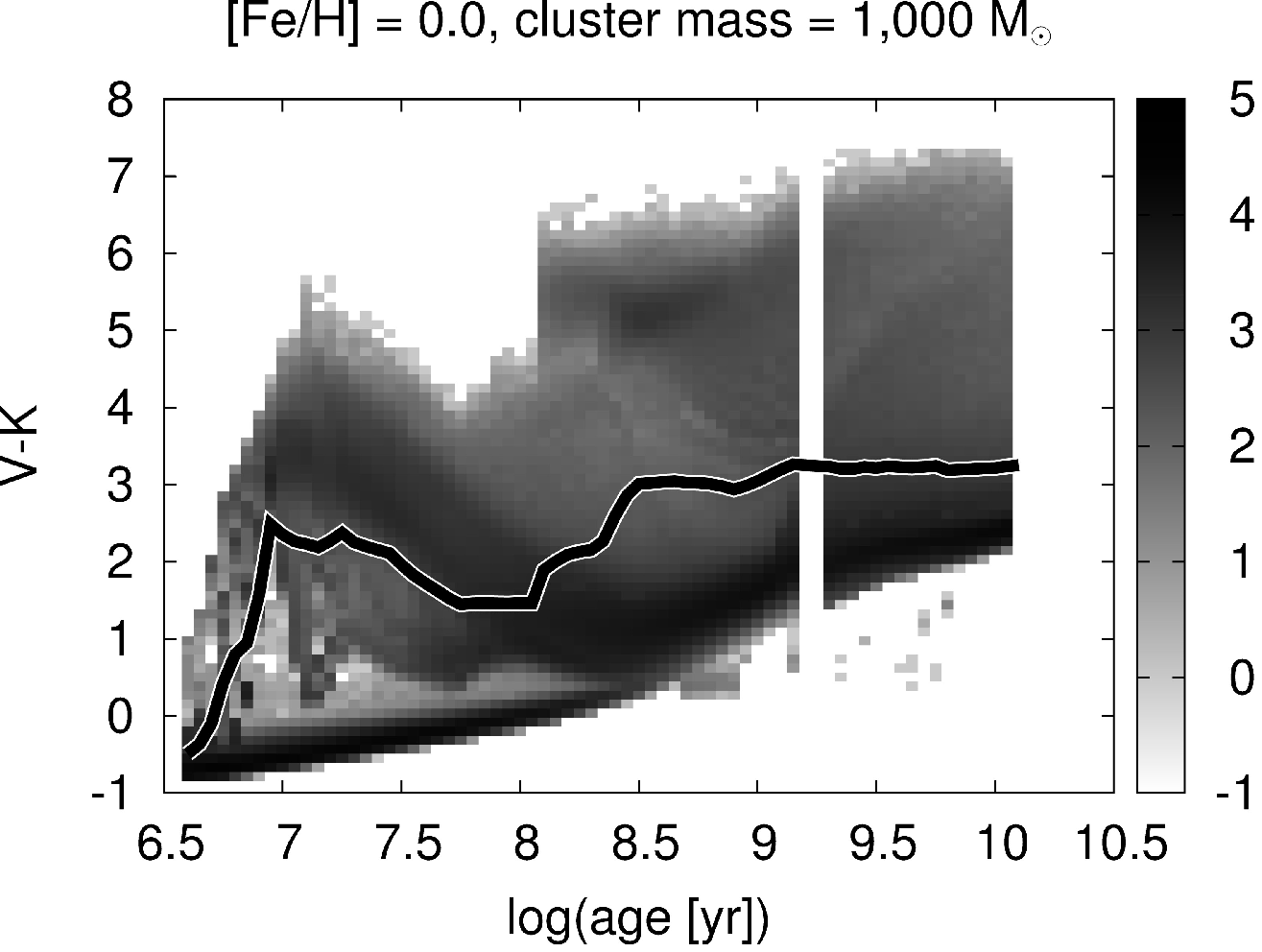} \\
  \end{tabular}
\end{center}
\caption{Examples of time evolution of various colors, presented smoothed to enhance clarity. Shown are cluster samples with cluster masses of $10^5$ \msunmsun (top row), $10^4$ \msunmsun (middle), and $10^3$ \msunmsun (bottom). Properties shown are the magnitudes in the Johnson $U-B$ color (left column), Johnson $B-V$ (middle), and Johnson $V$ $-$ Bessel-Brett $K$ (right). Greyscale represents the log(number of models per bin). The models have solar metallicity and no foreground extinction. The solid lines (with white shadow) represent the corresponding fully-sampled models. For cluster masses of $10^4$ \msun, 4 models of single clusters have been added to give an impression of their evolution.}
\label{fig:example_colors}
\end{figure*}

\begin{figure*}
\begin{center}
  \begin{tabular}{ccc}
   \includegraphics[angle = 270,width = 0.3\linewidth]{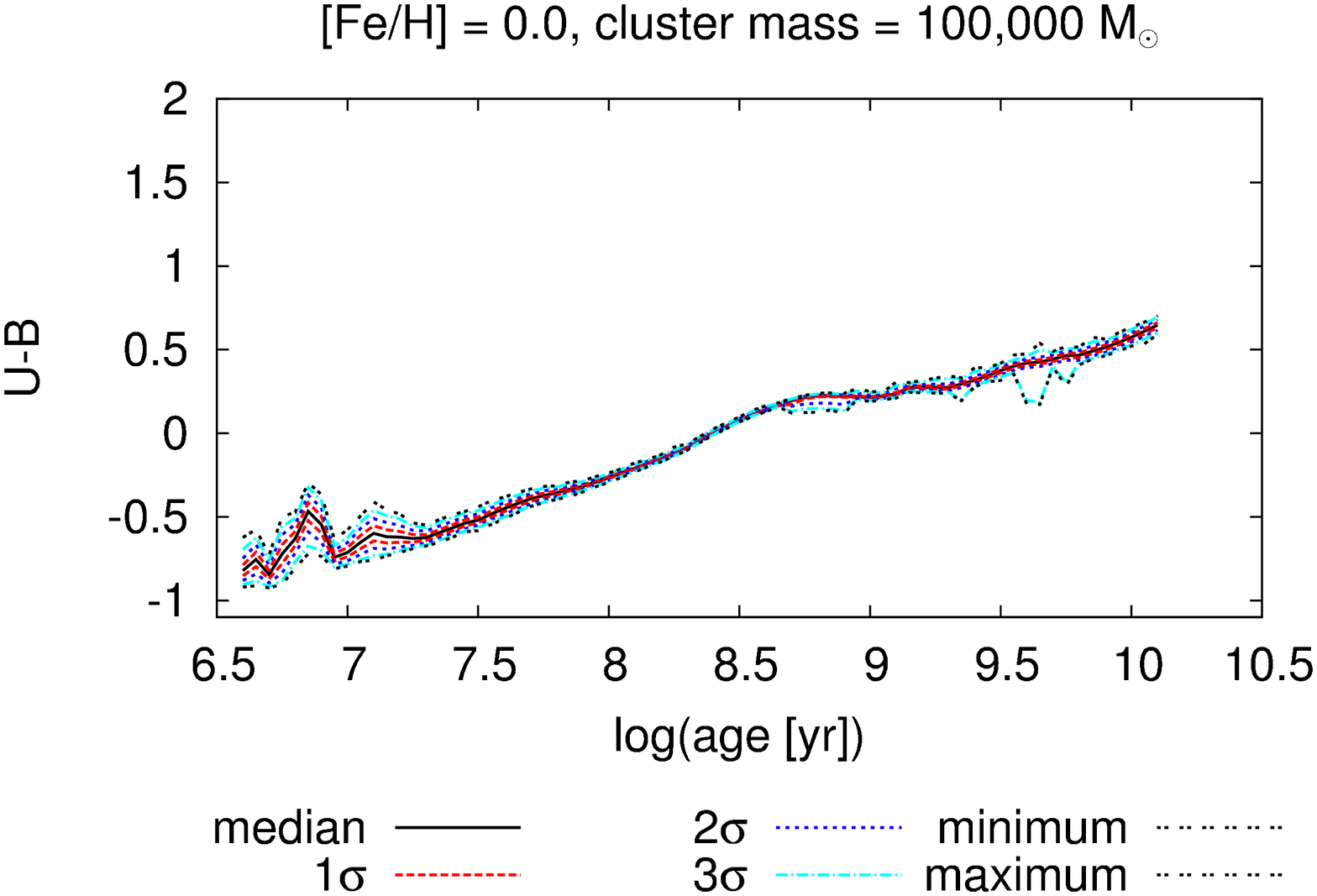} & 
   \includegraphics[angle = 270,width = 0.3\linewidth]{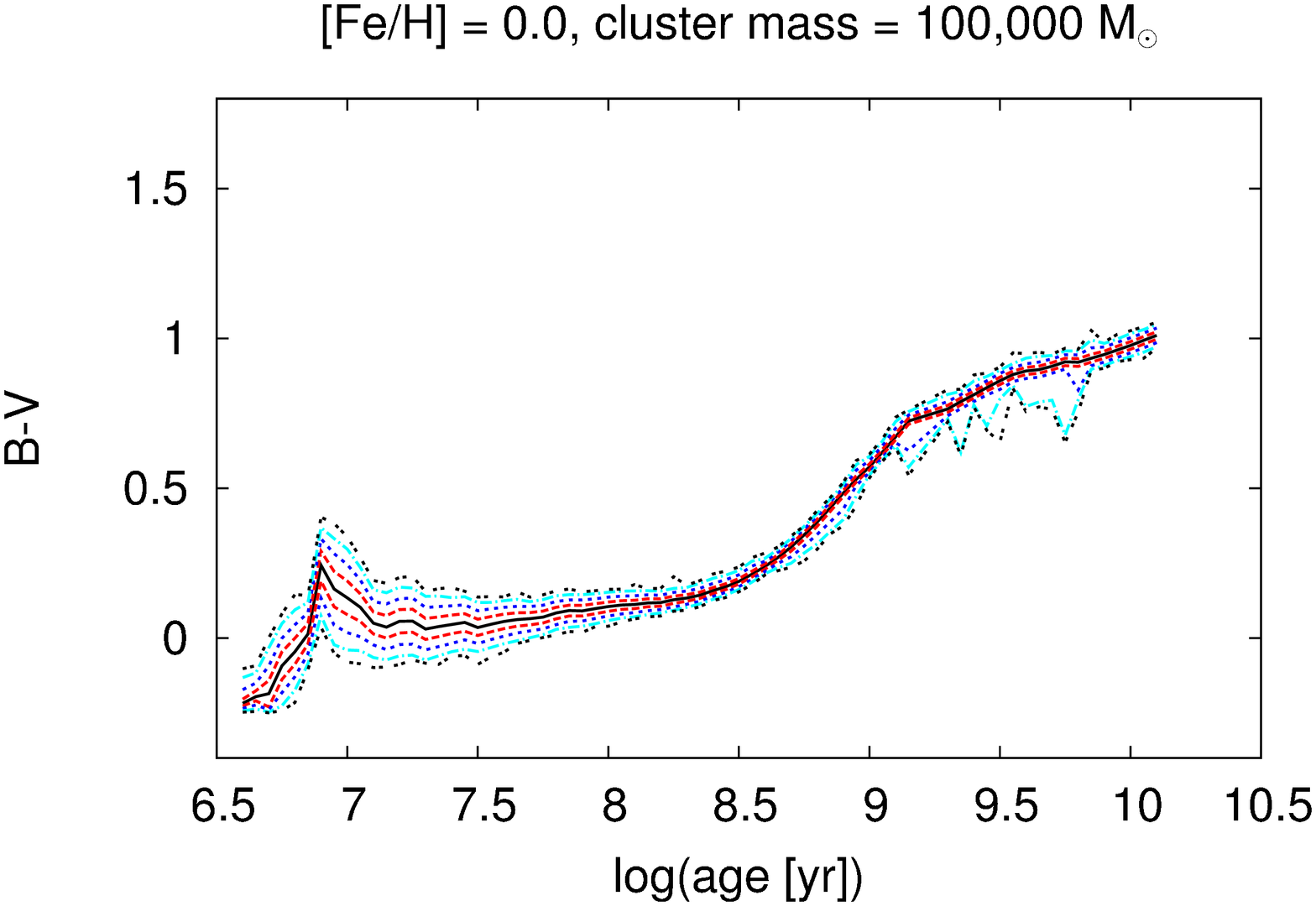} & 
   \includegraphics[angle = 270,width = 0.3\linewidth]{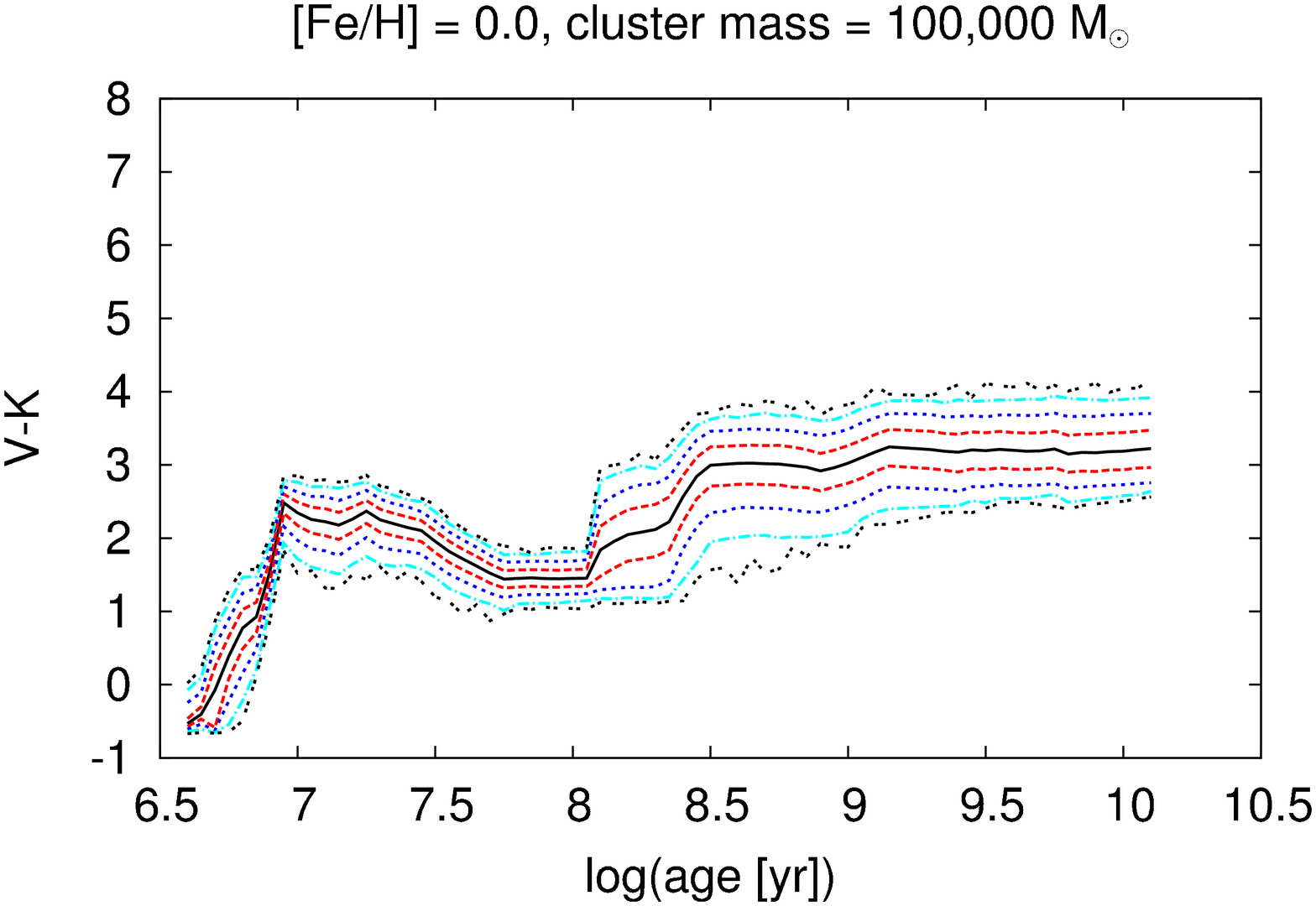} \\
   \includegraphics[angle = 270,width = 0.3\linewidth]{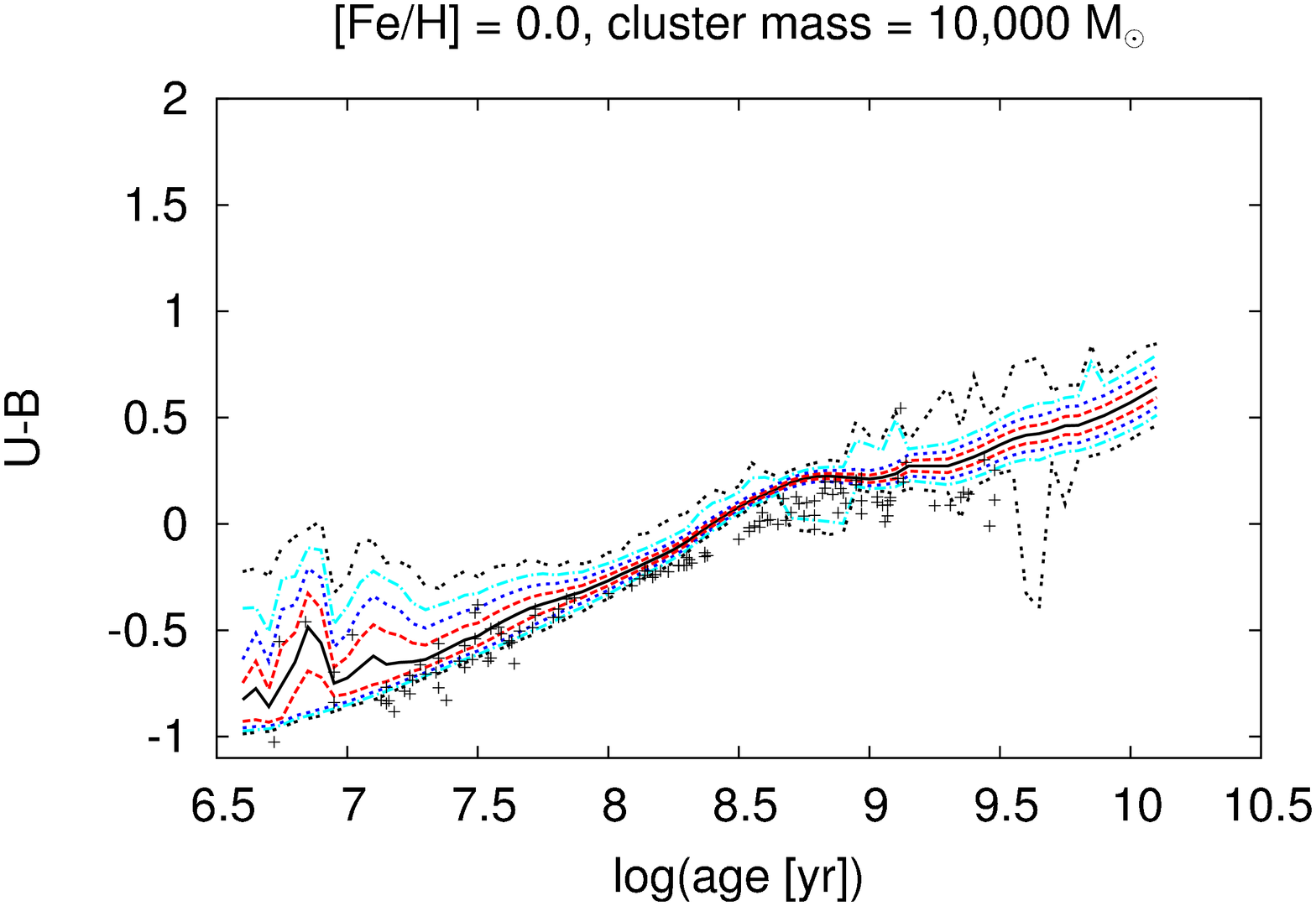} & 
   \includegraphics[angle = 270,width = 0.3\linewidth]{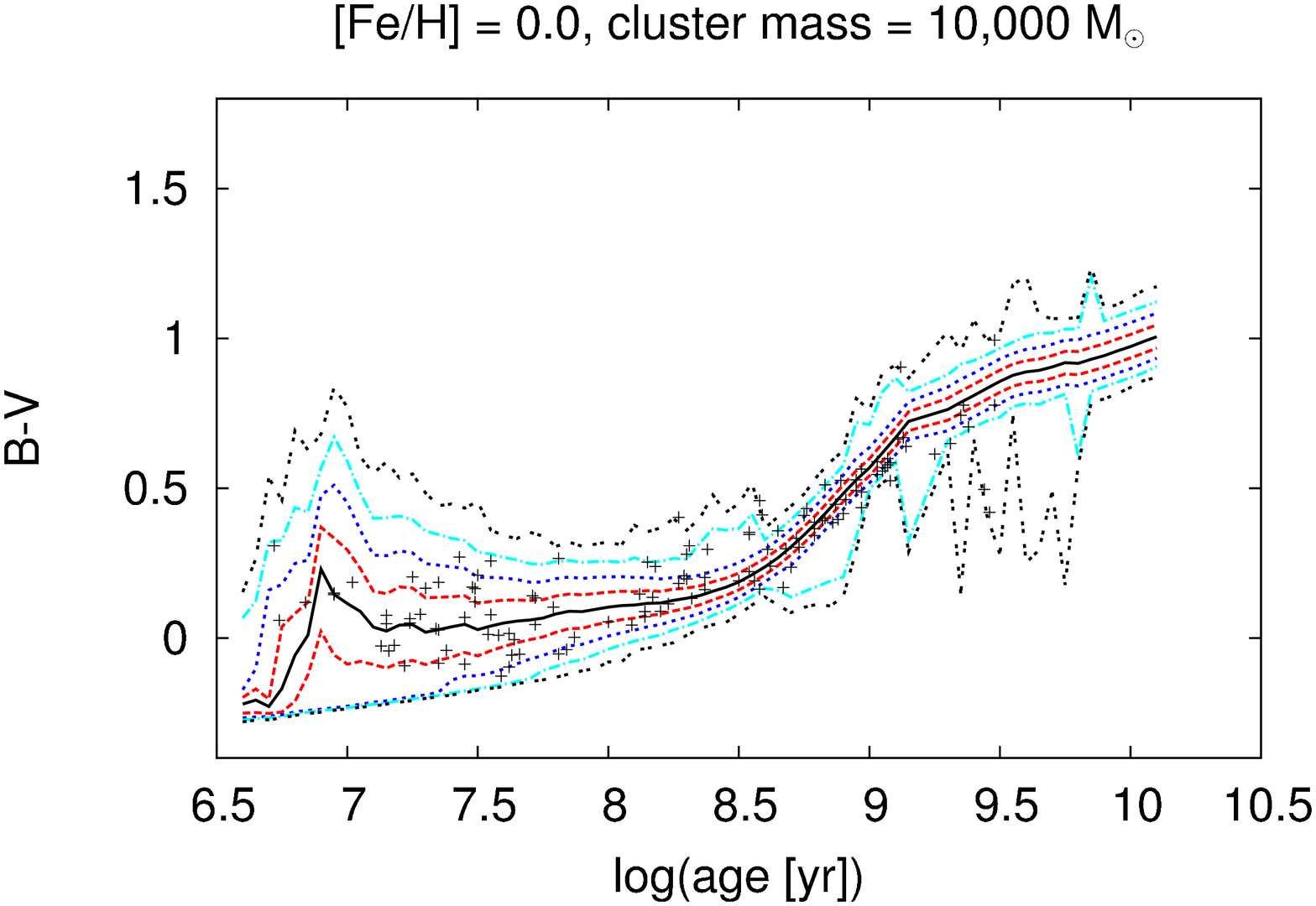} & 
   \includegraphics[angle = 270,width = 0.3\linewidth]{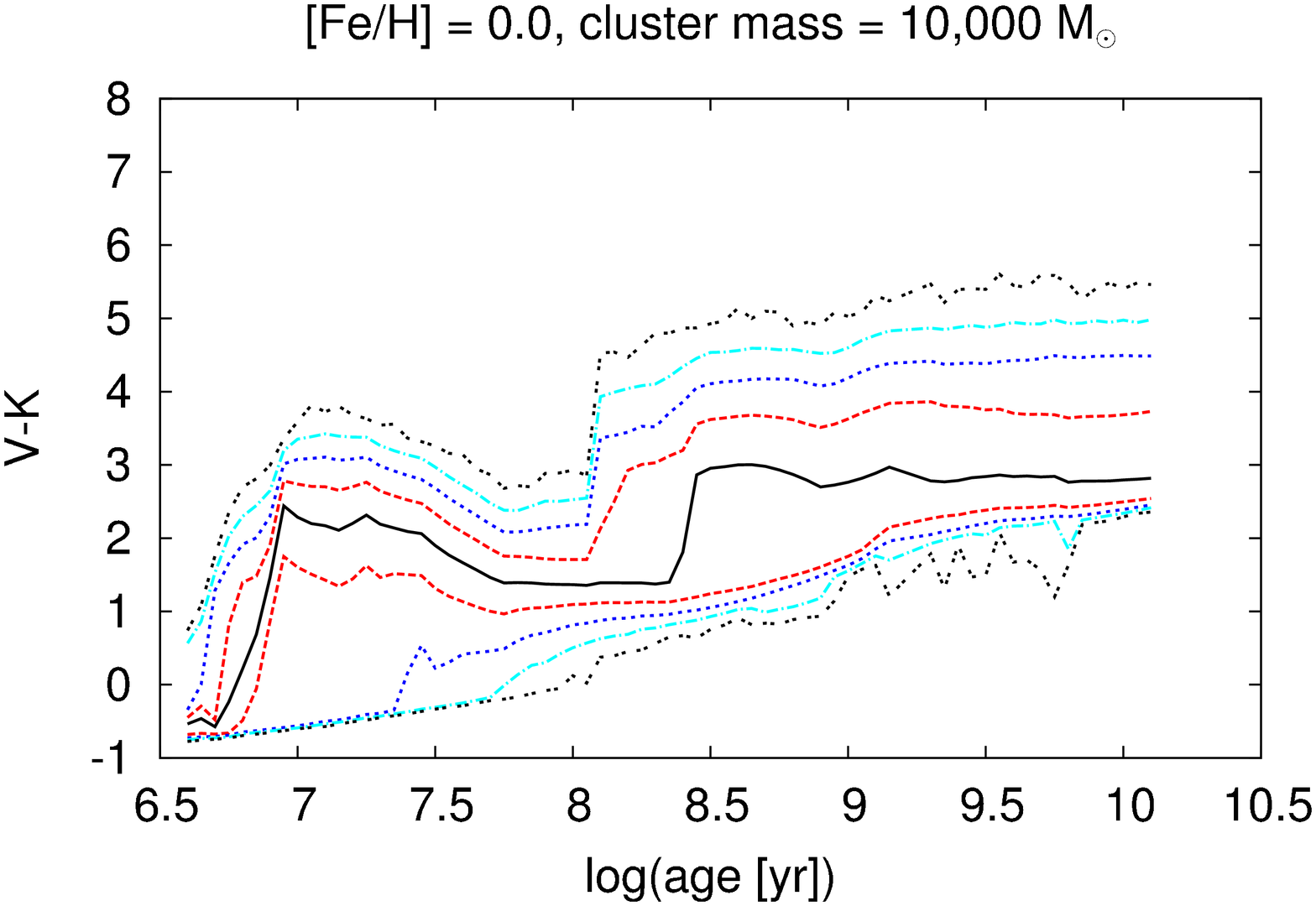} \\
   \includegraphics[angle = 270,width = 0.3\linewidth]{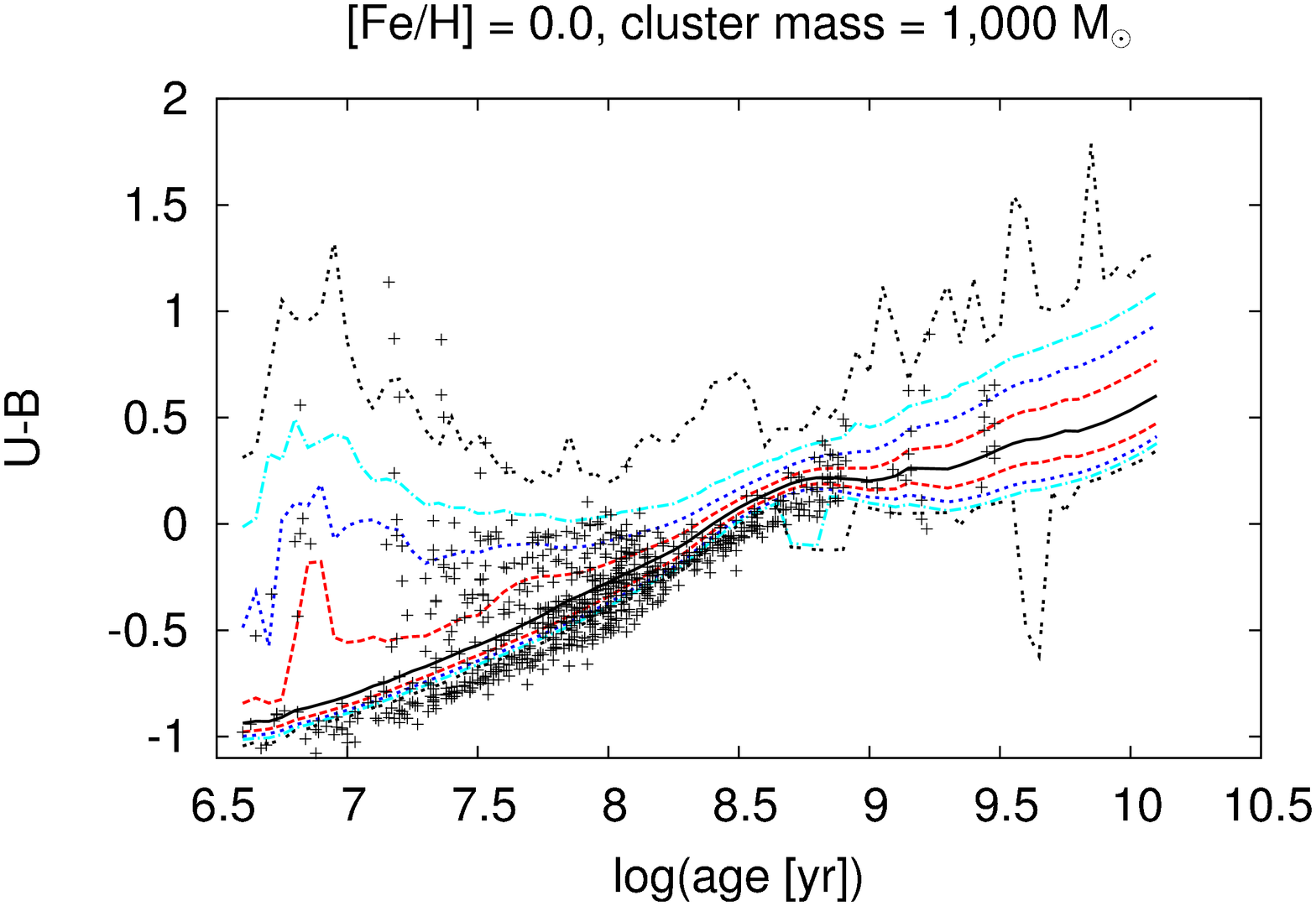} & 
   \includegraphics[angle = 270,width = 0.3\linewidth]{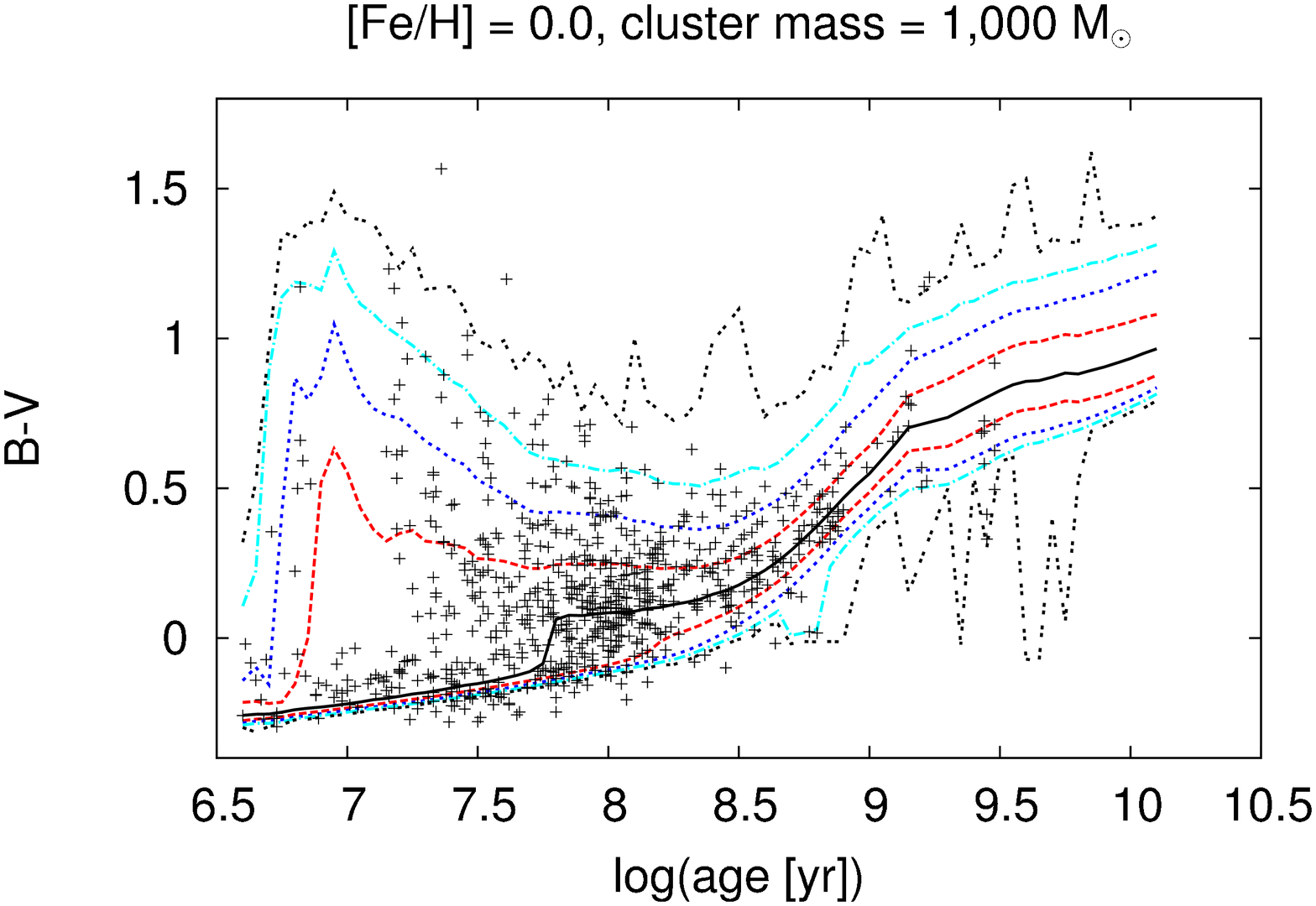} & 
   \includegraphics[angle = 270,width = 0.3\linewidth]{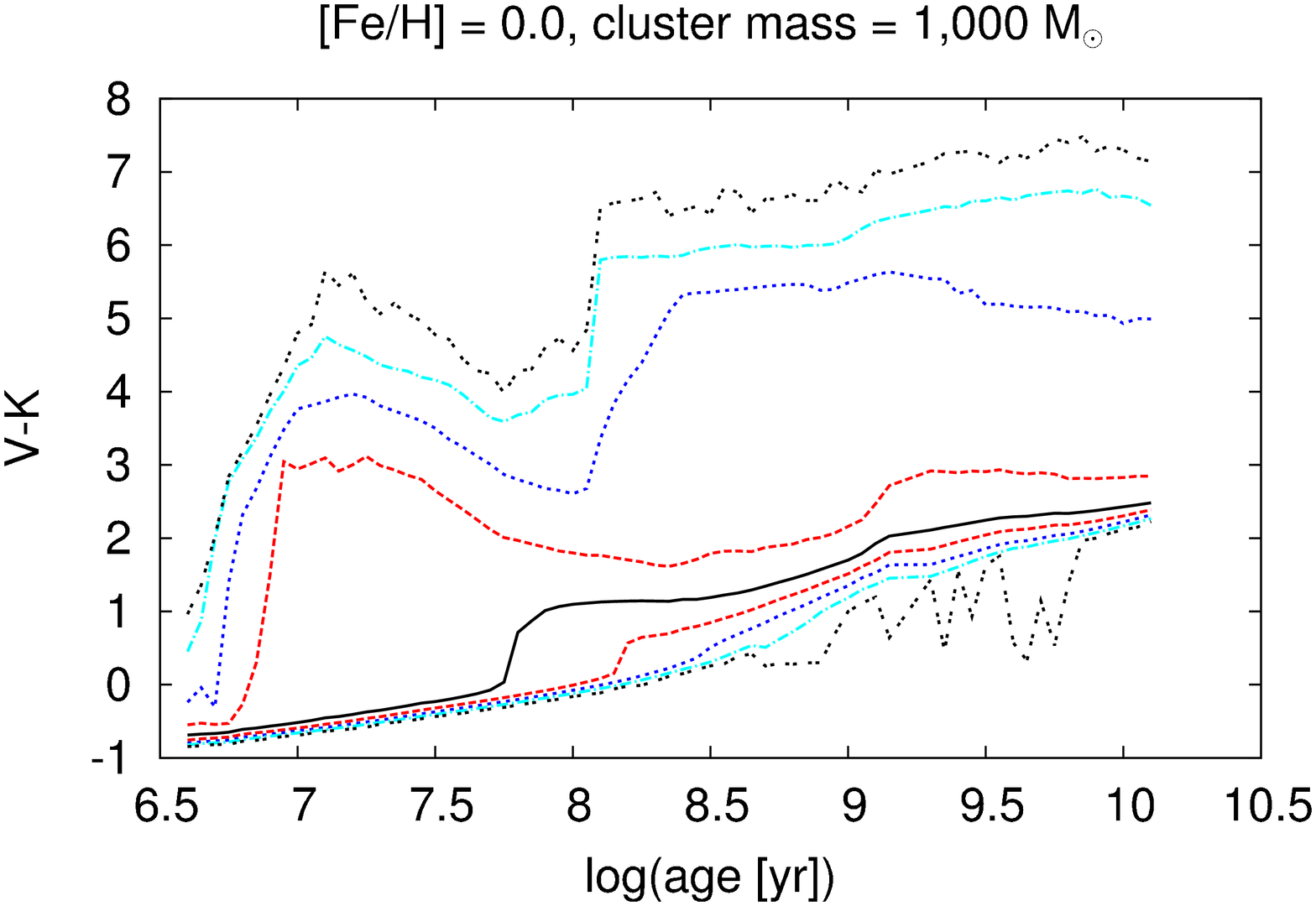} \\
  \end{tabular}
\end{center}
\caption{Examples of time evolution of various colors. Shown are cluster samples with cluster masses of $10^5$ \msunmsun (top row), $10^4$ \msunmsun (middle), and $10^3$ \msunmsun (bottom). Properties shown are the magnitudes in the Johnson $U-B$ color (left column), Johnson $B-V$ (middle), and Johnson $V$ $-$ Bessel-Brett $K$ (right). Shown are the median, 1$\sigma$, 2$\sigma$, and 3$\sigma$ quantiles as well as the minimum and maximum values for every input age. The models have solar metallicity and no foreground extinction. Panels for $10^3$ and $10^4$ \msunmsun show corresponding observations of integrated stellar cluster colors from \citet{2003AJ....126.1836H} and ages for ranges of derived cluster mass (derived mass $\le 4 \times 10^3$ \msunmsun for model mass of $10^3$ \msun, $4 \times 10^3 <$ derived mass $\le 15 \times 10^3$ \msunmsun for model mass of $10^4$ \msun; ages and masses derived by \citealt{2012ApJ...751..122P}). As the observations do not extend to NIR passbands, $V-K$ lacks the associated data.}
\label{fig:example_colors2}
\end{figure*}

The time evolution of some regularly used standard colors is presented in Figs. \ref{fig:example_mags} $-$ \ref{fig:example_colors2}. To indicate the density of model predictions, greyscale colors have been used in logarithmic intervals in Figs. \ref{fig:example_mags} and \ref{fig:example_colors}. The black solid lines (surrounded by white shading) represent the fully-sampled models for comparison. In color evolution plots for $10^4$ \msunmsun clusters, 4 white lines show the time evolution of 4 individual cluster models. Figs. \ref{fig:example_mags2} and \ref{fig:example_colors2} show the same distributions, based on the quantiles for each input age, representing the median value, 1$\sigma$/2$\sigma$/3$\sigma$-equivalent ranges (each represented by the upper and lower quantiles, with the respective ranges in between) as well as the minimum and maximum value of the distributions. Depending on the cluster mass, even the median color and absolute magnitude change. These changes represent the combined effect of the varying impact of stochasticity as a function of cluster mass and to a lesser degree the finite number of calculated models. The GALEV web-page provides the scripts to make similar plots for the colors chosen and downloaded. For some distributions and input ages, the median values are located in sparsely populated regions between two ridges with large numbers of models. However, the median represents the best statistical property of these distributions in combination with the quantiles.

Compared to earlier research, we provide the stochastic cluster models for public access and for a significantly extended parameter range (see Sect. \ref{sec:massclean}). Compared to MASSCLEAN (\citealt{2009AJ....138.1724P}), the only other available set of stochastic IMF models, we extend the provided metallicity coverage (15 metallicities compared to 2 metallicities provided by MASSCLEAN), magnitudes (184 filters instead of 13), and provide models with foreground extinction $E(B-V)$ for a larger number of models (see Table \ref{tab:modelnumbers}) for 9 cluster masses (instead of 4). More detailed comparison is presented in Sect. \ref{sec:massclean}.

Fig. \ref{fig:example_mags} $-$ \ref{fig:example_colors2} show the main characteristics of the models: 
\be
\item The distributions of model predictions with stochastic IMF sampling show a wide spread around the predictions with fully-sampled IMF description;
\item The distributions show substructure based on the stochastic occurrence of a small number of bright stars (e.g. supergiants, AGB stars). Individual bright stars cause certain regions of the model predictions to be far off the main distribution;
\item We compared the following quantiles: 1$\sigma$ for total cluster mass \Mcl = $10^3$ \msun, 2$\sigma$ for \Mcl = $10^4$ \msun, and 3$\sigma$ for \Mcl = $10^5$ \msun. For blue colors (such as $U-B$ and $B-V$) these quantiles are comparable (with 1$\sigma$ for $10^3$ \msunmsun being slightly the largest). For red colors (such as $V-I$ and $V-K$) and ages $>$ 100 Myr the 1$\sigma$ for \Mcl = $10^3$ \msunmsun quantiles are more narrow than the other quantiles. The quantiles bluer than the median behave more smoothly than the quantiles redder than the median, as stochastic outliers in the stellar population are more likely to have redder colors (such as red supergiants and AGB stars);
\item The individual models, shown for clusters with \Mcl = $10^4$ \msun, indicate strong jumps between successive age steps, tracing the evolution and death of individual stars;
\item The spread of model predictions around the median prediction increases with decreasing cluster mass. This effect is also seen in observations (see Fig. \ref{fig:example_colors}). The cluster masses and ages shown were derived by analyzing the observed SEDs with stochastic MASSCLEAN models (\citealt{2012ApJ...751..122P});
\item Observed cluster colors, for derived cluster ages and masses, follow the distributions of our model predictions. Observed $U-B$ colors are slightly bluer than the model predictions. However, the observed $U$-band was not a standard Johnson $U$-band filter while the model predictions are for the standard Johnson $U$-band;
\item Deviations of stochastic model predictions from fully-sampled model predictions in different colors are not directly correlated, but follow wide-spread distributions as well (see color-color plots in Fig. \ref{fig:example_mags}, right column).
\ee

Each stochastic IMF has been used to create stochastic models for 15 metallicities: 
$Z$ = 0.0001, 0.0002, 0.0004, 0.0006, 0.0008, 0.001, 0.002, 0.004, 0.006, 0.008, 0.010, 0.015, 0.020, 0.025, 0.030, equivalent to \feh =  $-$2.3, $-$2.0, $-$1.7, $-$1.5, $-$1.4, $-$1.3, $-$1.0, $-$0.7, $-$0.5, $-$0.4, $-$0.3, $-$0.125, 0.0, $+$0.1, $+$0.18 dex, resp. Models are provided in 184 filters (see Table \ref{tab:filter}) for foreground extinction values $E(B-V)$ = 0.0 $-$ 2.0 mag (in steps of $\Delta$$E(B-V)$ = 0.2 mag). While magnitudes for the full set of filters  given in Table \ref{tab:filter} are provided for the ``main model set'', only magnitudes in the bold-faced filters are provided for additional models. These additional models are only available for zero foreground extinction, i.e. $E(B-V)$ = 0.0 mag. ``Main models'' and ``additional models'' are provided for the same isochrones and metallicities, while differences occur in the number of values provided for foreground extinction and magnitudes. The relative number of models varies, depending on cluster mass (see Table \ref{tab:modelnumbers}). All models are available online at http$://$data.galev.org$/$models$/$anders13 .

\begin{table*}
\caption{Filters provided for the main set of model predictions. Only filters of the bold-faced filter systems are provided for additional models. The models are available at http$://$data.galev.org$/$models$/$anders13 .} 
\begin{center}
\begin{tabular}{l l}
\hline
Filter system & Filter\\
\hline
\bf{Johnson} & {\it U B V R I}\\
\bf{Cousins} & {\it R I}\\
\bf{Bessel/Brett} & {\it J H K L}\\
\bf{Stroemgren} & {\it u v b y}\\
\bf{Washington} & {\it C M T1 T2}\\

SDSS (\citealt{2010AJ....139.1628D}) & {\it u g r i z}\\
\bf{2MASS} & {\it J H K$_s$}\\
\bf{VISTA} & {\it Z Y J H K$_s$}\\
UKIRT $-$ WFCAM & {\it Z Y J H K}\\

HST $-$ WFPC2 & F160BW F170W F185W F218W F255W F300W F336W F380W F410M F439W F450W\\
  & F467M F547M F555W F569W F606W F622W F675W F702W F791W F814W\\
HST $-$ NICMOS NIC1 & F090M F110M F110W F140W F145M F160W F165M F170M\\
HST $-$ NICMOS NIC2 & F110W F160W F171M F180M F187W F204M F205W  F207M F222M F237M\\
HST $-$ NICMOS NIC3 & F110W F150W F160W F175W F205M F222M F240M\\
HST $-$ ACS $-$ HRC & F220W F250W F330W F435W F475W F550M F555W F606W F625W\\
  & F775W F814W F850LP\\ 
HST $-$ ACS $-$ WFC & F435W F475W F550M F555W F606W F625W F775W F814W F850LP\\
HST $-$ WFC3 $-$ UVIS & F200LP F218W F225W F275W F300X F336W F350LP F390M F390W\\
  & F410M F438W F467M F475W F475X F547M F555W F600LP F606W F621M F625W\\
  & F689M F763M F775W F814W F845M F850LP\\
HST $-$ WFC3 $-$ IR & F098M F105W F110W F125W F127M F139M F140W F153M F160W\\

\bf{GALEX} & FUV NUV\\
\bf{Spitzer $-$ IRAC} & 3.5$\mu$m 4.5$\mu$m 5.7$\mu$m 8.0$\mu$m\\
WISE & {\it W1}(3.4$\mu$m) {\it W2}(4.6$\mu$m) {\it W3}(12$\mu$m) {\it W4}(22$\mu$m)\\

ESO VLT $-$ FORS & Bessel-{\it U} Bessel-{\it B} Bessel-{\it V} Bessel-{\it R} Bessel-{\it I} Gunn-{\it u} Gunn-{\it v} Gunn-{\it g} Gunn-{\it r} Gunn-{\it z}\\
  & Special-{\it U} Special-{\it R}\\
ESO VLT $-$ ISAAC & {\it J J$_s$ H K$_s$ L}\\
ESO VLT $-$ HAWK$-$I & {\it Y J H K}\\
ESO 2.2m WFI & {\it U38 U50 B99 B123 V89 Rc162}\\
SUBARU $-$ Suprimecam & {\it B V R I g r i z}\\

\hline
\end{tabular}
\label{tab:filter}
\end{center}
\end{table*}

\begin{table}
\caption{Number of stochastic cluster models provided, for both the full list of magnitudes and the additional models for a restricted list of magnitudes. $^1$: USML = 42.7 \msunmsun (\citealt{2007ApJ...671.1550P}).}
\begin{center}
\begin{tabular}{l r r}
\hline
Cluster mass (\msun)& $\#$models & additional $\#$models \\
\hline
$10^3$              & 114,000 & 158,000\\
$10^3$  \& USML$^1$ & 47,000 & 156,000\\
$2 \times 10^3$     & 76,000 & 149,000\\
$5 \times 10^3$     & 89,000 & 140,000\\
$10^4$            & 72,000 & 137,000\\
$2 \times 10^4$   & 63,000 & 127,000\\
$5 \times 10^4$   & 60,000 & 71,000\\
$10^5$            & 53,000 & 60,000\\
$2 \times 10^5$   & 49,000 & 46,000\\
\hline
\end{tabular}
\label{tab:modelnumbers}
\end{center}
\end{table}

\subsection{Comparison with MASSCLEAN models}
\label{sec:massclean}

The only other set of available model predictions with stochastically-sampled IMFs are the MASSCLEAN models of \citet{2009AJ....138.1724P,2010ApJ...713L..21P}. \citet{2012ApJ...751..122P} presents an application to stellar clusters in the Large Magellanic Cloud. MASSCLEAN combines stochastically-sampled IMFs with stellar magnitudes available from the isochrones. Therefore, it is limited to the choice of filters and spectral library used for the isochrone generation. The GALEV code allows easy exchange of the isochrones, spectral libraries, and set of filter response curves. This facilitates significant flexibility, which will be explored in future releases.

The currently available version is MASSCLEAN 2.012. Compared with MASSCLEAN, our GALEV models extend the capabilities significantly:
\be
\item MASSCLEAN provides 2 metallicities (\feh $=$ $-$0.4 and \feh $=$ 0.0 dex), while we provide 15 metallicities ($-$2.3 $\le$ \feh $\le$ $+$0.18 dex);
\item MASSCLEAN provides magnitudes for 13 filters ($UBVRIJHK$ and ESO/VISTA $ZYJHK_s$), while we provide magnitudes for 184 filters (including filters of major surveys and all broad-band \& medium-band Hubble Space Telescope [HST] filters, see Table \ref{tab:filter});
\item MASSCLEAN provides a limited number of models ($\approx$ 5000 models for 4 cluster masses), while we provide a significantly larger number of models for 9 cluster masses (see Table \ref{tab:modelnumbers}).
\ee

The MASSCLEAN models had already shown behavior very similar to the stochastic models presented here: a systematic spread of the model magnitudes and colors around the limit for stellar clusters of infinite mass (i.e. using the fully-sampled IMF description), with the spread increasing with decreasing cluster mass.

\section{Determination of parameters}
\label{sec:paradet}

The derived integrated photometry of the stochastically-sampled models was used to re-determine the physical properties of the stellar cluster (i.e. cluster age, mass, foreground extinction $E(B-V)$, and metallicity \feh) to study the analysis accuracy of observed cluster photometry. The SED of each stellar cluster was analyzed with fully-sampled IMF models, to mimic the approach regularly adopted for observed cluster photometry. For this analysis the program ``AnalySED'' (\citealt{2004MNRAS.347..196A}) was used.

Briefly, ``AnalySED'' performs the following steps:
\be
\item The cluster SED is compared with all model SEDs;
\item Scaling determines the cluster mass;
\item $\chi^2$ determination of the scaled cluster SED with respect to all model SEDs;
\item Each model SED is assigned a probability based on the $\chi^2$ value determined;
\item If the total probability $< 10^{-30}$ (i.e. all $\chi^2$ values $>$ 70) the fit is considered as ``poorly-fitting'';
\item Scaling of the individual probabilities to achieve a total probability of unity;
\item The model SED (and its associated cluster parameters) with the highest probability is considered the best fit;
\item Models with decreasing probabilities are used to determine the parameter uncertainties, until the summed probability reaches 68.26 \% (1$\sigma$ uncertainty range).
\ee

The ability to determine a best-fitting model and the cluster parameters of the best fit depend on various factors (see e.g. \citealt{2004MNRAS.347..196A,2013A&A...550A..20D}):
\bi
\item The combination of filters used;
\item The photometric uncertainties;
\item The ranges of metallicity and foreground extinction explored in the fitting process.
\ei

\section{Results}
\label{sec:results}

In this section we explore the accuracy of parameter determination (i.e. cluster age, mass, foreground extinction $E(B-V)$, and metallicity \feh) of stochastically-sampled clusters. In Sect. \ref{sec:fitpar} we will investigate the impact of photometric uncertainties $\sigma$, and allowed ranges of metallicity \& extinction for fitting dust-free input clusters with solar metallicity of $10^3$ \msun, $10^4$ \msun, and $10^5$ \msun. The results will be used in Sect. \ref{sec:massesfilters} to explore the effects for a wide range of cluster parameters, and filter combinations.

\subsection{Impact of fitting parameters}
\label{sec:fitpar}

We first study the special case of dust-free input clusters with solar metallicity (\feh=0 and $E(B-V)$=0) to illustrate the method and highlight findings applicable to stochastic clusters in general. A more general study of clusters with non-solar metallicity and finite foreground extinction is discussed below in Sect. \ref{sec:massesfilters}.

First, we investigate which fraction of stochastic models can be fitted at all with the fully-sampled models. A summary is provided in Table \ref{tab:numbers}. We investigated three sets of allowed ranges in metallicity and foreground extinction: 
\bi
\item Range1: Restricting to the input values (\feh = 0.0 dex and $E(B-V)$ = 0.0 mag);
\item Range2: Allowing both extinction and metallicity to vary over a narrow range around the input values ($-$0.125 $\le$ \feh $\le$ $+$0.1 dex and 0.0 $\le$ $E(B-V)$ $\le$ 0.2 mag);
\item Range3: Allowing both extinction and metallicity to vary over the full range around the input values ($-$2.3 $\le$ \feh $\le$ $+$0.18 dex and 0.0 $\le$ $E(B-V)$ $\le$ 2.0 mag).
\ei

Table \ref{tab:numbers} indicates three trends: 
\be
\item The fraction of poorly-fitted clusters increases with decreasing cluster mass. The lower the cluster mass is, the stronger the impact of stochastic IMF sampling, relative to the total stellar population in a stellar cluster (with related enhanced changes in the cluster SED). This effect is well-known and described, e.g. in \citet{2009AJ....138.1724P,2011A&A...529A..25S}.
\item For a given cluster mass, the allowed fitting ranges in metallicity and extinction strongly influence the number of poorly-fitted clusters: the smaller the allowed ranges around the input values are, the higher the number of poorly-fitted clusters. The age range with the highest fraction of fitted clusters depends on input cluster mass and fitting constraints. These results suggest large deviations between the original cluster parameters and the fit results, which will be studied below. 
\item For clusters with a total mass of $10^4$ \msunmsun the impact of the assumed photometric uncertainties was studied: The smaller the photometric uncertainty, the more significantly the derived $\chi^2$ value is affected by SED mismatches caused by the stochastic IMF sampling. This effect leads to an enhanced number of poorly-fitted clusters.
\ee

\begin{table*}
\caption{Fraction of models (in \%) found poorly-fitted during the AnalySED fitting, using $UBVRIJHK$ magnitudes. The stochastic clusters were created with \feh = 0.0 dex, foreground extinction $E(B-V)$ = 0.0 mag, and an assumed photometric uncertainty of $\Delta$mag $ = $ 0.1 mag for all magnitudes (unless specified). The percentages in columns 3 $-$ 6 represent the fractions of poorly-fitted models with input ages in four age ranges. The number of models in these age ranges are 12.7\%, 28.2\%, 28.2\%, and 30.9\% of the total number of models. Three different fitting ranges were investigated: Range1: \feh = 0.0 dex \& $E(B-V)$ = 0.0 mag (i.e. the true model values), range2: \feh = $-$0.125 $-$ $+$0.1 dex \& $E(B-V)$ = 0.0 $-$ 0.2 mag (i.e. small ranges around the true model values), and range3: \feh = $-$2.3 $-$ $+$0.18 dex \& $E(B-V)$ = 0.0 $-$ 2.0 mag (i.e. wide range in metallicity and extinction).} 
\begin{center}
\begin{tabular}{l l r r r r }
\hline
cluster mass (\msun) & fitting ranges & age $\le$ 10 Myr & 10 Myr $<$ age $\le$ 100 Myr & 100 Myr $<$ age $\le$ 1 Gyr & age $>$ 1 Gyr \\
\hline
$10^5$ & range1  & 0.0 & 0.0 & 0.0 & 0.0 \\
$10^5$ & range2  & 0.0 & 0.0 & 0.0 & 0.0 \\
$10^5$ & range3  & 0.0 & 0.0 & 0.0 & 0.0 \\

$10^4$           & range1  & 0.0 & 0.0 & 15.1 & 23.4 \\
$10^4$, 0.001 mag & range2  & 99.9 & 99.9 & 99.9 & 99.9 \\
$10^4$, 0.01 mag  & range2  & 74.5 & 62.5 & 83.3 & 89.3 \\
$10^4$, 0.1 mag   & range2  & 0.0 & 0.0 & 12.0 & 19.4 \\
$10^4$, 0.2 mag   & range2  & 0.0 & 0.0 & 6.0 & 5.6 \\
$10^4$           & range3  & 0.0 & 0.0 & 0.0 & 0.0 \\

$10^3$ & range1  & 1.6 & 1.3 & 20.0 & 54.8 \\
$10^3$ & range2  & 0.9 & 1.0 & 16.2 & 44.9 \\
$10^3$ & range3  & 0.0 & 0.0 & 2.5 & 0.2 \\
\hline
\end{tabular}
\label{tab:numbers}
\end{center}
\end{table*}

\begin{table*}
\caption{Cluster parameter determination accuracy based on AnalySED fitting, using $UBVRIJHK$ magnitudes and fully-sampled fitting models. The stochastic clusters were created with \feh = 0.0 dex, foreground extinction $E(B-V)$ = 0.0 mag, and an assumed photometric uncertainty of $\Delta$mag $ = $ 0.1 mag for all magnitudes. The values in columns 3 $-$ 6 represent the median and quantile 1$\sigma$ of fitted log(ages) with input log(ages), fitted log(mass) with input log(mass), and fitted $E(B-V)$ with input $E(B-V)$ in four input age ranges. Fitting ranges as in Table \ref{tab:numbers}.} 
\begin{center}
\begin{tabular}{l l l r r r r }
\hline
cluster mass (\msun) & fitting ranges & parameter & age $\le$ 10 Myr & 10 Myr $<$ age $\le$ 100 Myr & 100 Myr $<$ age $\le$ 1 Gyr & age $>$ 1 Gyr \\
\hline
$10^5$ & range1  & age & 0.0 $\pm$ 0.10 & 0.0 $\pm$ 0.25 & 0.0 $\pm$ 0.20 & 0.0 $\pm$ 0.20 \\ 
$10^5$ & range2  & age & 0.0 $\pm$ 0.10 & 0.0 $\pm$ 0.15 & 0.05 $\pm$ 0.15 & 0.0 $\pm$ 0.30 \\
$10^5$ & range3  & age & 0.0 $\pm$ 0.50 & $-$0.05 $\pm$ 0.20 & $-$0.05 $\pm$ 0.20 & 0.0 $\pm$ 0.45 \\

$10^4$ & range1  & age & 0.0 $\pm$ 0.80 & 0.0 $\pm$ 1.00 & $-$0.1 $\pm$ 0.70 & $-$0.15 $\pm$ 0.65 \\
$10^4$ & range2  & age & 0.0 $\pm$ 0.30 & $-$0.05 $\pm$ 0.80 & $-$0.1 $\pm$ 0.65 & 0.0 $\pm$ 0.40 \\
$10^4$ & range3  & age & 0.15 $\pm$ 0.45 & $-$0.05 $\pm$ 0.35 & $-$0.1 $\pm$ 0.70 & 0.0 $\pm$ 1.15 \\

$10^3$ & range1  & age & $-$0.1 $\pm$ 0.35 & $-$0.55 $\pm$ 2.20 & $-$0.6 $\pm$ 1.15 & $-$0.3 $\pm$ 0.95 \\
$10^3$ & range2  & age & $-$0.1 $\pm$ 0.35 & $-$0.6 $\pm$ 2.15 & $-$0.6 $\pm$ 1.20 & $-$0.05 $\pm$ 1.05 \\
$10^3$ & range3  & age & 0.15 $\pm$ 0.35 & $-$0.5 $\pm$ 0.80 & $-$0.05 $\pm$ 2.20 & 0.15 $\pm$ 1.10 \\
\hline
$10^5$ & range1  & mass & 0.0 $\pm$ 0.13 & 0.0 $\pm$ 0.21 & 0.0 $\pm$ 0.18 & 0.0 $\pm$ 0.26 \\
$10^5$ & range2  & mass & 0.03 $\pm$ 0.13 & 0.0 $\pm$ 0.16 & 0.0 $\pm$ 0.14 & $-$0.01 $\pm$ 0.22 \\
$10^5$ & range3  & mass & 0.09 $\pm$ 0.72 & $-$0.01 $\pm$ 0.19 & 0.0 $\pm$ 0.10 & $-$0.01 $\pm$ 0.24 \\

$10^4$ & range1  & mass & 0.04 $\pm$ 1.18 & $-$0.02 $\pm$ 0.96 & $-$0.08 $\pm$ 0.60 & $-$0.15 $\pm$ 0.76 \\
$10^4$ & range2  & mass & 0.05 $\pm$ 0.97 & $-$0.03 $\pm$ 0.81 & $-$0.07 $\pm$ 0.59 & $-$0.05 $\pm$ 0.53 \\
$10^4$ & range3  & mass & 0.27 $\pm$ 0.84 & 0.0 $\pm$ 0.50 & 0.0 $\pm$ 0.50 & 0.04 $\pm$ 0.56 \\

$10^3$ & range1  & mass & $-$0.39  $\pm$  1.34 & $-$0.75  $\pm$  2.08 & $-$0.49  $\pm$  1.19 & $-$0.33  $\pm$  1.18 \\
$10^3$ & range2  & mass & $-$0.29  $\pm$  1.24 & $-$0.74  $\pm$  2.09 & $-$0.48  $\pm$  1.22 & $-$0.19  $\pm$  1.06 \\
$10^3$ & range3  & mass &  0.22  $\pm$  0.57 & $-$0.63  $\pm$  1.13 & $-$0.28  $\pm$  1.38 & $-$0.09  $\pm$  1.69 \\
\hline
$10^5$ & range2  & $E(B-V)$ & 0.1 $\pm$ 0.20 & 0.1 $\pm$ 0.20 & 0.1 $\pm$ 0.20 & 0.1 $\pm$ 0.20 \\
$10^5$ & range3  & $E(B-V)$ & 0.1 $\pm$ 0.25 & 0.1 $\pm$ 0.20 & 0.15 $\pm$ 0.25 & 0.15 $\pm$ 0.30 \\

$10^4$ & range2  & $E(B-V)$ & 0.05 $\pm$ 0.20 & 0.1 $\pm$ 0.20 & 0.2 $\pm$ 0.20 & 0.0 $\pm$ 0.20 \\
$10^4$ & range3  & $E(B-V)$ & 0.1 $\pm$ 0.40 & 0.2 $\pm$ 0.40 & 0.3 $\pm$ 0.40 & 0.3 $\pm$ 0.70 \\

$10^3$ & range2  & $E(B-V)$ & 0.0 $\pm$ 0.15 & 0.15 $\pm$ 0.20 & 0.2 $\pm$ 0.20 & 0.0 $\pm$ 0.15 \\
$10^3$ & range3  & $E(B-V)$ & 0.05 $\pm$ 0.10  & 0.1 $\pm$ 0.50 & 0.35 $\pm$ 0.35 & 0.2 $\pm$ 0.80 \\
\hline
\end{tabular}
\label{tab:parameter}
\end{center}
\end{table*}

Therefore, we investigate the spread of derived cluster parameters around the input values to quantify this effect. The spread is given based on quantiles. For every model the difference between derived parameter and input parameter is determined (for ages and masses the logarithm of the quantity is used). We consider the parameter difference between the 84.135\% and the 15.865\% quantile as the ``quantile 1$\sigma$'' measure. The median and ``quantile 1$\sigma$'' values for the rederived cluster ages, masses, and foreground extinctions $E(B-V)$ are presented in Table \ref{tab:parameter} for various cluster masses and fitting range restrictions. In Table \ref{tab:uncertainty} the effect of assigned photometric uncertainties is investigated. Median values $>$ 0.0 indicate the rederived parameter being larger than the input value.

The following general trends are observed, based on the SEDs of stochastically-sampled clusters deviating from the model cluster SEDs using a fully-sampled IMF:
\be
\item Fitting results are more widely distributed for decreasing cluster mass, and all cluster parameters are affected;
\item The deviations do not only correspond to a wide spread of fitting results (i.e. ``quantile 1$\sigma$'' values), but the median of the fitting results also deviates from the input values;
\item When fitting parameter ranges are extended, the additional parameter space (i.e. increased ranges for metallicity and foreground extinction) is significantly used for the fitting. However, regarding derived ages and masses the differences between the 3 sets of fitting constraints are not very strong and do not appear like a homogeneous trend. Even knowing and using the correct values for metallicity and extinction does not reduce the uncertainty ranges significantly, the stochasticity is the dominating effect;
\item Increasing photometric uncertainties allow a larger number of stochastic clusters to be fitted with fully-sampled models, although with decreasing fitting accuracy.
\ee

We tested the corresponding derived spread when quantified using the Gaussian standard deviation 1$\sigma$. In general, the derived Gaussian standard deviations were smaller than the equivalent quantile 1$\sigma$. The ``quantile 1$\sigma$'' measure can be applied to much more general cases: Since the derived parameter distributions we studied are significantly non-Gaussian, using a Gaussian standard deviation can lead to misleading results. The ``quantile 1$\sigma$'' can be generally applied to all distributions.

\begin{figure*}
\begin{center}
  \begin{tabular}{ccc}
   \includegraphics[angle = 270,width = 0.3\linewidth]{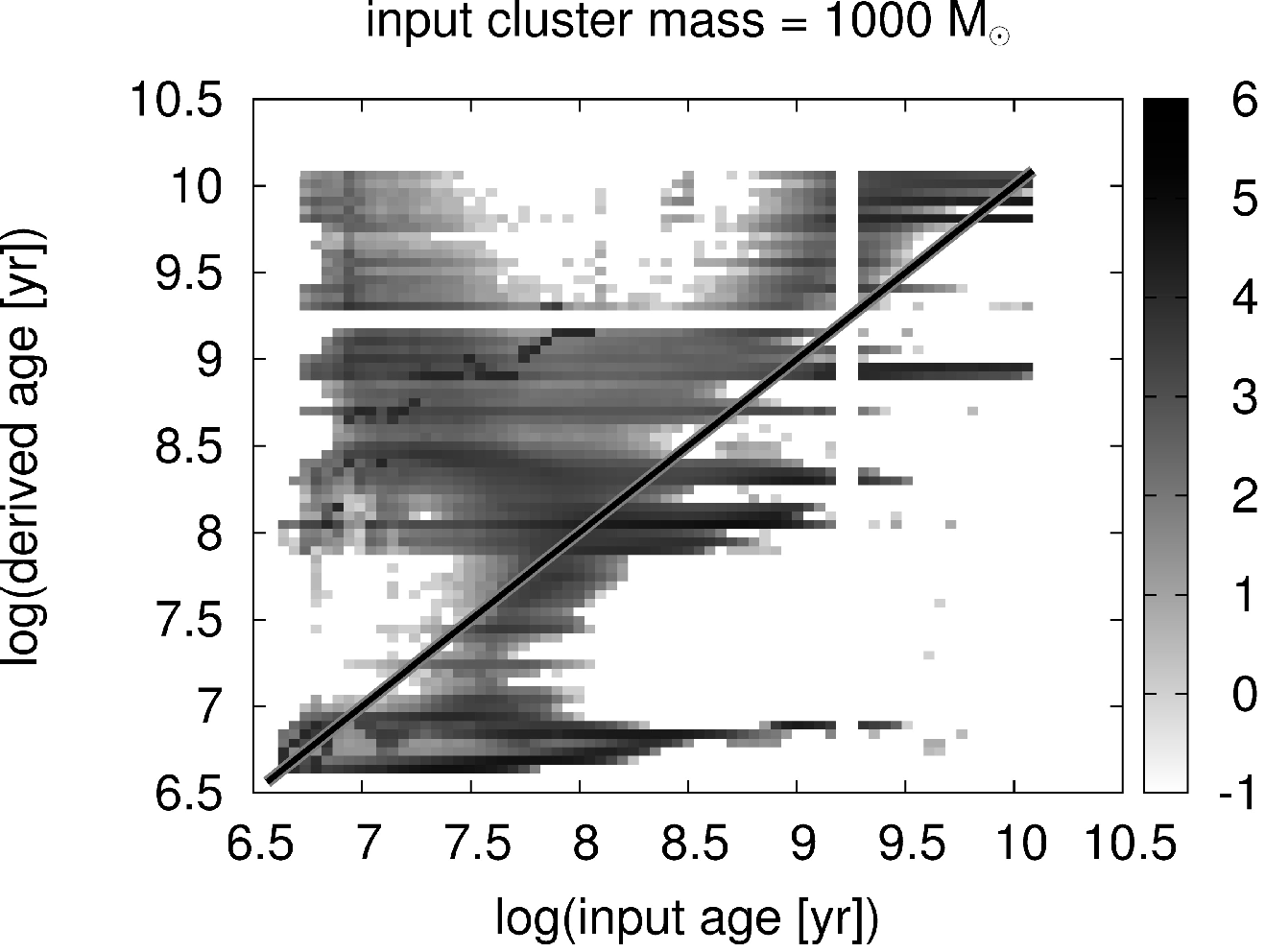} &
   \includegraphics[angle = 270,width = 0.3\linewidth]{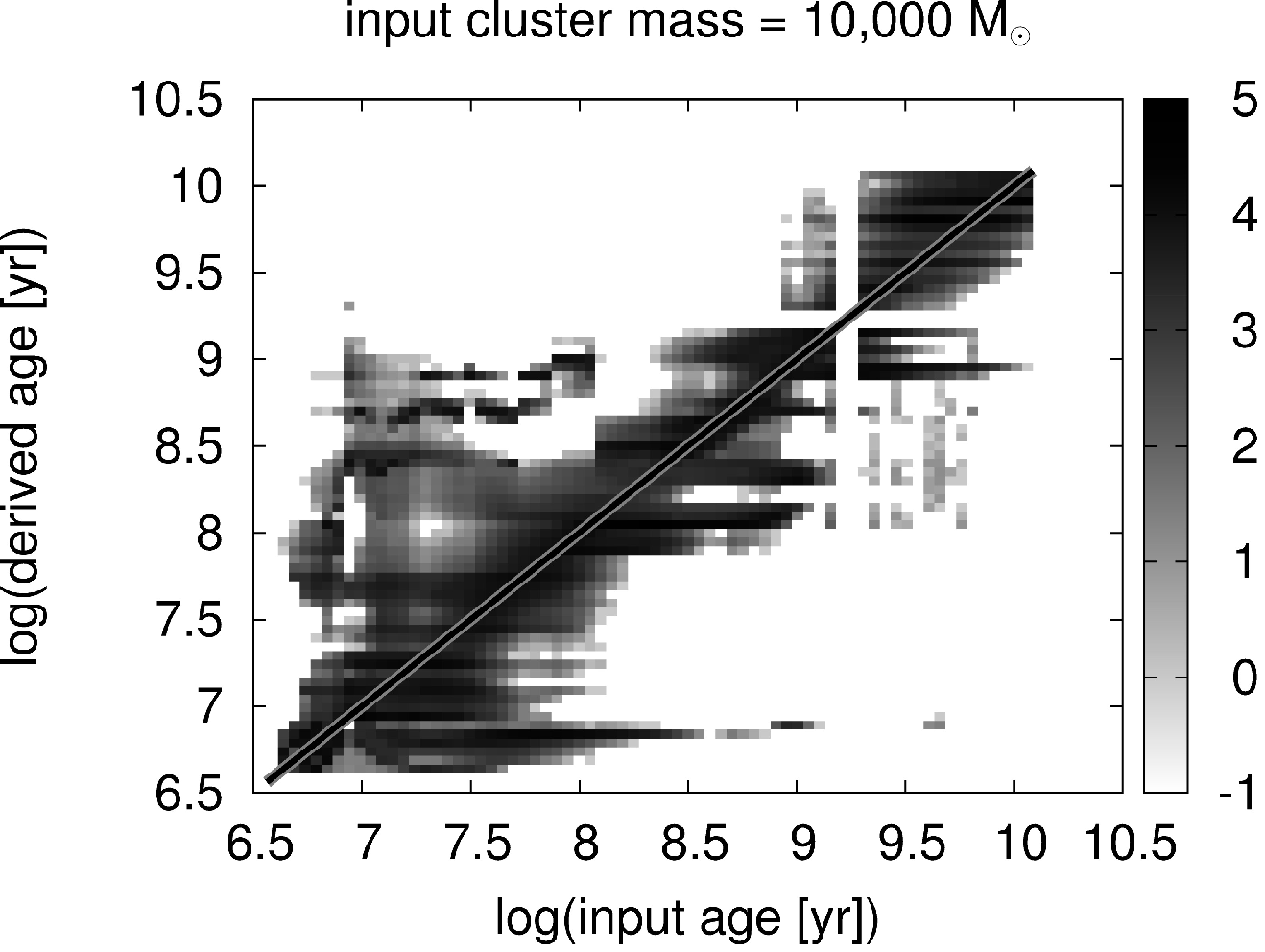} & 
   \includegraphics[angle = 270,width = 0.3\linewidth]{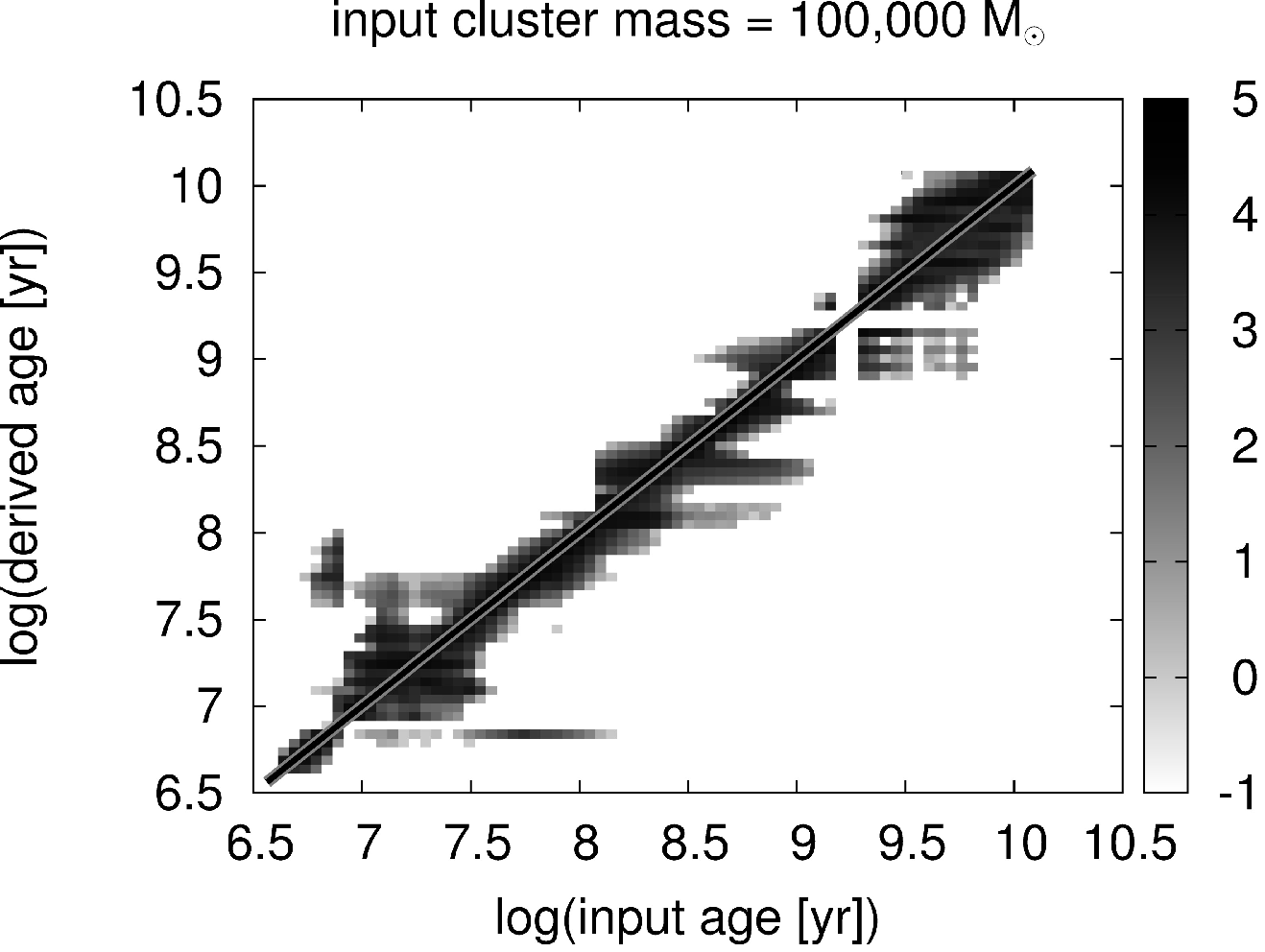} \\
   \includegraphics[angle = 270,width = 0.3\linewidth]{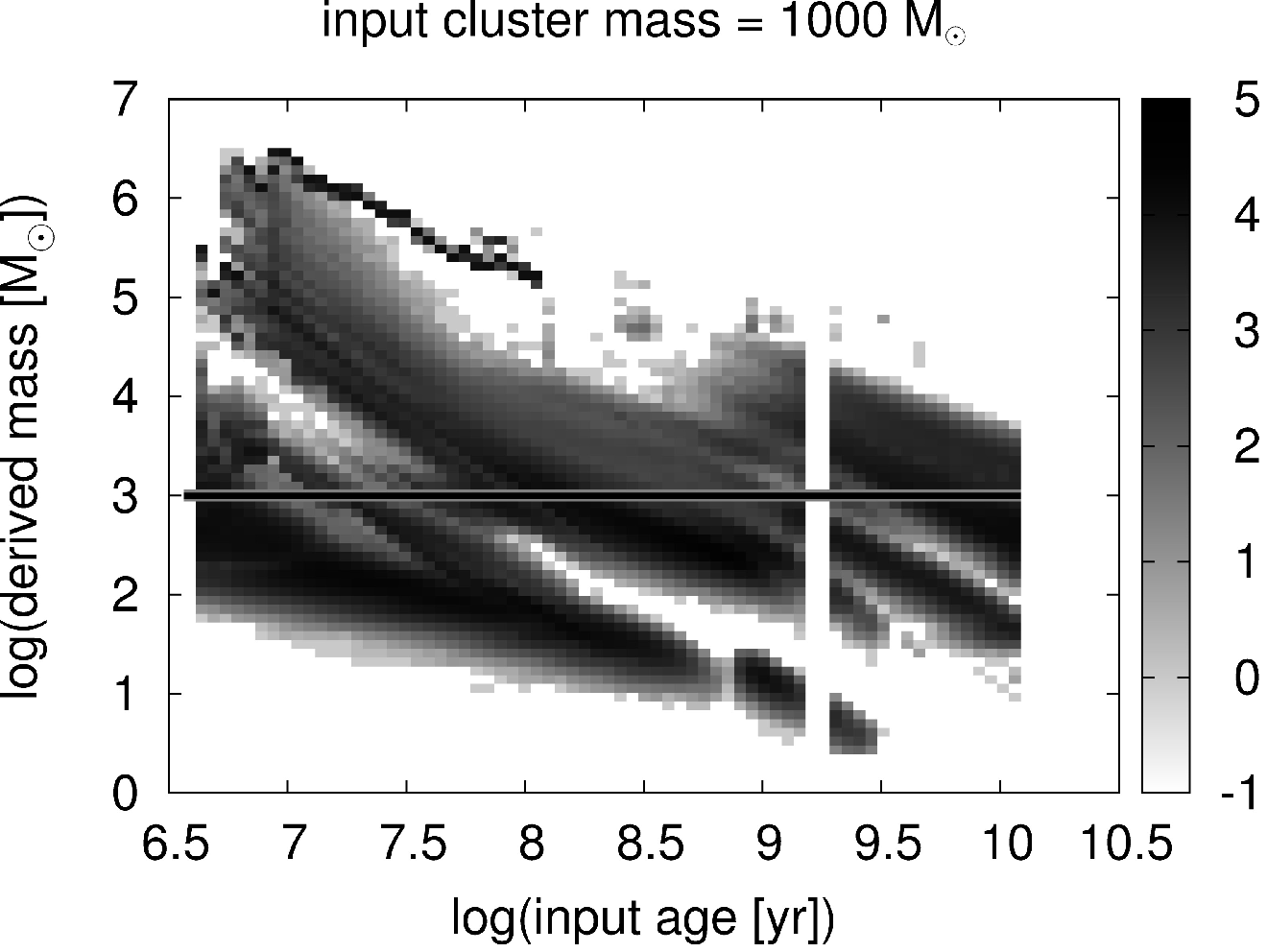} & 
   \includegraphics[angle = 270,width = 0.3\linewidth]{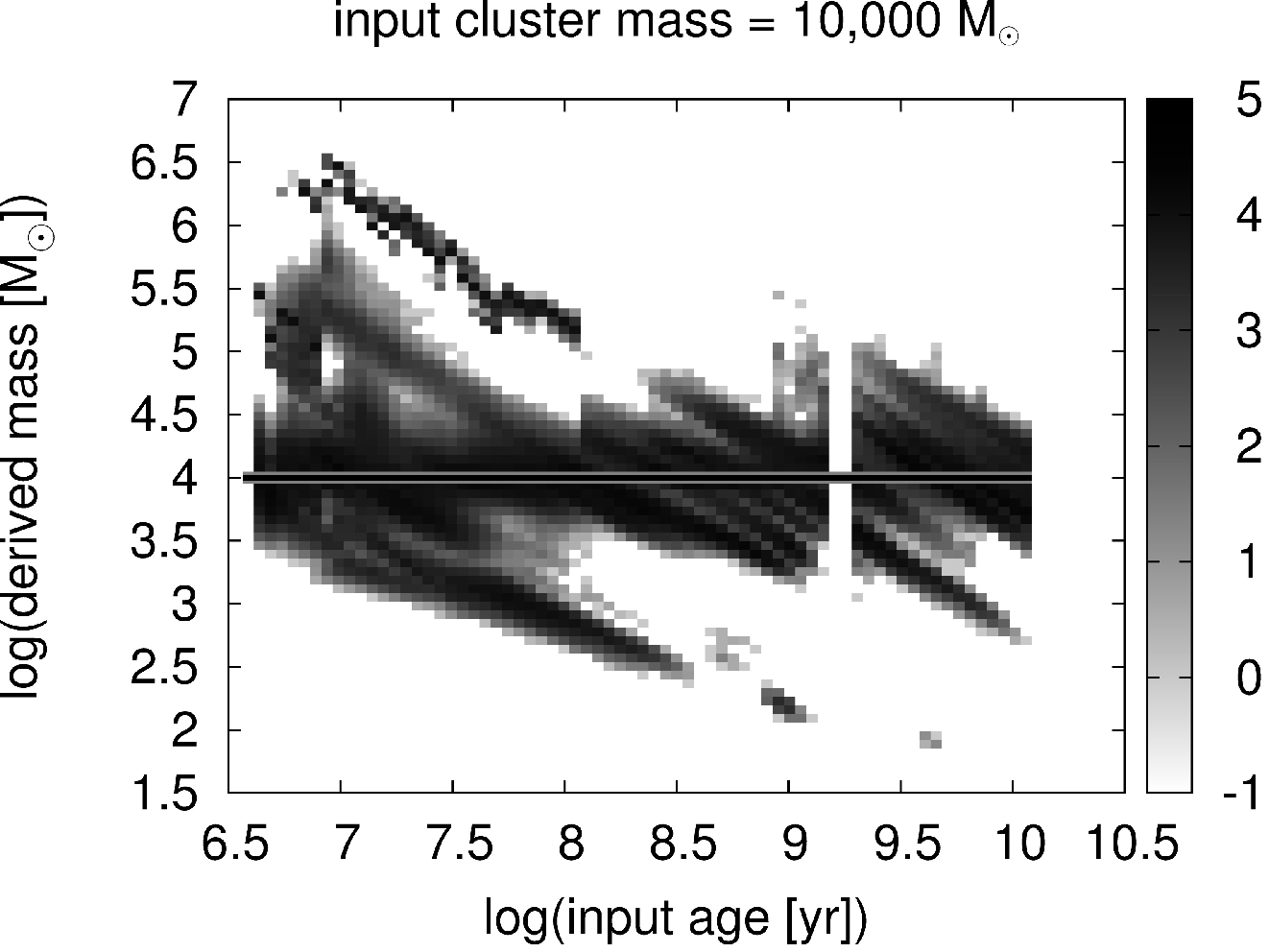} & 
   \includegraphics[angle = 270,width = 0.3\linewidth]{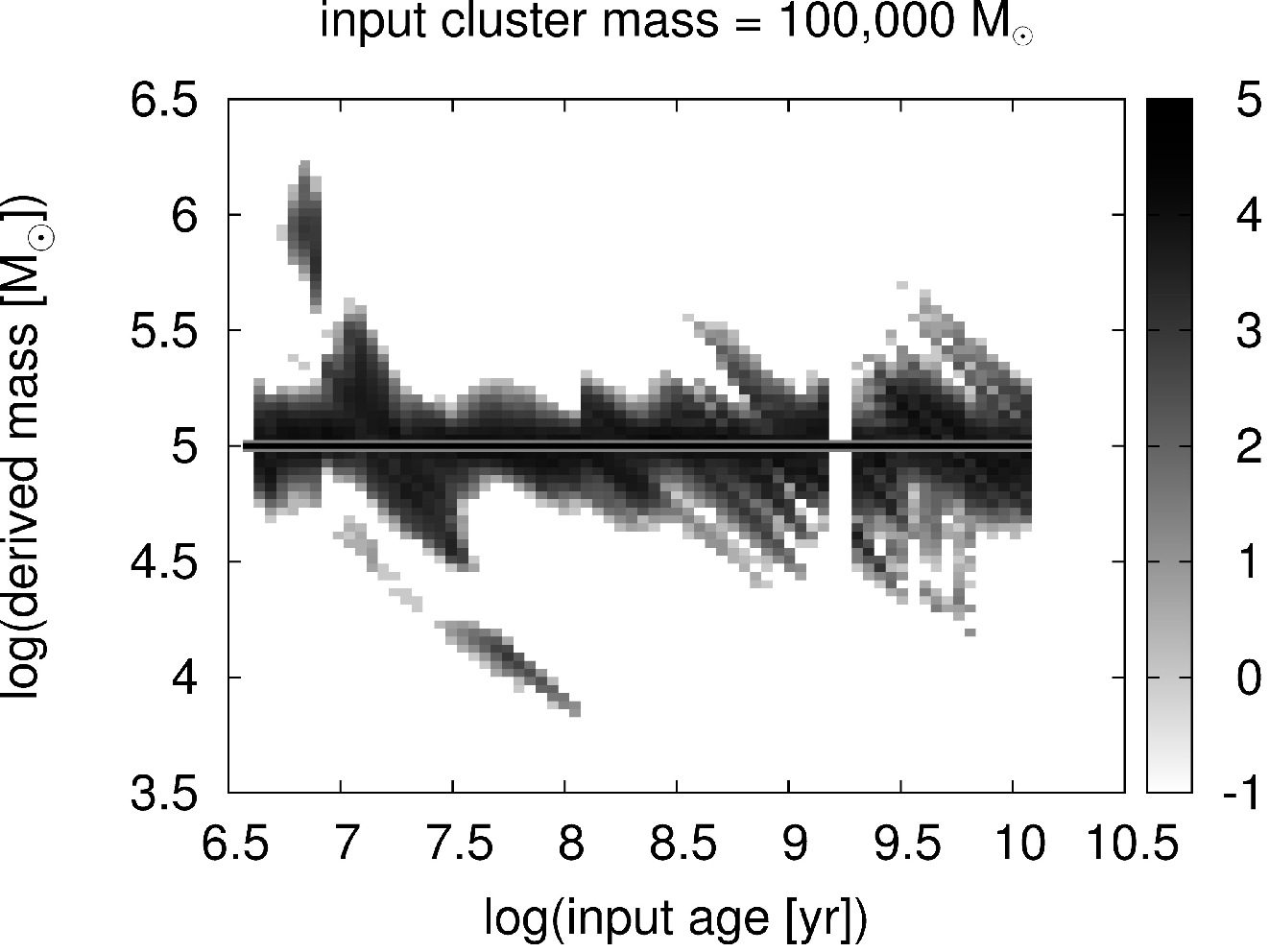} \\
  \end{tabular}
\end{center}
\caption{Examples of fitting accuracy. Shown are cluster samples with $UBVRIJHK$ magnitudes, photometric uncertainties of $\sigma$ = 0.1 mag, solar metallicity, and no foreground extinction. Fitting was done for the narrow parameter range (i.e. $-$0.125 $\le$ \feh $\le$ $+$0.1 dex and 0.0 $\le$ $E(B-V)$ $\le$ 0.2 mag). Presented are the rederived ages (top row) and masses (bottom) for cluster masses of $10^3$ \msunmsun (left column), $10^4$ \msunmsun (middle), and $10^5$ \msunmsun (right). The straight black lines mark the input values.}
\label{fig:firstfits}
\end{figure*}

In Fig. \ref{fig:firstfits} we show the rederived ages (top row) and masses (bottom row) for a range of cluster masses. All clusters have photometry in the standard Johnson/Cousins/Bessell-Brett system with magnitudes in the $UBVRIJHK$ bands, input values of \feh = 0.0, and $E(B-V)$ = 0.0, fitted with the narrow parameter range (\feh = $-$0.125 $-$ $+$0.1 dex, $E(B-V)$ = 0.0 $-$ 0.2 mag). From left to right, the clusters have initial masses of $10^3$ \msun, $10^4$ \msun, and $10^5$ \msun. 
From the stochastic models and the analytic models two ages have been removed: log(age) = 9.2 and log(age) = 9.25 since for these ages the spectra are dominated by a few very bright AGB stars. For these ages all stochastic models roughly coincide for any color, in contrast to all other ages. In addition, the color evolutions show a strong discontinuity for these ages.

Earlier research by \citet{2011A&A...529A..25S} found the following impact of stochastic IMF sampling on derived cluster parameters:
\be
\item For low-mass clusters with masses $\le 10^3$ \msun, they find a wide distribution of wrongly derived ages for stochastic clusters (up to a factor of 100 between input and derived ages);
\item For clusters with masses $> 10^3$ \msun, they find a spread of derived ages around the input ages, with the spread decreasing with increasing cluster mass;
\item They do not perform stochastic tests to quantify the spread of derived parameters.
\ee

These earlier results qualitatively coincide with our results: Derived ages and masses show wide distributions around the input values. This spread becomes smaller with increasing cluster mass. However, even for clusters with initial mass = $10^5$ \msunmsun we find quantile 1$\sigma$ uncertainties for log(age) in the range of 0.1 $-$ 0.3 dex and for log(mass) = 0.13 $-$ 0.22 dex, depending on input age. Systematic offsets are visible for young ages (e.g. derived ages in the range of 100 Myr $-$ 10 Gyr for input ages $<$ 30 Myr), i.e. when few bright stars are at the upper main-sequence or in rapidly evolving states. For clusters with initial masses of $10^3$ \msun, a tendency to underestimate the metallicity is observed, which becomes weaker for clusters with initial masses of $10^4$ \msun. The impact of input cluster metallicity and foreground extinction on the fitting results will be investigated in the next section. The derived extinction is almost uniformly distributed over the full available range.

\begin{table*}
\caption{Cluster parameter determination accuracy based on AnalySED fitting, using $UBVRIJHK$ magnitudes and fully-sampled fitting models. The stochastic clusters were created with cluster masses of $10^4$ \msun, \feh = 0.0 dex, foreground extinction $E(B-V)$ = 0.0 mag, and a range of photometric uncertainties $\sigma$ (for all magnitudes). The values in columns 3 $-$ 6 represent the median and quantile 1$\sigma$ of fitted log(ages) with input log(ages), fitted log(mass) with input log(mass), and fitted $E(B-V)$ with input $E(B-V)$ in four age ranges. Fitting ranges are narrow around input values (\feh =  $-$0.125 $-$ $+$0.1 dex and $E(B-V)$ =  0.0 $-$ 0.2 mag). } 
\begin{center}
\begin{tabular}{l l r r r r }
\hline
$\sigma$ & parameter & age $\le$ 10 Myr & 10 Myr $<$ age $\le$ 100 Myr & 100 Myr $<$ age $\le$ 1 Gyr & age $>$ 1 Gyr \\
\hline
0.001 mag & age & 0.0 $\pm$ 0.05 & 0.0 $\pm$ 0.10     & 0.0 $\pm$ 0.00    & 0.0 $\pm$ 0.00 \\
0.01 mag  & age & 0.0 $\pm$ 0.15 & $-$0.05 $\pm$ 0.35 & $-$0.05 $\pm$ 0.10 & 0.05 $\pm$ 0.35 \\
0.1 mag   & age & 0.0 $\pm$ 0.30 & $-$0.05 $\pm$ 0.80 & $-$0.1 $\pm$ 0.65 & 0.0 $\pm$ 0.40 \\
0.2 mag   & age & 0.0 $\pm$ 0.30 & $-$0.05 $\pm$ 0.85 & 0.0 $\pm$ 0.70    & 0.0 $\pm$ 0.45 \\
\hline
0.001 mag & mass & 0.0 $\pm$ 0.05  & 0.01 $\pm$ 0.13    & $-$0.01 $\pm$ 0.05 & 0.05 $\pm$ 0.06 \\
0.01 mag  & mass & 0.0 $\pm$ 0.30  & $-$0.02 $\pm$ 0.37 & $-$0.01 $\pm$ 0.13 & 0.08 $\pm$ 0.31 \\
0.1 mag   & mass & 0.05 $\pm$ 0.97 & $-$0.03 $\pm$ 0.81 & $-$0.07 $\pm$ 0.59 & $-$0.05 $\pm$ 0.53 \\
0.2 mag   & mass & 0.02 $\pm$ 0.42 & $-$0.07 $\pm$ 0.94 & $-$0.05 $\pm$ 0.76 & $-$0.03 $\pm$ 1.17 \\
\hline
0.001 mag & $E(B-V)$ & 0.0 $\pm$ 0.05  & 0.05 $\pm$ 0.15  & 0.05 $\pm$ 0.10 & 0.1 $\pm$ 0.15 \\
0.01 mag  & $E(B-V)$ & 0.1 $\pm$ 0.15  & 0.15 $\pm$ 0.20  & 0.15 $\pm$ 0.20 & 0.15 $\pm$ 0.20 \\
0.1 mag   & $E(B-V)$ & 0.05 $\pm$ 0.20 & 0.1 $\pm$ 0.20   & 0.2 $\pm$ 0.20  & 0.0 $\pm$ 0.20 \\
0.2 mag   & $E(B-V)$ & 0.0 $\pm$ 0.20  & 0.15 $\pm$ 0.20  & 0.2 $\pm$ 0.20  & 0.05 $\pm$ 0.20 \\
\hline
\end{tabular}
\label{tab:uncertainty}
\end{center}
\end{table*}

\subsection{Cluster parameters and filter combinations}
\label{sec:massesfilters}

For further comparison we concentrate on the fitting results when small parameter ranges around the input parameters are allowed, i.e. $-$0.125 $\le$ \feh $\le$ $+$0.1 dex and 0.0 $\le$ $E(B-V)$ $\le$ 0.2 mag.

\begin{table*}
\caption{Cluster parameter determination accuracy based on AnalySED fitting, using a variety of cluster masses. The stochastic clusters were created with an $UBVRIJHK$ SED, \feh = 0.0 dex, foreground extinction $E(B-V)$ = 0.0 mag, and an assumed photometric uncertainty of $\Delta$mag $ = $ 0.1 mag for all magnitudes. The table contains the same values as in Table \ref{tab:parameter} for four input age ranges, based on the narrow fitting ranges.} 
\begin{center}
\begin{tabular}{l l r r r r }
\hline
cluster mass (\msun) & parameter & age $\le$ 10 Myr & 10 Myr $<$ age $\le$ 100 Myr & 100 Myr $<$ age $\le$ 1 Gyr & age $>$ 1 Gyr \\
\hline
$2 \times 10^5$ & age & 0.00 $\pm$ 0.05    & 0.00 $\pm$ 0.15    & 0.0 $\pm$ 0.10     & 0.0 $\pm$ 0.25 \\ 
$10^5$          & age & 0.00 $\pm$ 0.10    & 0.00 $\pm$ 0.15    & $-$0.05 $\pm$ 0.15 & 0.0 $\pm$ 0.30 \\ 
$5 \times 10^4$ & age & 0.00 $\pm$ 0.20    & 0.0 $\pm$ 0.25     & $-$0.05 $\pm$ 0.20 & 0.0 $\pm$ 0.30 \\
$2 \times 10^4$ & age & 0.00 $\pm$ 0.25    & 0.0 $\pm$ 0.50     & $-$0.05 $\pm$ 0.40 & 0.0 $\pm$ 0.40 \\ 
$10^4$          & age & 0.00 $\pm$ 0.30    & $-$0.05 $\pm$ 0.80 & $-$0.10 $\pm$ 0.65 & 0.0 $\pm$ 0.40 \\ 
$5 \times 10^3$ & age & $-$0.05 $\pm$ 0.25 & $-$0.15 $\pm$ 1.00 & $-$0.30 $\pm$ 0.85 & 0.0 $\pm$ 0.65 \\ 
$2 \times 10^3$ & age & $-$0.10 $\pm$ 0.30 & $-$0.50 $\pm$ 1.15 & $-$0.55 $\pm$ 1.10 & $-$0.05 $\pm$ 0.85 \\ 
$10^3$          & age & $-$0.1 $\pm$ 0.35  & $-$0.60 $\pm$ 2.15 & $-$0.60 $\pm$ 1.20 & $-$0.05 $\pm$ 1.05 \\ 
$10^3$, USML    & age & $-$0.15 $\pm$ 0.35 & $-$0.65 $\pm$ 1.30 & $-$0.60 $\pm$ 1.20 & $-$0.05 $\pm$ 1.05 \\ 
\hline
$2 \times 10^5$ & mass & 0.02 $\pm$ 0.10    & 0.0 $\pm$ 0.13     & 0.0 $\pm$ 0.09     & $-$0.01 $\pm$ 0.17 \\
$10^5$          & mass & 0.03 $\pm$ 0.13    & 0.0 $\pm$ 0.16     & 0.0 $\pm$ 0.14     & $-$0.01 $\pm$ 0.22 \\
$5 \times 10^4$ & mass & 0.03 $\pm$ 0.20    & $-$0.01 $\pm$ 0.22 & 0.00 $\pm$ 0.21    & $-$0.01 $\pm$ 0.28 \\
$2 \times 10^4$ & mass & 0.05 $\pm$ 0.83    & $-$0.01 $\pm$ 0.47 & $-$0.02 $\pm$ 0.38 & $-$0.02 $\pm$ 0.40 \\
$10^4$          & mass & 0.05 $\pm$ 0.97    & $-$0.03 $\pm$ 0.81 & $-$0.07 $\pm$ 0.59 & $-$0.05 $\pm$ 0.53 \\
$5 \times 10^3$ & mass & $-$0.01 $\pm$ 0.62 & $-$0.13 $\pm$ 1.06 & $-$0.25 $\pm$ 0.75 & $-$0.09 $\pm$ 0.71 \\
$2 \times 10^3$ & mass & $-$0.20 $\pm$ 0.85 & $-$0.54 $\pm$ 1.27 & $-$0.42 $\pm$ 1.12 & $-$0.15 $\pm$ 0.91 \\
$10^3$          & mass & $-$0.29 $\pm$ 1.24 & $-$0.74 $\pm$ 2.09 & $-$0.48 $\pm$ 1.22 & $-$0.19 $\pm$ 1.06 \\
$10^3$, USML    & mass & $-$0.31 $\pm$ 0.90 & $-$0.76 $\pm$ 1.42 & $-$0.45 $\pm$ 1.21 & $-$0.17 $\pm$ 1.04 \\
\hline
$2 \times 10^5$ & $E(B-V)$ & 0.10 $\pm$ 0.20 & 0.10 $\pm$ 0.15 & 0.10 $\pm$ 0.20 & 0.10 $\pm$ 0.20 \\
$10^5$          & $E(B-V)$ & 0.10 $\pm$ 0.20 & 0.10 $\pm$ 0.20 & 0.10 $\pm$ 0.20 & 0.10 $\pm$ 0.20 \\
$5 \times 10^4$ & $E(B-V)$ & 0.10 $\pm$ 0.20 & 0.10 $\pm$ 0.20 & 0.10 $\pm$ 0.20 & 0.05 $\pm$ 0.20 \\
$2 \times 10^4$ & $E(B-V)$ & 0.10 $\pm$ 0.20 & 0.10 $\pm$ 0.20 & 0.15 $\pm$ 0.20 & 0.00 $\pm$ 0.20 \\
$10^4$          & $E(B-V)$ & 0.05 $\pm$ 0.20 & 0.10 $\pm$ 0.20 & 0.20 $\pm$ 0.20 & 0.00 $\pm$ 0.20 \\
$5 \times 10^3$ & $E(B-V)$ & 0.00 $\pm$ 0.20 & 0.15 $\pm$ 0.20 & 0.20 $\pm$ 0.20 & 0.00 $\pm$ 0.20 \\
$2 \times 10^3$ & $E(B-V)$ & 0.00 $\pm$ 0.20 & 0.20 $\pm$ 0.20 & 0.20 $\pm$ 0.20 & 0.00 $\pm$ 0.15 \\
$10^3$          & $E(B-V)$ & 0.00 $\pm$ 0.15 & 0.15 $\pm$ 0.20 & 0.20 $\pm$ 0.20 & 0.00 $\pm$ 0.15 \\
$10^3$, USML    & $E(B-V)$ & 0.00 $\pm$ 0.05 & 0.20 $\pm$ 0.20 & 0.20 $\pm$ 0.20 & 0.00 $\pm$ 0.15 \\
\hline
\end{tabular}
\label{tab:parameter}
\end{center}
\end{table*}

\begin{table*}
\caption{Accuracy of cluster parameter determination based on AnalySED fitting, using a variety of SEDs. The stochastic 
clusters were created with solar metallicity, foreground extinction $E(B-V)$ = 0.0 mag, and an assumed photometric uncertainty of $\Delta$mag $ = $ 0.1 mag for all magnitudes. The table contains the same values as in Table \ref{tab:parameter} for four input age ranges, based on the narrow fitting ranges. ``$FUV \rightarrow K$'' is GALEX filters FUV+NUV \& $UBVRIJHK$.}
\begin{center}
\begin{tabular}{l l l r r r r }
\hline
cluster mass (\msun) & parameter & SED & age $\le$ 10 Myr & 10 Myr $<$ age $\le$ 100 Myr & 100 Myr $<$ age $\le$ 1 Gyr & age $>$ 1 Gyr \\
\hline
$10^5$     & age  & $FUV \rightarrow K$        &   0.00    $\pm$ 0.05  &  0.00 $\pm$ 0.10  &  0.00  $\pm$ 0.10    & 0.00 $\pm$ 0.25  \\
$10^5$     & age  & $UBVIK$                    &   0.00    $\pm$ 0.10  &  0.00 $\pm$ 0.15  & $-$0.04  $\pm$ 0.15  & 0.00 $\pm$ 0.30  \\
$10^5$     & age  & $UBVI$                     &   0.00    $\pm$ 0.10  & 0.00 $\pm$ 0.20   &  0.00  $\pm$ 0.05    & $-$0.05 $\pm$ 0.25  \\

$10^4$     & age  & $FUV \rightarrow K$        &   0.00    $\pm$ 0.25  & $-$0.05  $\pm$ 0.40  &  $-$0.05 $\pm$ 0.25  &  0.05 $\pm$  0.25 \\
$10^4$     & age  & $UBVIK$                    &   0.00    $\pm$ 0.30  & $-$0.05  $\pm$ 0.75  &  $-$0.10 $\pm$ 0.65  &  0.00 $\pm$  0.45 \\
$10^4$     & age  & $UBVI$                     &   0.00    $\pm$ 0.35  & $-$0.10  $\pm$ 0.80  & 0.00 $\pm$ 0.24      & $-$0.09 $\pm$  0.45 \\

$10^3$     & age  & $FUV \rightarrow K$        &   $-$0.15    $\pm$ 0.35  & $-$0.60  $\pm$ 1.05  & $-$0.05  $\pm$ 0.90   &  0.10  $\pm$ 0.25  \\
$10^3$     & age  & $UBVIK$                    &   $-$0.10    $\pm$ 0.35  & $-$0.55  $\pm$ 2.11  &  $-$0.60  $\pm$ 1.14  & $-$0.25  $\pm$ 1.00  \\
$10^3$     & age  & $UBVI$                     &   $-$0.10    $\pm$ 0.50  & $-$0.55  $\pm$ 1.95  & $-$0.10  $\pm$ 0.70   & $-$0.25  $\pm$ 0.75  \\
\hline
$10^5$     & mass  & $FUV \rightarrow K$        &  0.02   $\pm$ 0.13  & $-$0.01 $\pm$ 0.11  & 0.00 $\pm$ 0.10  &  0.02 $\pm$ 0.14 \\
$10^5$     & mass  & $UBVIK$                    &  0.02   $\pm$ 0.13  & 0.00 $\pm$ 0.17     & 0.00 $\pm$ 0.14  & $-$0.01 $\pm$ 0.21 \\
$10^5$     & mass  & $UBVI$                     &  0.03   $\pm$ 0.13  & $-$0.01 $\pm$ 0.19  & 0.00 $\pm$ 0.05  & $-$0.02 $\pm$ 0.19 \\

$10^4$     & mass  & $FUV \rightarrow K$        &  0.03   $\pm$ 0.48  & $-$0.04 $\pm$ 0.55  &  $-$0.01 $\pm$ 0.29  &  0.03 $\pm$ 0.23 \\
$10^4$     & mass  & $UBVIK$                    &  0.04   $\pm$ 0.64  & $-$0.04 $\pm$ 0.78  & $-$0.06 $\pm$ 0.59   & $-$0.04 $\pm$ 0.56 \\
$10^4$     & mass  & $UBVI$                     &  0.04   $\pm$ 0.55  & $-$0.09 $\pm$ 0.95  & 0.00 $\pm$ 0.25      & $-$0.04 $\pm$ 0.53 \\

$10^3$     & mass  & $FUV \rightarrow K$        &   $-$0.36   $\pm$ 0.65  & $-$0.72 $\pm$ 1.26  & $-$0.30 $\pm$ 0.78  &  0.08  $\pm$ 0.38 \\
$10^3$     & mass  & $UBVIK$                    &   $-$0.27   $\pm$ 1.29  & $-$0.73 $\pm$ 2.11  & $-$0.47 $\pm$ 1.15  & $-$0.42  $\pm$ 1.15 \\
$10^3$     & mass  & $UBVI$                     &   $-$0.26   $\pm$ 2.32  & $-$0.70 $\pm$ 2.02  & $-$0.08 $\pm$ 0.81  & $-$0.25  $\pm$ 1.06 \\
\hline
\end{tabular}
\label{tab:parameter_addmass}
\end{center}
\end{table*}

\begin{table*}
\caption{Accuracy of cluster parameter determination based on AnalySED fitting, using a variety of cluster metallicities. The stochastic clusters were created with cluster mass of $10^4$ \msun, foreground extinction $E(B-V)$ = 0.0 mag, an assumed photometric uncertainty of $\Delta$mag $ = $ 0.1 mag for all magnitudes, and the narrow fitting range (for \feh = $-$2.3 dex only the next higher metallicity was used). The table contains the same values as in Table \ref{tab:parameter} for four input age ranges.} 
\begin{center}
\begin{tabular}{l l r r r r }
cluster metallicity (dex) & parameter & age $\le$ 10 Myr & 10 Myr $<$ age $\le$ 100 Myr & 100 Myr $<$ age $\le$ 1 Gyr & age $>$ 1 Gyr \\
\hline
$-$2.3      & age &  0.0  $\pm$ 0.30   & $-$0.2 $\pm$ 0.55   & $-$0.2 $\pm$ 1.75   & $-$0.1 $\pm$ 2.60\\
$-$1.7      & age &  0.0  $\pm$ 0.30   & $-$0.15 $\pm$ 0.85  & $-$0.15 $\pm$ 0.85  & 0.05 $\pm$ 0.45 \\
$-$1.0      & age &  0.05  $\pm$ 0.40  & $-$0.1 $\pm$ 0.70   & $-$0.15 $\pm$ 1.65  & 0.0 $\pm$ 0.40 \\
$-$0.4      & age &  0.0  $\pm$ 0.35   & $-$0.1 $\pm$ 1.15   & $-$0.05 $\pm$ 0.65  & $-$0.05 $\pm$ 0.65 \\
\hline
$-$2.3      & mass &  0.05  $\pm$ 0.85  & $-$0.13 $\pm$ 0.60  & $-$0.13 $\pm$ 1.60   & $-$0.09 $\pm$ 2.32 \\
$-$1.7      & mass &  0.06  $\pm$ 0.51  & $-$0.09 $\pm$ 0.90  & $-$0.06 $\pm$ 0.62  & 0.04 $\pm$ 0.52 \\
$-$1.0      & mass &  0.13  $\pm$ 0.91  & $-$0.05 $\pm$ 1.59  & $-$0.06 $\pm$ 1.59  & 0.0 $\pm$ 0.47 \\
$-$0.4      & mass &  0.05  $\pm$ 0.88  & $-$0.06 $\pm$ 1.18  & $-$0.05 $\pm$ 0.55  & $-$0.06 $\pm$ 0.69 \\
\hline
$-$2.3      & $E(B-V)$ & 0.0   $\pm$ 0.15  & 0.05 $\pm$ 0.20 & 0.15 $\pm$ 0.20 & 0.05 $\pm$ 0.20\\
$-$1.7      & $E(B-V)$ & 0.0   $\pm$ 0.15  & 0.1 $\pm$ 0.20  & 0.2 $\pm$ 0.20 & 0.05 $\pm$ 0.20\\
$-$1.0      & $E(B-V)$ & 0.0   $\pm$ 0.20  & 0.15 $\pm$ 0.20 & 0.2 $\pm$ 0.20 & 0.05 $\pm$ 0.20\\
$-$0.4      & $E(B-V)$ & 0.1   $\pm$ 0.20  & 0.15 $\pm$ 0.20 & 0.2 $\pm$ 0.20 & 0.1 $\pm$ 0.20\\
\hline
\end{tabular}
\label{tab:parameter_metallicity}
\end{center}
\end{table*}

\begin{table*}
\caption{Accuracy of cluster parameter determination based on AnalySED fitting, using a variety of cluster foreground extinction. The stochastic clusters were created with cluster mass of $10^4$ \msun, metallicity \feh $=$ 0.0 dex, an assumed photometric uncertainty of $\Delta$mag $ = $ 0.1 mag for all magnitudes, and the narrow fitting range (i.e. \feh =  $-$0.125 $-$ $+$0.1 dex and $E(B-V)$ = $\pm$ 0.15 mag around the input value). The table contains the same values as in Table \ref{tab:parameter} and the fraction of poorly fitted clusters for four input age ranges.}
\begin{center}
\begin{tabular}{l l r r r r }
cluster $E(B-V)$ (mag) & parameter & age $\le$ 10 Myr & 10 Myr $<$ age $\le$ 100 Myr & 100 Myr $<$ age $\le$ 1 Gyr & age $>$ 1 Gyr \\
\hline
0.4  & \% poor fit &  17.3     &  35.5     &  54.6     &  64.5     \\
0.8  & \% poor fit &  48.5     &  48.5     &  54.7     &  58.7     \\
1.2  & \% poor fit &  48.5     &  48.5     &  49.6     &  48.8     \\
1.6  & \% poor fit &  79.6     &  71.2     &  57.9     &  48.5     \\
2.0  & \% poor fit &  99.9     &  100.0    &  95.6     &  56.4     \\
\hline
0.4  & age & 0.10  $\pm$  1.45     & 0.55  $\pm$ 1.90     & 0.35  $\pm$ 0.75     & 0.4  $\pm$ 0.60     \\
0.8  & age & 0.15  $\pm$  1.05     & 0.2  $\pm$ 1.05      & 0.1  $\pm$ 0.60      & 0.1  $\pm$ 0.40     \\
1.2  & age & 0.25  $\pm$  0.70     & 0.2  $\pm$ 0.75      & $-$0.1  $\pm$ 0.60   & $-$0.2  $\pm$ 0.85  \\
1.6  & age & $-$0.10  $\pm$ 0.40   & $-$1.0  $\pm$ 0.85   & $-$1.3  $\pm$ 1.20   & $-$1.35  $\pm$ 1.10 \\
2.0  & age & $-$0.25  $\pm$ 0.10   &    $--$              & $-$2.25  $\pm$ 0.10  & $-$2.55  $\pm$ 1.15 \\
\hline
0.4  & mass & $-$0.43  $\pm$ 1.51   & 0.03  $\pm$ 1.89      & $-$0.13  $\pm$ 0.61   & 0.06  $\pm$ 0.53    \\
0.8  & mass & $-$0.20  $\pm$ 1.34   & $-$0.13  $\pm$ 0.76   & $-$0.15  $\pm$ 0.42   & $-$0.1  $\pm$ 0.37  \\
1.2  & mass & $-$0.10  $\pm$ 1.31   & $-$0.11  $\pm$ 0.49   & $-$0.24  $\pm$ 0.49   & $-$0.34  $\pm$ 0.38 \\
1.6  & mass & 0.39  $\pm$ 0.38      & $-$0.58  $\pm$ 0.74   & $-$0.82  $\pm$ 1.08   & $-$0.67  $\pm$ 0.75 \\
2.0  & mass & 1.09  $\pm$ 0.20      &  $--$                 & $-$0.90  $\pm$ 0.13   & $-$1.3  $\pm$ 1.17  \\
\hline
0.4  & $E(B-V)$ & 0.05  $\pm$  0.25    &  0.1 $\pm$ 0.25       & 0.1  $\pm$ 0.25      & 0.1  $\pm$ 0.05     \\
0.8  & $E(B-V)$ & 0.1   $\pm$  0.15    &  0.1 $\pm$ 0.10       & 0.15  $\pm$ 0.25     & 0.15  $\pm$ 0.10    \\
1.2  & $E(B-V)$ & $-$0.1  $\pm$  0.15  &  $-$0.05 $\pm$ 0.10   & 0.0  $\pm$ 0.20      & 0.0  $\pm$ 0.20     \\
1.6  & $E(B-V)$ & 0.1  $\pm$  0.00     &  $-$0.1  $\pm$ 0.00   & $-$0.1  $\pm$ 0.00   & $-$0.05  $\pm$ 0.10 \\
2.0  & $E(B-V)$ & $-$0.2  $\pm$  0.00  &  $--$                 & $-$0.2  $\pm$ 0.00   & $-$0.2  $\pm$ 0.0   \\
\hline
\end{tabular}
\label{tab:parameter_extinction}
\end{center}
\end{table*}

Table \ref{tab:parameter} provides the resulting parameter spreads from fitting clusters with a range of cluster masses. In general, the spread of derived parameters around the input values as well as the median offsets become smaller with increasing cluster mass. Increasing the cluster mass from $10^3$ \msunmsun to $10^5$ \msunmsun reduces the offset by approx. a factor 10 and the spread by approx. a factor of 5 for the derived log(age) and log(mass). However, even for the highest cluster mass investigated (i.e. $2 \times 10^5$ \msun) a non-negligible spread is found: Quantile 1$\sigma$ spreads found are 0.05 $-$ 0.25 dex for log(age) and 0.09 $-$ 0.17 dex for log(mass). These results are shown in Fig. \ref{fig:firstfits}.

Table \ref{tab:parameter_addmass} shows the impact of the filter combination used on the accuracy of derived parameters. This investigation was done for clusters with total masses of $10^3$ \msunmsun (without USML), $10^4$ \msun, and $10^5$ \msun. We have extended the tests using $UBVRIJHK$ (see Table \ref{tab:parameter}) by adding UV filters of the GALEX satellite ($FUV$ and $NUV$) as well as restricting to $UBVIK$ or $UBVI$. In general, we find the ``$FUV \rightarrow K$'' filter combination to give the tightest parameter distributions. Comparing $UBVIK$ and $UBVI$ fitting results shows small differences for ages $\le$ 100 Myr. However, for older ages, use of $UBVIK$ gives wider distributions and larger median offsets from the input values compared to using $UBVI$. For ages $\ge$ 100 Myr, stellar populations contain some AGB stars, which are bright in the near-infrared (e.g. in the $K$-band). The number of bright AGB stars is significantly affected by stochastic IMF sampling, and hence the stochastic IMF sampling affects the integrated $K$-band magnitude significantly (see e.g. Figs. \ref{fig:example_mags} and \ref{fig:example_mags2} for the time evolution of integrated $K$-band magnitudes). In summary, we conclude that both stellar masses and ages for low-mass clusters (i.e. $10^3$ \msun) are hard to determine more accurately than to within 0.5 $-$ 0.7 dex, largely independent of the choice of filters. Exceptions are the derived ages for input ages $<$ 10 Myr (when the models evolve slowly or are dominated by red supergiants) and for input ages $>$ 1 Gyr using the $FUV \rightarrow K$ SED (when clusters have very low emission in the GALEX FUV \& NUV filters).

Table \ref{tab:parameter_metallicity} shows the accuracy of derived parameters for a range in cluster metallicity: \feh = $-$2.3, $-$ 1.7, $-$1.0, and $-$0.4 dex. This investigation was done for clusters with total masses of $10^4$ \msun, an SED using $UBVRIJHK$ magnitudes, and $E(B-V)$ = 0.0. The narrow fitting range was used, except for \feh = $-$2.3 dex (only \feh = $-$2.3 and \feh = $-$2.0 dex were used, as \feh = $-$2.3 dex is the lowest metallicity). The spread for the derived parameters changes inhomogeneously for the cluster metallicities, age ranges, and parameters investigated. Only clusters with input ages $>$ 1 Gyr and \feh = $-$2.3 dex seem to have a consistently larger spread, compared with other parameter combinations. Therefore, the full range of modeled cluster metallicities is provided.

We tested the impact of varying the foreground extinction of the stochastically-sampled clusters on the derived cluster ages and masses, presenting the results in Table \ref{tab:parameter_extinction}. The stochastic clusters were created with cluster mass of $10^4$ \msun, metallicity \feh $=$ 0.0 dex, and a narrow fitting range (i.e. \feh =  $-$0.125 $-$ $+$0.1 dex and $E(B-V)$ = $\pm$ 0.15 mag around the input value) was used. The foreground extinction $E(B-V)$ was varied between 0.4 mag and 2.0 mag. The fraction of poorly fitted clusters increases significantly with increasing foreground extinction. At $E(B-V)=$ 2.0 mag only a small fraction is fitted accurately (i.e. $<$ 5 \% for ages $\le$ 1 Gyr, $<$ 45 \% for ages larger than 1 Gyr). Both median values and quantile 1$\sigma$ uncertainties deviate from the results without foreground extinction, resulting in larger offsets from the input values.

\section{Conclusions}
\label{sec:conclusions}

Stochastic effects to sample the stellar IMF realistically have not yet been sufficiently taken into account for population synthesis modelling. As we have shown in this paper, the impact on color predictions and parameter determinations using stellar cluster models with a fully-sampled IMF is highly significant. The strongest impact on model predictions is seen for parameter ranges and magnitudes/colors, dominated by a small number of bright stars. For young clusters (age $\le$ 15 Myr) such stars are supergiant stars, while for older clusters (100 Myr $\le$ age $\le$ 1 Gyr) the light is dominated by AGB stars. In both cases, the strongest impact is seen in the near-infrared. In addition, the cumulative effects of deviations of the stochastic IMF from the underlying fully-sampled IMF cause more general deviations.

The effect on cluster parameter determination is a clear function of cluster mass and parameter fitting techniques: the impact of stochastic IMF sampling increases with a decreasing number of stars dominating the light. In general, this is related to the total cluster mass and the choice of filters used, as filters are affected by varying amounts (e.g. depending on the stellar population and cluster age).

Due to the large number of models provided, classical techniques of fitting observations (such as determining the $\chi^2$ of all model SEDs compared with the SED of an observed cluster) become unfeasible. For an observed cluster, trying to estimate the cluster parameters that would give rise to such an SED is a challenging problem for a number of reasons. The data do not follow a well-defined distribution and there may be many modes in the range of the parameters. Our future work will try to tackle this issue by applying the method of \citet{PriebeMarquette}. They described a semi-parametric approach to data which builds on the advantages of both mixture model analysis and kernel density estimation. We propose restricting the range of possible parameters through modeling the distributions using this technique, so that an observer could better estimate the range of parameters which would lead to such a cluster.

Our first application to stellar cluster observations are the {\it HST} observations of NGC 5253. Analysis of these clusters using model SEDs with a fully-sampled IMF has been presented in \citet{2013MNRAS.431.2917D}, which indicated cluster masses prone to stochastic IMF sampling. Therefore, we will re-analyze the observational data using the stochastic models presented in this paper (de Grijs et al., {\sl in prep.}).

All model predictions for the range of cluster masses, ages, metallicities \feh, foreground extinction $E(B-V)$, and filters presented are available at\\ http$://$data.galev.org$/$models$/$anders13. The web-page also contains the scripts to produce plots of density distributions as well as quantiles for retrieved magnitudes and colors. Such plots are essential to judge the reliability of cluster analyses for the parameter range and the colors the user chooses.

{\bf Acknowledgments} 

P.A. and R.d.G. acknowledge funding by the National Natural Science Foundation of China (NSFC, grant number 11073001). P.A. acknowledges funding by the China Postdoctoral Science Foundation. R.K. gratefully acknowledges Financial support from STScI theory grant HST-AR-12840.01-A. Support for Program number HST-AR-12840.01-A was provided by NASA through a grant from the Space Telescope Science Institute, which is operated by the Association of Universities for Research in Astronomy, Incorporated, under NASA contract NAS5-26555. P.A. wishes to thank S. Gao and J.-Z. Zhao for technical support, N. Bissantz and E. Silva-Villa for helpful discussions.

\bibliographystyle{apj}
\bibliography{StochasticClusters}

\end{document}